\documentclass[fleqn,10pt,twocolumn]{wlscirep}
\usepackage[utf8]{inputenc}
\usepackage[T1]{fontenc}
\usepackage{subfig}
\usepackage{caption}
\usepackage{mwe} 
\usepackage{amsmath}
\usepackage{soul}
\usepackage{amssymb}
\usepackage{bm}
\usepackage{ulem}
\usepackage{makecell}
\usepackage[version=4]{mhchem}
\usepackage{multirow}
\usepackage{tikz}

\usepackage[page]{appendix}

\let\oldAA\AA
\renewcommand{\AA}{\text{\normalfont\oldAA}}

\newcommand{\fig}{Fig.}
\newcommand{\Fig}{Figure}
\newcommand{\figref}[1]{\fig~\ref{#1}}
\newcommand{\figsref}[1]{Figs.~\ref{#1}}
\newcommand{\Figref}[1]{\Fig~\ref{#1}}
\newcommand{\Figsref}[1]{Figures~\ref{#1}}

\newcommand{\tabref}[1]{Table~\ref{#1}}
\newcommand{\Tabref}[1]{Table~\ref{#1}}

\renewcommand{\eqref}[1]{Eq.~(\ref{#1})}

\DeclareMathOperator*{\argmax}{arg\,max}

\graphicspath{{figures/}{figures/}}

\makeatletter
\def\input@path{{figures/}}
\makeatother

\usepackage[switch]{lineno} 

\title{Uncertainty-biased molecular dynamics for learning uniformly accurate interatomic potentials}

\author[1, 2, $^\ast$]{Viktor Zaverkin}
\author[3, 4]{David Holzm{\"u}ller}
\author[1]{Henrik Christiansen}
\author[1]{Federico Errica}
\author[1]{Francesco Alesiani}
\author[1]{Makoto Takamoto}
\author[1, 5]{Mathias Niepert}
\author[2]{Johannes K{\"a}stner}

\affil[1]{NEC Laboratories Europe GmbH, Kurf{\"u}rsten-Anlage 36, 69115 Heidelberg, Germany}
\affil[2]{Institute for Theoretical Chemistry, University of Stuttgart, Pfaffenwaldring 55, 70569 Stuttgart, Germany}
\affil[3]{Institute for Stochastics and Applications, University of Stuttgart, Pfaffenwaldring 57, 70569 Stuttgart, Germany}
\affil[4]{SIERRA, INRIA Paris, 2 rue Simone Iff, 75012 Paris, France}
\affil[5]{Institute for Artificial Intelligence, University of Stuttgart, Universit{\"a}tsstra{\ss}e 32, 70569 Stuttgart, Germany}
\affil[$\ast$]{viktor.zaverkin@neclab.eu}

\begin{abstract}
	Efficiently creating a concise but comprehensive data set for training machine-learned interatomic potentials (MLIPs) is an under-explored problem. Active learning, which uses biased or unbiased molecular dynamics (MD) to generate candidate pools, aims to address this objective. Existing biased and unbiased MD-simulation methods, however, are prone to miss either rare events or extrapolative regions---areas of the configurational space where unreliable predictions are made. This work demonstrates that MD, when biased by the MLIP's energy uncertainty, simultaneously captures extrapolative regions and rare events, which is crucial for developing uniformly accurate MLIPs. Furthermore, exploiting automatic differentiation, we enhance bias-forces-driven MD with the concept of bias stress. We employ calibrated gradient-based uncertainties to yield MLIPs with similar or, sometimes, better accuracy than ensemble-based methods at a lower computational cost. Finally, we apply uncertainty-biased MD to alanine dipeptide and MIL-53(Al), generating MLIPs that represent both configurational spaces more accurately than models trained with conventional MD.
\end{abstract}

\begin{document}

\flushbottom
\maketitle

\thispagestyle{empty}

\section*{Introduction \label{sec:intro}}

Computational techniques are invaluable for exploring complex configurational and compositional spaces of molecular and material systems. The accuracy and efficiency, however, depend on the chosen computational methods. Ab initio molecular dynamics (MD) simulations using density-functional theory (DFT) provide accurate results but are computationally demanding. Atomistic simulations with classical force fields offer a faster alternative but often lack accuracy. Thus, developing accurate and computationally efficient interatomic potentials is a key challenge successfully addressed by machine-learned interatomic potentials (MLIPs).\cite{Butler2018, Smith2020, Chanussot2021, Xie2023, Gubaev2023} An essential component of any MLIP is the accurate encoding of the atomic system by a local representation, which depends on configurational (atomic positions) and compositional (atomic types) degrees of freedom.\cite{Langer2022} Recently, a wide range of MLIPs have been introduced, comprising linear and kernel-based models,\cite{Rupp2012, Faber2018, Shapeev2016, Drautz2019} Gaussian approximation,\cite{Bartok2010, Bartok2013} and neural network (NN) interatomic potentials,\cite{Behler2007, Artrith2016, Smith2017, Zaverkin2020, Zaverkin2021b} including graph NNs,\cite{Schuett2017, Schuett2021, Batzner2022, Batatia2022, Gasteiger2022gemnet, Liao2023equiformerv2, Passaro2023} all demonstrating remarkable success in atomistic simulations.

The effectiveness of MLIPs, however, crucially relies on training data sufficiently covering configurational and compositional spaces.\cite{Friederich2021, Unke2021} Without such training data, MLIPs cannot faithfully reproduce the underlying physics. An open challenge, therefore, is the generation of comprehensive training data sets for MLIPs, covering relevant configurational and compositional spaces and ensuring that resulting MLIPs are uniformly accurate across these spaces. This objective must be realized while reducing the number of expensive DFT evaluations, which provide reference energies, atomic forces, and stresses. This challenge is further complicated by the limited knowledge of physical conditions, such as temperature and pressure, at which configurational changes occur. Setting temperatures and pressures excessively high can result in atomic system degradation before exploring the relevant phase space.

\begin{figure*}[t!]
	\centering
	\includegraphics[width=\textwidth]{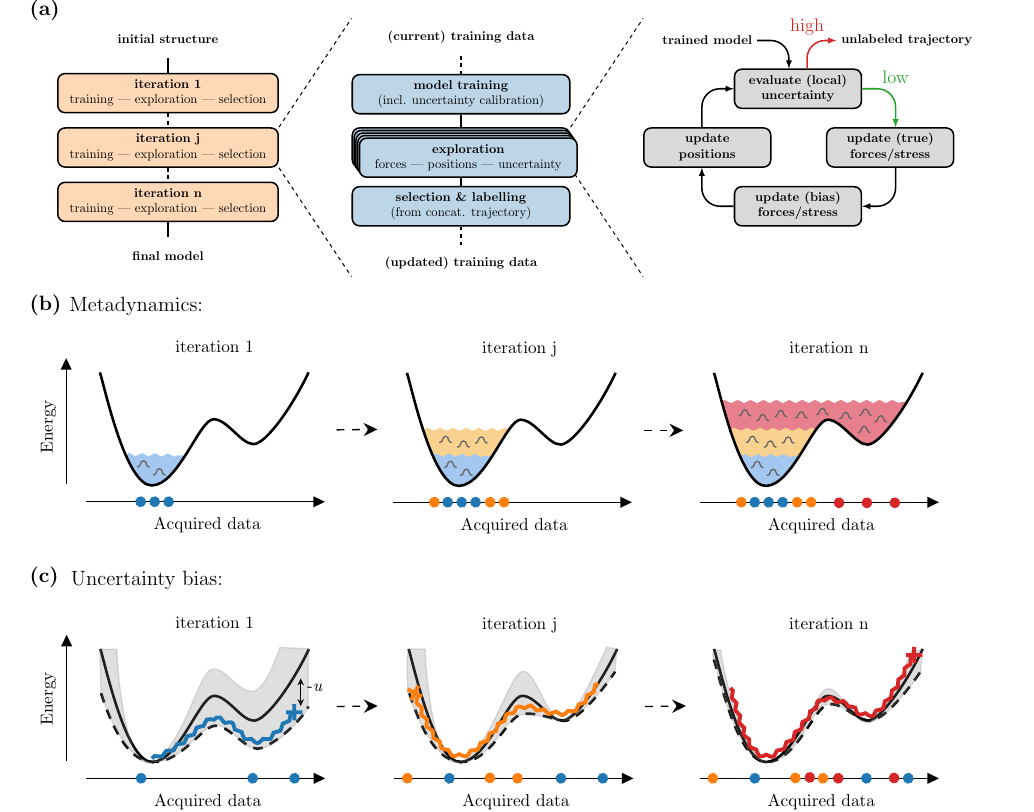}
	\caption{A schematic overview of an AL algorithm for MLIP training. Training structures are selected from data gathered during biased or unbiased MD simulations. \textbf{(a)} An AL experiment begins with training an MLIP in the first iteration using a small set of randomly perturbed initial configurations. The current MLIP is employed in each iteration to run parallel MD simulations. Each simulation continues until it reaches a predefined uncertainty threshold. Then, a batch of configurations is selected from all trajectories. Reference energies and forces of these samples are evaluated using a DFT solver, updating the training data set. The updated data set is employed for training the MLIP in the next iteration. \textbf{(b)} Adaptive biasing strategies like metadynamics enhance the exploration of the configurational space. In metadynamics, exploration along manually defined CVs is facilitated by adding Gaussian functions to a history-dependent bias (areas filled by blue, orange, and red colors). However, even for well-defined CVs, exploring the configurational space of interest may require long simulation times due to the diffusive motion along these CVs. \textbf{(c)} Uncertainty-biased MD aims to minimize uncertainty $u$ (grey shaded area) related to the actual error, thereby facilitating the exploration of the configurational space. In uncertainty-biased MD, we subtract the MLIP's energy uncertainty from the predicted energy (continuous black line) and run MD simulations using the altered energy surface (dashed black line). Curved lines denote distinct MD trajectories. Unlike metadynamics, uncertainty-biased MD operates without defining CVs and drives MD simulations toward high uncertainty regions in each iteration.}
	\label{fig:scheme}
\end{figure*}

To address this challenge, iterative active learning (AL) algorithms are used to improve the accuracy of MLIPs by providing an augmented data set;\cite{Li2015, Podryabinkin2017, Gastegger2017, ZhangLinfeng2019, Vandermause2020, Shuaibi2021, Briganti2023, Wang2023finetuna} see \figref{fig:scheme} (a). They select the data most informative to the model, i.e., atomic configurations with high energy and force uncertainties, as estimated by the model. This data is drawn from configurational and compositional spaces explored during, e.g., MD simulations. Reference DFT energies, atomic forces, and stresses are evaluated for the selected configurations. Furthermore, energy and force uncertainties indicate the onset of extrapolative regions---regions where unreliable predictions are made---prompting the termination of MD simulations and the evaluation of reference DFT values. In this AL setting, covering the configurational space and exploring extrapolative configurations might require running longer MD simulations and defining physical conditions for observing slow configurational changes (rare events).

Alternatively, enhanced sampling methods can significantly speed up the exploration of the configurational space by using adaptive biasing strategies such as metadynamics;\cite{hub94, Laio2002, Barducci2008, Demuynck2017, Yoo2021, Yang2022, Vandenhaute2023} see \figref{fig:scheme} (b). However, metadynamics requires manually selecting a few collective variables (CVs) that are assumed to describe the system. The limited number of CVs restricts exploration, as they might miss relevant transitions and parts of the configurational space. In contrast, MD simulations biased toward regions of high uncertainty can enhance the discovery of extrapolative configurations.\cite{Kulichenko2023, vanerOord2022} A related work utilizes uncertainty gradients for adversarial training of MLIPs.\cite{Schwalbe-Koda2021, Carrete2023} To obtain MLIPs that are uniformly accurate across the relevant configurational space, however, simultaneous exploration of rare events and extrapolative configurations is necessary. The extent to which uncertainty-biased MD can achieve this objective remains an unexplored research area.

This work demonstrates the capability of uncertainty-biased MD to explore the configurational space, including fast exploration of rare events and extrapolative regions; see \figref{fig:scheme} (c). We achieve this by exploring the CVs of alanine dipeptide---a widely used model for protein backbone structure. To assess the coverage of the CV space, we introduce a measure using a tree-based weighted recursive space partitioning. Furthermore, we extend existing uncertainty-biased MD simulations by automatic differentiation (AD) and propose a biasing technique that utilizes bias stresses obtained by differentiating the model's uncertainty with respect to infinitesimal strain deformations. We assess the efficiency of the proposed technique by running MD simulations in isothermal–isobaric ($NpT$) statistical ensemble and exploring cell parameters of MIL-53(Al)---a flexible metal-organic framework (MOF) featuring closed- and large-pore stable states. Both benchmark systems are often used in studies assessing enhanced sampling and data generation methods.\cite{Laio2002, Demuynck2017, Schwalbe-Koda2021, Vandenhaute2023}

A key ingredient of AL algorithms with dynamically generated candidate pools is a sensitive metric for detecting the onset of extrapolative regions. These regions are typically associated with large errors in MLIP predictions. However, MLIP uncertainties often underestimate actual errors,\cite{Kuleshov2018, Pernot2022} resulting in the exploration of unphysical regions, negatively affecting MLIP training. Thus, calibrated uncertainties are crucial for generating high-quality MLIPs with AL, which involves configurations explored during MLIP-based MD,\cite{Tran2020, Hu2022, Pernot2022} but might be unnecessary in AL tasks that rely on relative uncertainties.\cite{Zaverkin2021a, Zaverkin2022d, Holzmueller2022} In our setting, we demonstrate that conformal prediction (CP) helps align the largest force error with its corresponding uncertainty value. This approach effectively makes MLIPs not underestimate force errors, which is important for preventing MD from exploring unphysical configurations. Thus, CP-based uncertainty calibration helps set reasonable uncertainty thresholds without limiting the exploration of the configurational space. In contrast, conventional approaches drive MD away from high-uncertainty regions, which can hinder exploration.\cite{Schran2020b}

Contrary to existing work,\cite{Kulichenko2023, vanerOord2022} which relies on ensembles of MLIPs for uncertainty quantification, we propose using ensemble-free uncertainties of NN-based MLIPs derived from gradient features.\cite{Zaverkin2021a, Zaverkin2022d, Holzmueller2022} These features can be interpreted as the sensitivity of a model's output to parameter changes. Recent studies demonstrate that gradient-based uncertainties perform comparably to ensemble-based counterparts in AL.\cite{Zaverkin2022d, Holzmueller2022, Kirsch2023} Furthermore, they yield the exact posterior in the case of linear models.\cite{Shapeev2016, Drautz2019} We demonstrate that gradient features can define uncertainties of total and atom-based properties, such as energy and atomic forces. To make gradient-based uncertainties computationally efficient, we employ the sketching technique\cite{Woodruff2014} and reduce the dimensionality of gradient features. For many NN-based MLIPs, gradient-based approaches can significantly reduce the computational cost of uncertainty quantification and accelerate the time-consuming MD simulations compared to ensemble-based methods. However, the latter can be made computationally efficient, e.g., through parallelization or employing specific settings with non-trainable descriptors and gradient-free force uncertainties.\cite{Carrete2023}

We further enhance configurational space exploration and improve the computational efficiency of AL by employing batch selection algorithms.\cite{Zaverkin2022d, Holzmueller2022} These algorithms simultaneously select multiple atomic configurations from trajectories generated during parallel MD simulations. Batch selection algorithms enforce the informativeness and diversity of the selected atomic structures. Thus, they ensure the construction of maximally diverse training data sets.

\section*{Results} \label{sec:results}

\subsection*{Overview}

In the following, we first demonstrate the necessity of uncertainty calibration on an example of MIL-53(Al) to constrain MD to physically reasonable regions of the configurational space. Then, we present two complementary analyses demonstrating the improved data efficiency of MLIPs obtained by our AL approach, developing MLIPs for alanine dipeptide and MIL-53(Al). Furthermore, we investigate how uncertainty-biased MD enhances the exploration of the configurational space, utilizing bias forces and stress. To benchmark our results, we draw a comparison with MD run at elevated temperatures and pressures as well as metadynamics simulations. The details on the ensemble-free uncertainties (distance- and posterior-based ones) and uncertainty-biased MD can be found in Methods.

\subsection*{Calibrating uncertainties with conformal prediction}

\begin{figure*}[t!]
	\centering
	\includegraphics[width=\textwidth]{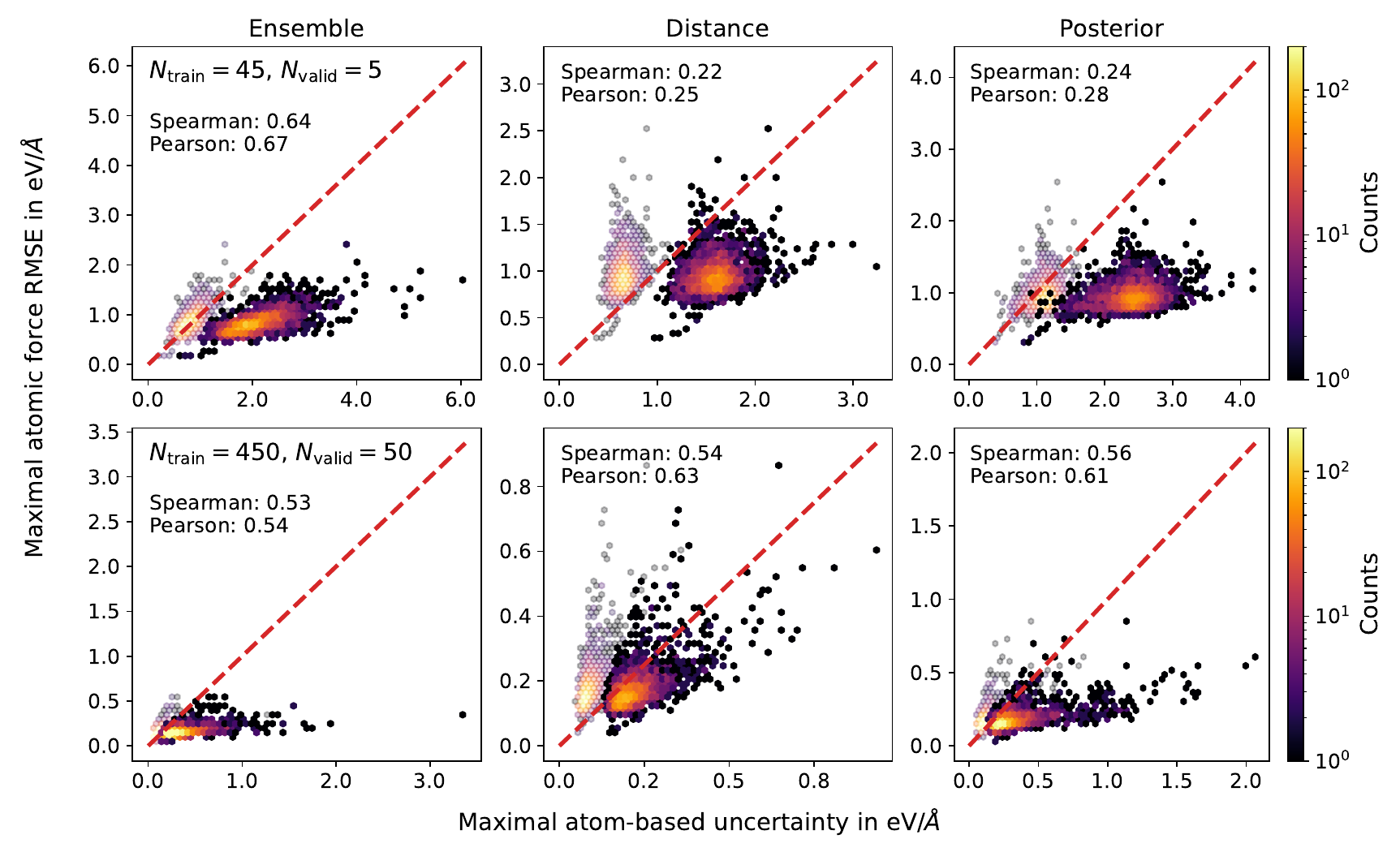}\hfill
	\caption{Correlation of maximal atom-based uncertainties with maximal atomic force RMSEs for MIL-53(Al). The results are presented for the test data set from Ref.~\citenum{Vandenhaute2023}. All uncertainty quantification methods are calibrated using CP and atomic force RMSEs. The top row shows the results of MLIPs trained using 45 atomic configurations, while five are used for early stopping and uncertainty calibration. The bottom row shows the results obtained with $450$ and $50$ MIL-53(Al) configurations, respectively. The training and validation data are taken from Ref.~\citenum{Vandenhaute2023}. Transparent hexbin points represent uncertainties calibrated with $\alpha=0.5$ (low confidence; see Methods), while opaque ones denote uncertainties calibrated with $\alpha=0.05$ (high confidence). Calibrating uncertainties with a high confidence level helps align the largest actual error with the corresponding uncertainty, shifting the hexbin points to or below the red diagonal line. This alignment is crucial for identifying unreliable predictions and prompting the termination of MD simulations.}
	\label{fig:mil53_uncertainty}
 \end{figure*}

Total and atom-based uncertainties are typically poorly calibrated,\cite{Pernot2022} meaning that they often underestimate actual errors. The underestimation of atomic force errors is particularly dangerous when dynamically generating candidate pools, as it may result in exploring unphysical configurations with extremely large errors in predicted forces. These unphysical configurations often cause convergence issues in reference DFT calculations. Additionally, poor calibration complicates defining an appropriate uncertainty threshold for prompting the termination of MD simulations and the evaluation of reference DFT energies, atomic forces, and stresses. To address this issue, we utilize inductive CP, which computes a re-scaling factor based on predicted uncertainties and prediction errors on a calibration set. The confidence level $1-\alpha$ in CP is defined such that the probability of underestimating the error is at most $\alpha$ on data drawn from the same distribution as the calibration set. For more details, see Methods.

\Figref{fig:mil53_uncertainty} demonstrates the correlation of maximal atom-based uncertainties, $\max_i u_i$, with maximal atomic force RMSEs, $\max_i \sqrt{\frac{1}{3} \sum_{k=1}^3 (\Delta F_{i, k})^2}$, for the MIL-53(Al) test data set from Ref.~\citenum{Vandenhaute2023} based on numerous first principles MD trajectories at 600~K. We chose maximal atomic force RMSE as our target metric to identify extrapolative regions due to its high sensitivity to unphysical local atomic environments. In MLIP-based atomistic simulations, we model it using maximal atom-based uncertainty. Employing quantiles or averages of atomic force RMSE could extend simulation time by reducing sensitivity to extreme values; however, exploring these alternatives is left for future work.

In \figref{fig:mil53_uncertainty}, transparent hexbins represent uncertainties calibrated with a lower confidence ($\alpha=0.5$; see Methods), while opaque ones depict those calibrated with a higher confidence ($\alpha=0.05$). The presented uncertainties are derived from gradient features or an ensemble of three MLIPs and calibrated using CP with atomic force RMSEs.\cite{Hu2022} For posterior- and distance-based uncertainties, which are unitless, the re-scaling with CP ensures that the resulting uncertainties are provided in correct units, i.e., eV/\AA. Ensemble-based uncertainty quantification already provides correct units, which CP preserves. Equivalent results for alanine dipeptide, including the correlation between average uncertainties and average force RMSEs, can be found in the Supplementary Information.

\Figref{fig:mil53_uncertainty} (top) demonstrates results for MLIPs trained on 45 MIL-53(Al) configurations, while five samples were used for early stopping and uncertainty calibration. \Figref{fig:mil53_uncertainty} (bottom) shows the results for MLIPs trained and validated on 450 and 50 MIL-53(Al) configurations, respectively. In both experiments, the training and validation samples were selected from the data sets provided by Ref.~\citenum{Vandenhaute2023}. The first 50 samples correspond to randomly perturbed structures, while the remaining 450 are generated using metadynamics combined with incremental learning.\cite{Vandenhaute2023} The latter is an iterative algorithm that improves MLIPs by training on configurations generated sequentially over time, using the last frame of atomistic simulations.

We observe that uncertainties calibrated with a lower confidence level often underestimate actual errors. In this case, MD can explore unphysical regions before reaching the uncertainty threshold, especially in cases with a weak correlation between uncertainties and actual errors. By employing CP with higher confidence, we help align the largest prediction error with the corresponding uncertainty, thereby improving its ability to identify the onset of extrapolative regions. This alignment becomes apparent in \figref{fig:mil53_uncertainty}, where CP shifts the hexbin points to be on or below the diagonal.

In \figref{fig:mil53_uncertainty} (top), we find that even training and calibrating models with a few randomly perturbed atomic configurations is sufficient for robust identification of unreliable predictions. This result is crucial as we rely on such data sets to initialize our AL experiments, eliminating the need for predefined data sets.\cite{Kulichenko2023, vanerOord2022} Furthermore, we observe that, for MIL-53(Al), calibrated uncertainties from model ensembles tend to overestimate the actual error to a greater extent than gradient-based approaches. While this may not be critical when exploring unphysical configurations, it can prematurely terminate MD simulations. This trend is consistent across all training and calibration data sizes. Lastly, the results provided here and in the Supplementary Information demonstrate that all uncertainty methods perform comparably regarding Pearson and Spearman correlation coefficients.

\subsection*{Performance of bias-forces-driven active learning}

\begin{figure*}[t!]
	\includegraphics[width=\textwidth]{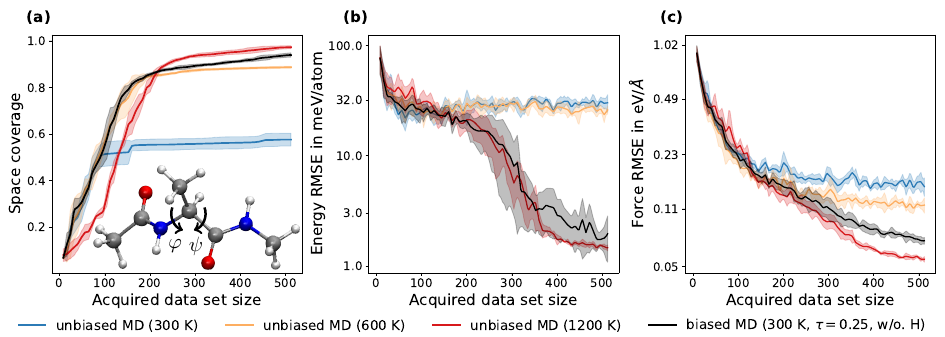}\hfill
	\includegraphics[width=\textwidth]{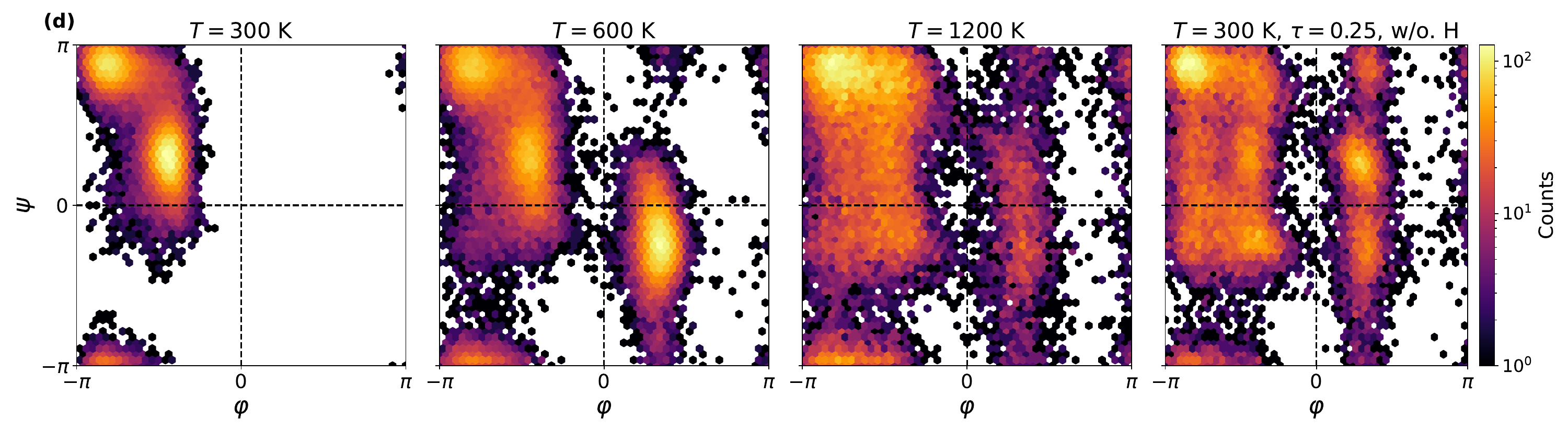}
	\caption{Comparison of AL approaches employing biased and unbiased MD simulations to generate the candidate pool of atomic configurations for alanine dipeptide. Results are provided for the posterior-based uncertainty quantification derived from sketched gradient features. Unlike unbiased MD simulations, which rely on atom-based uncertainties to terminate MD simulations, biased MD simulations use total and atom-based uncertainties to bias MD simulations and prompt their termination, respectively. We use three metrics to assess the performance of our AL approaches: \textbf{(a)} Coverage of the CV space; \textbf{(b)} Energy RMSE; and \textbf{(c)} Force RMSE. All RMSEs are evaluated on the alanine dipeptide test data set; see Methods. Shaded areas denote the standard deviation across five independent runs. The alanine dipeptide molecule, including its CVs, is shown as an inset in~\textbf{(a)}. The color code of the inset molecule is C grey, O red, N blue, and H white. \textbf{(d)} Ramachandran plots demonstrating the CV spaces explored by the four AL experiments. Biased MD simulations achieve exceptional performance, close to those of MD conducted at 1200~K, without knowledge of temperatures that accelerate transitions between stable states. The CV space covered by uncertainty-biased MD simulations at 300~K matches that of unbiased simulations at 1200~K, significantly outperforming the coverage achieved by unbiased MD at 300~K and 600~K.}
	\label{fig:diala_data_efficiency}
\end{figure*}

\begin{table*}[t!]
	\caption{CV space coverage, atomic energy (E-) and atomic force (F-) RMSEs, as well as position (Pos.) and uncertainty (Unc.) auto-correlation times (ACTs) for alanine dipeptide experiments conducted with posterior-based uncertainties. E- and F-RMSEs are reported for MLIPs obtained at the end of each experiment, while CV space coverage and ACTs are computed using the entire trajectory obtained throughout the experiment. E-RMSE is given in meV/atom, while F-RMSE is in eV/\AA. All E-RMSE and F-RMSE values are computed for the test data set obtained from a long MD trajectory at 1200~K; see Methods. ACTs are provided in ps. For biased MD, we compare two cases: one with (w.) biasing hydrogen atoms and one without (w/o.). We also compare biased and unbiased MD with experiments that involve the random selection (random sel.) strategy for acquiring training data. The best performance is highlighted in bold, and the second-best performance is underlined.
	\label{tab:ala2-results}
	}
	\begin{center}
	\begin{tabular}{lccccc}
	\toprule 
	Experiment		    			         & CV space cov.			& E-RMSE 					& F-RMSE 						& Pos. ACT\textsuperscript{\emph{a}} 	& Unc. ACT\textsuperscript{\emph{a}}   \\
    \midrule
    random sel. (300~K)				         & 0.58 $\pm$ 0.03			& 34.09 $\pm$ 6.29		    & 0.191 $\pm$ 0.019		        & --		                            & --		                            \\
	random sel. (600~K)				         & 0.76 $\pm$ 0.04			& 31.44 $\pm$ 4.77		    & 0.143 $\pm$ 0.015		        & --		                            & --		                            \\
    random sel. (1200~K)			         & \emph{0.95 $\pm$ 0.01}	& 19.83 $\pm$ 4.62          & 0.116 $\pm$ 0.017		        & --		                            & --		                            \\
	unbiased MD (300~K)				         & 0.58 $\pm$ 0.03			& 30.29 $\pm$ 5.47			& 0.149 $\pm$ 0.019				& 2.07 $\pm$ 0.11				        & 327.11 $\pm$ 8.69				        \\
	unbiased MD (600~K)    			         & 0.89 $\pm$ 0.00			& 26.03 $\pm$ 2.23			& 0.116 $\pm$ 0.012				& 1.23 $\pm$ 0.02				        & 257.88 $\pm$ 22.01			        \\
	unbiased MD (1200~K)   			         & \textbf{0.97 $\pm$ 0.01}	& \textbf{1.47 $\pm$ 0.09} 	& \textbf{0.055 $\pm$ 0.002}	& \emph{0.74 $\pm$ 0.02}		        & \emph{21.41	$\pm$ 4.91}		        \\
	biased MD (300~K, $\tau=0.25$, w. H) 	 & 0.87 $\pm$ 0.02 			& 5.09 $\pm$ 5.40 			& 0.082 $\pm$ 0.016				& 2.08 $\pm$ 0.13				        & \textbf{19.38 $\pm$ 7.42} 	        \\
	biased MD (300~K, $\tau=0.25$, w/o. H)   & 0.94 $\pm$ 0.01			& \emph{1.97 $\pm$ 0.88}  	& \emph{0.071 $\pm$ 0.003}	    & \textbf{0.69 $\pm$ 0.04}		        & 52.79	$\pm$ 19.40				        \\
	\bottomrule 
	\end{tabular}
	\end{center}
    \footnotesize{\textsuperscript{\emph{a}} ACTs computed for experiments with the random selection (random sel.) strategy are excluded from the analysis because different approaches may introduce systematic biases, making the comparison unreliable.}
\end{table*}

Exploring the configurational space of complex molecular systems, particularly those with multiple stable states, is essential for developing accurate and robust MLIPs. We apply bias-forces-driven MD combined with AL to develop MLIPs for alanine dipeptide in vacuum. This dipeptide exhibits two stable conformers characterized by the backbone dihedral angles $\phi$ and $\psi$ (see \figref{fig:diala_data_efficiency}): the $\mathrm{C}_\mathrm{7eq}$ state with $\phi \approx -1.5$~rad and $\psi\approx 1.19$~rad and the $\mathrm{C}_\mathrm{ax}$ state with $\phi \approx 0.9$~rad and $\psi\approx -0.9$~rad.\cite{Bolhuis2000} We use unbiased MD as the baseline for generating candidate pools in two scenarios: AL with candidates selected from unbiased MD trajectories based on their uncertainty (and diversity) and candidates sampled from them at random. The performance of MLIPs is assessed employing the test data obtained from a long MD trajectory at 1200~K; see Methods. We employ the AMBER ff19SB force field for reference energy and force calculations,\cite{Tian2020} as implemented in the TorchMD package using PyTorch.\cite{Doerr2021, Paszke2019}

\Figref{fig:diala_data_efficiency} demonstrates the performance of MLIPs obtained for alanine dipeptide depending on the number of acquired configurations. \Tabref{tab:ala2-results} presents error metrics evaluated for MLIPs at the end of each experiment. Here, we provide results for the posterior-based uncertainty and uncertainty-biased MD at 300~K. The Supplementary Information presents equivalent results for other uncertainty methods and temperatures. \Figref{fig:diala_data_efficiency} (a) presents the coverage of the CV space defined by $\phi$ and $\psi$, calculated using all MD trajectories up to the current AL step. We measure the coverage of the respective space by a tree-based weighted recursive space partitioning; see Methods. AL experiments combined with unbiased MD at 1200~K serve as the upper-performance limit for MLIPs in the case of alanine dipeptide, achieving the highest coverage of $0.97$ after acquiring 512 configurations. Increasing temperature even further while using interatomic potentials, which allow for bond breaking and formation, may lead to the degradation of the molecule. Uncertainty-biased MD simulations at 300~K result in slightly lower coverage values, surpassing the coverages achieved by unbiased MD at 300~K and 600~K.

Furthermore, biased MD at 300~K outperforms unbiased dynamics at 1200~K, efficiently covering the CV space before acquiring $\sim$200 configurations. This observation is attributed to the gradual increase in driving forces induced by the uncertainty bias, resulting in a more gradual distortion of the atomic structure. In contrast, high-temperature unbiased simulations perturb the system more strongly and rapidly enter extrapolative regions without exploring relevant configurational changes. Thus, high-temperature simulations may also cause the degradation of the investigated atomic systems, unlike uncertainty-biased dynamics applied at mild physical conditions.

Figures \ref{fig:diala_data_efficiency} (b) and (c) present energy and force RMSEs evaluated on the alanine dipeptide test data set; see Methods. Consistent with the findings in \figref{fig:diala_data_efficiency} (a), AL approaches combined with biased MD at 300~K outperform their unbiased counterparts at 300~K and 600~K once they acquire $\sim$100 configurations. Biased AL experiments achieve energy RMSE of 1.97 meV/atom, close to those observed in high-temperature MD simulations, surpassing others by a factor of more than 13. A similar trend is observed for force RMSE. Biased AL experiments achieve an RMSE of 0.071~eV/\AA, outperforming their counterparts at 300~K and 600~K by factors of 2.1 and 1.6, respectively.

These results demonstrate the efficiency of uncertainty-biased dynamics in exploring the configurational space and developing accurate and robust MLIPs. Moreover, generating training data that sufficiently covers the configurational space by combining AL with biased MD does not significantly increase the computational demand compared to conventional AL with unbiased MD; see the Supplementary Information. Lastly, MLIPs trained with candidates selected based on their uncertainty (and diversity) from biased and unbiased MD trajectories systematically outperform MLIPs trained with candidates selected at random; see \tabref{tab:ala2-results}.

Biased AL experiments achieve exceptional performance without knowledge of temperatures that accelerate transitions between stable states; see \figref{fig:diala_data_efficiency} (d). Identifying these temperatures requires running MD simulations at different conditions to explore the configurational space without degrading the atomic system. In contrast, given the mild physical conditions such as temperatures of 300~K and 600~K, biased MD simulations outperform their unbiased counterparts at 300~K and 600~K and achieve comparable performance to experiments at 1200~K for $\tau \lesssim 0.5$ and $0.2 \lesssim \tau \lesssim 0.4$, respectively. The available range of biasing strength values may be more restricted at more extreme conditions. Adding uncertainty bias to MD at 1200~K results in an even stronger system perturbation than during unbiased MD without yielding any improvement. For additional details, see the Supplementary Information.

Our results offer evidence of rare event exploration (the exploration of both stable states of alanine dipeptide) through uncertainty-biased dynamics. The following section will present a detailed analysis of the exploration rates. Additionally, we have identified how to further improve our biased MD simulations by making biasing strengths species dependent; see the Supplementary Information. The results presented in this section, achieved with a biasing strength of zero for hydrogen atoms, outperform settings where all atoms are biased equally, with improvements by a factor of 1.08 in coverage and 1.15 in force RMSE; see \tabref{tab:ala2-results}. Thus, a more sophisticated data-driven redistribution of biasing strengths can further enhance the performance of bias-forces-driven MD simulations. However, learning species-dependent biasing strengths necessitates defining a suitable loss function that promotes the fast exploration of phase space,\cite{Christiansen2023} which falls beyond the scope of this work.

\subsection*{Exploration rates for collective variables of alanine dipeptide}

\begin{figure*}[t!]
	\includegraphics[width=\textwidth]{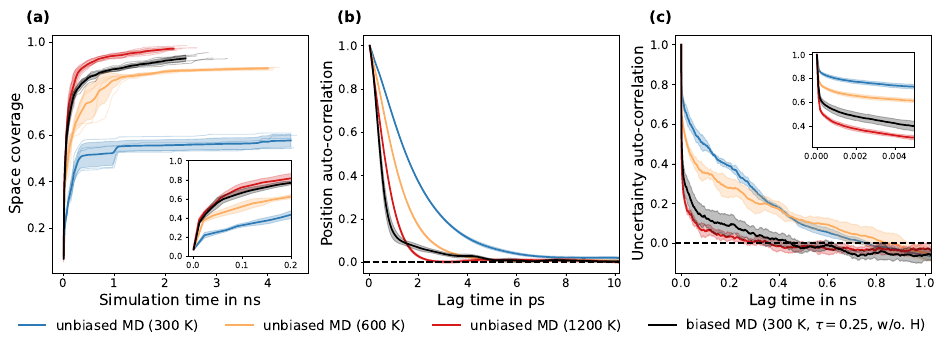}\hfill
	\includegraphics[width=\textwidth]{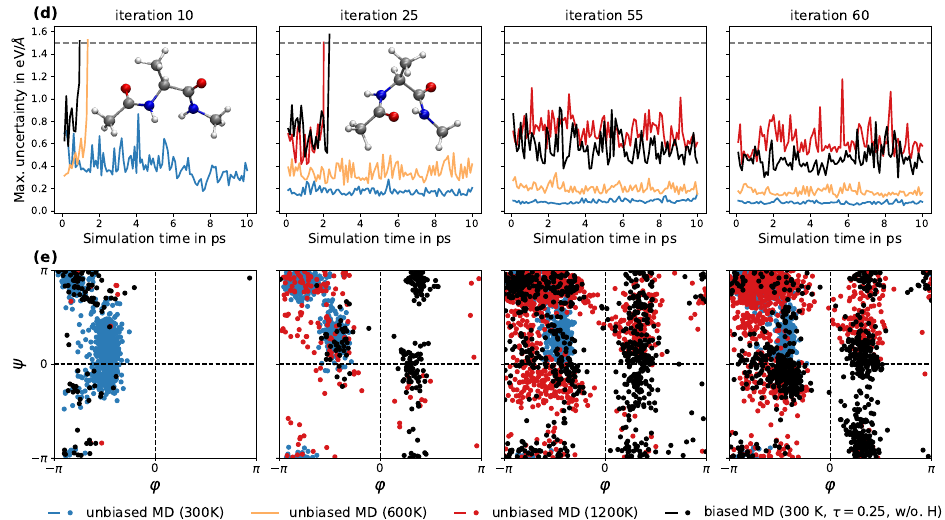}
	\caption{Evaluation of CV space exploration rates for biased and unbiased MD simulations of alanine dipeptide. Here, MD simulations generate candidate pools of atomic configurations for AL algorithms. Results are provided for the posterior-based uncertainty quantification derived from sketched gradient features. Unlike unbiased MD simulations, which rely on atom-based uncertainties to terminate MD simulations, biased MD simulations use total and atom-based uncertainties to bias MD simulations and prompt their termination, respectively. We use three metrics to asses the exploration rates: \textbf{(a)} Coverage of the CV space; \textbf{(b)} Auto-correlation functions of atomic positions; and \textbf{(c)} Auto-correlation functions of atom-based uncertainties. Shaded areas denote the standard deviation across five independent runs. \textbf{(d)} Time evolution of the maximal atom-based uncertainty within an AL iteration and the entire experiment. Time evolution is shown for one of the eight MD simulations. The dashed gray line represents the uncertainty threshold of 1.5~eV/\AA. The insets show configurations that reached the uncertainty threshold for uncertainty-biased MD. \textbf{(e)} Ramachandran plots illustrate the exploration of the CV space over AL iterations and the entire experiment. Ramachandran plots are presented for unbiased MD simulations at 300~K and 1200~K and biased MD simulations at 300~K. Simulation time refers to the effective number of MD steps ($\times$ 0.5~fs) required to reach the final coverage, while lag time denotes the time interval between two successive MD frames. Biased MD simulations at 300~K exhibit at least two times higher exploration rates than their unbiased counterparts at 300~K and 600~K. Their exploration rates are comparable to those of unbiased MD simulations at 1200~K, with the advantage of gradually distorting the molecule, reducing the risk of its degradation.}
	\label{fig:diala_enhanced_sampling}
\end{figure*}

We have observed that uncertainty-biased MD simulations effectively explore the configurational space of alanine dipeptide, defined by its CVs. \Figref{fig:diala_enhanced_sampling} evaluates the extent to which the introduced bias forces in MD simulations accelerate their exploration. In \figref{fig:diala_enhanced_sampling} (a), we present the coverage of the CV space as a function of simulation time, i.e., of the effective number of MD steps. The figure demonstrates that uncertainty-biased AL experiments at 300~K outperform unbiased experiments at 300~K and 600~K. They achieve the same coverage in considerably shorter simulation times, thereby enhancing exploration rates by a factor of larger than two. At the same time, biased MD simulations yield results comparable to those obtained from unbiased MD simulations at 1200~K. Thus, uncertainty-biased MD explores configurational space at a similar rate to unbiased MD at 1200~K.

The exploration rates estimated from \figref{fig:diala_enhanced_sampling} (a) provide an approximate measure of how uncertainty-biased dynamics accelerate the exploration of configurational space. To offer a more thorough assessment, we examine auto-correlation functions (ACFs) computed for both position and uncertainty spaces in \figsref{fig:diala_enhanced_sampling} (b) and (c). Here, a faster decay corresponds to a faster exploration of the respective space. We compute ACFs using MD trajectories from all AL iterations. Additionally, we calculate the auto-correlation time (ACT) for each experiment. For the definition of ACF and ACT, see Methods. \Tabref{tab:ala2-results} presents ACTs for all AL experiments. Smaller ACTs correspond to a faster decay of ACFs, indicating a faster exploration of the respective spaces.

ACTs demonstrate that uncertainty-biased MD at 300~K explores position and uncertainty spaces two to six times faster than unbiased MD at 300~K and 600~K. Compared to unbiased MD at 1200~K, it achieves comparable exploration rates in the position space and rates lower by a factor of two for the uncertainty space. Biasing hydrogen atoms reduces the uncertainty ACT compared to experiments with zero hydrogen biasing strength but increases the position ACT by a factor of three. Thus, stronger atomic bond distortions, resulting in fast exploration of extrapolative regions, can explain a shorter uncertainty ACT of unbiased MD at 1200 K. While this effect can be unfavorable for promoting the exploration of rare events in biased MD, incorporating small, non-zero biasing strengths for hydrogen atoms may be necessary to ensure the robustness of MD simulations at elevated temperatures. Interestingly, we observe that uncertainty-biased MD explores both stable states in alanine dipeptide, even though 27 degrees of freedom (C, N, and O atoms) were effectively biased, demonstrating its remarkable efficiency.

To gain insight into the exploration of the CV space during AL, we refer to \figsref{fig:diala_enhanced_sampling} (d) and (e), which illustrate the time evolution of the maximal atom-based uncertainty and the CV space coverage for selected AL iterations. Biased MD systematically explores configurations with higher uncertainty values than unbiased MD at 300~K and 600~K. Furthermore, bias forces drive the exploration of both stable states of alanine dipeptide and promote transitions between them, similar to higher temperatures in unbiased MD. Later AL iterations in \figsref{fig:diala_enhanced_sampling} (d) and (e) demonstrate that MD driven by bias forces reduces the uncertainty level uniformly across the configurational space. Thus, given the correlation between uncertainties and actual errors, uncertainty-biased MD generates MLIPs uniformly accurate across the configurational space.

\subsection*{Performance of bias-stress-driven active learning}

\begin{figure*}[t!]
	\includegraphics[width=0.99\textwidth]{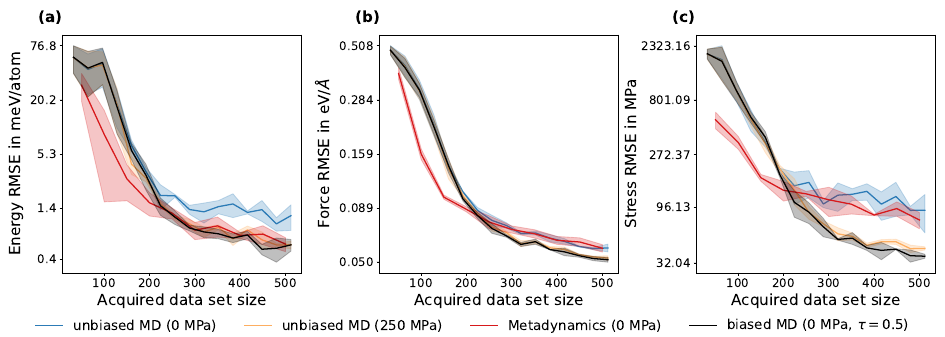}\hfill
	\includegraphics[width=0.99\textwidth]{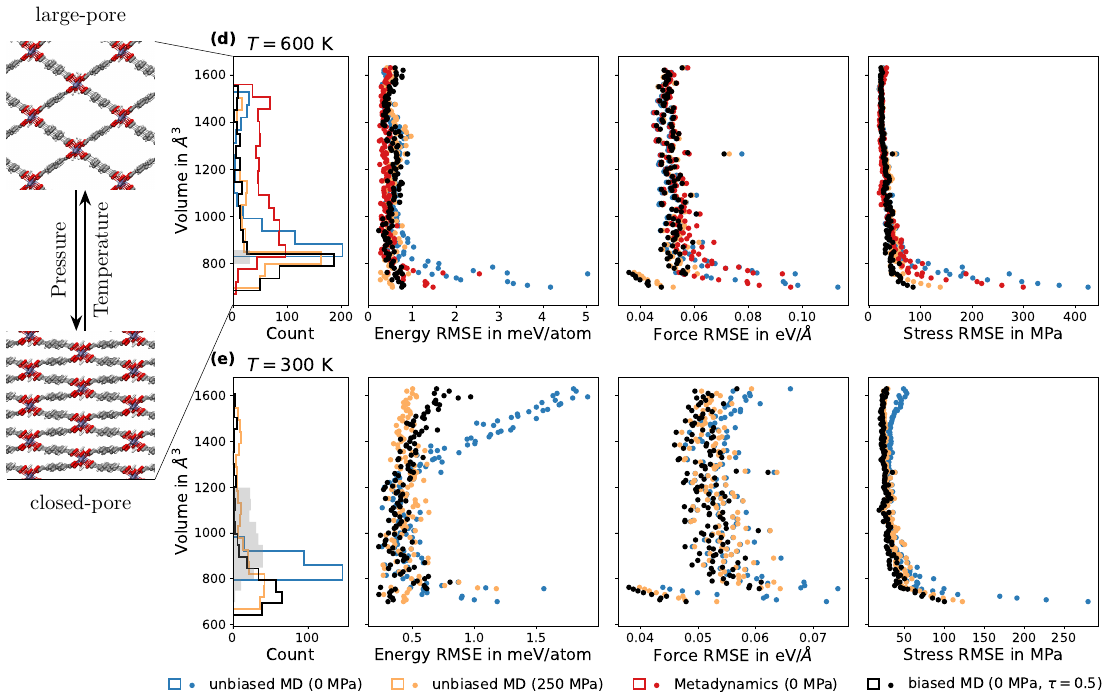}
	\caption{Comparison of AL approaches employing biased and unbiased MD simulations to generate the candidate pool of atomic configurations for MIL-53(Al). Results are provided for the posterior-based uncertainty quantification derived from sketched gradient features. Unlike unbiased MD simulations, which rely on atom-based uncertainties to terminate MD simulations, biased MD simulations use total and atom-based uncertainties to bias MD simulations and prompt their termination, respectively. We use three metrics to assess the performance of our AL approaches: \textbf{(a)} Energy RMSE; \textbf{(b)} Force RMSE; and \textbf{(c)} Stress RMSE. All RMSEs are evaluated on the MIL-53(Al) test data set.\cite{Vandenhaute2023} Shaded areas denote the standard deviation across three independent runs, except for metadynamics. For it, shaded areas denote standard deviation across three randomly initialized MLIPs. \textbf{(d)} Volume distribution for atomic configurations acquired during MD at 600~K, along with volume-dependent energy, force, and stress RMSEs. \textbf{(e)} Volume distribution for configurations acquired during MD at 300~K, along with volume-dependent energy, force, and stress RMSEs. We employ a temperature of 300~K to reduce the probability of exploring the large-pore state of MIL-53(Al). Bias-stress-driven MD simulations outperform metadynamics-based simulations with adaptive biasing of the cell parameters. Metadynamics aims to cover the volume space uniformly. In contrast, uncertainty-biased MD generates training data sets that uniformly reduce energy, force, and stress RMSEs. Additionally, biased MD simulations enhance the exploration of closed- and large-pore states of MIL-53(Al) shown in the inset of \textbf{(d)}.}
	\label{fig:mil53_performance}
 \end{figure*}

 \begin{table*}[t!]
	\caption{Atomic energy (E-), atomic force (F-), and stress (S-) RMSEs, as well as position (Pos.) and uncertainty (Unc.) auto-correlation times (ACTs) for MIL-53(Al) experiments conducted with posterior-based uncertainties. E-, F-, and S-RMSEs are reported for MLIPs obtained at the end of each experiment, while ACTs are computed using the entire trajectory sampled throughout the experiment. E-RMSE is given in meV/atom, F-RMSE in eV/\AA, and S-RMSE in MPa. All E-RMSE, F-RMSE, and S-RMSE values are computed for the test data set obtained based on first principles MD trajectories at 600~K; see Ref.~\citenum{Vandenhaute2023}. ACTs are provided in ps. We also compare biased and unbiased MD with experiments that involve the random selection (random sel.) strategy for acquiring training data. The best performance is highlighted in bold, and the second-best performance is underlined.
	\label{tab:mil53-results}
	}
	\begin{center}
	\begin{tabular}{lccccc}
	\toprule 
	Experiment		    				& E-RMSE						& F-RMSE 						& S-RMSE 						& Pos. ACT\textsuperscript{\emph{a}}        & Unc. ACT\textsuperscript{\emph{a}}     \\
	\midrule 
	\multicolumn{6}{c}{$T=600$~K}                                                                                                                                                                                            \\
	\midrule
    random sel. (0~MPa)					& 1.62 $\pm$ 0.52		        & 0.062 $\pm$ 0.002		        & 145.3 $\pm$ 35.49		        & --		                                & --		                             \\
    random sel. (250~MPa)				& 0.84 $\pm$ 0.09		        & 0.057 $\pm$ 0.001		        & 63.8 $\pm$ 15.82		        & --		                                & --		                             \\
	unbiased MD (0~MPa)					& 1.17 $\pm$ 0.36				& 0.058 $\pm$ 0.002				& 90.81 $\pm$ 32.82				& 10.60 $\pm$ 9.54				            & 88.05 $\pm$ 2.53				         \\
	unbiased MD (250~MPa)    			& \textbf{0.57 $\pm$ 0.05}		& \emph{0.052 $\pm$ 0.001}		& \emph{42.72 $\pm$ 1.37}		& \textbf{2.08 $\pm$ 0.58}		            & \emph{66.32 $\pm$ 2.02}		         \\
	Metadynamics (0~MPa)   				& \emph{0.58 $\pm$ 0.10}		& 0.058 $\pm$ 0.002 			& 74.83 $\pm$ 11.89 			& --							            & --							         \\
	biased MD (0~MPa, $\tau=0.5$) 		& \textbf{0.57 $\pm$ 0.08} 		& \textbf{0.051 $\pm$ 0.001} 	& \textbf{36.60 $\pm$ 1.46}		& \emph{2.75 $\pm$ 0.46}		            & \textbf{44.87 $\pm$ 14.08} 	         \\
	\midrule 
	\multicolumn{6}{c}{$T=300$~K}                                                                                                                                                                                            \\
	\midrule
    random sel. (0~MPa)					& 1.04 $\pm$ 0.26		        & 0.058 $\pm$ 0.001		        & 70.49 $\pm$ 6.61		        & --		                                & --		                             \\
    random sel. (250~MPa)				& 0.58 $\pm$ 0.08		        & 0.055 $\pm$ 0.002		        & 52.19 $\pm$ 2.22		        & --		                                & --		                             \\
	unbiased MD (0~MPa)					& 0.88 $\pm$ 0.20				& 0.056 $\pm$ 0.001				& 58.57 $\pm$ 5.94				& \emph{3.45 $\pm$ 4.06}		            & 99.25 $\pm$ 10.34				         \\
	unbiased MD (250~MPa)    			& \textbf{0.48 $\pm$ 0.01}		& \emph{0.054 $\pm$ 0.000}		& \emph{39.88 $\pm$ 1.76}		& \textbf{1.86 $\pm$ 0.14}		            & \emph{54.56 $\pm$ 4.82}				 \\
	biased MD (0~MPa, $\tau=0.5$) 		& \emph{0.49 $\pm$ 0.09} 		& \textbf{0.052 $\pm$ 0.001} 	& \textbf{33.89 $\pm$ 3.06}		& 42.92 $\pm$ 14.18				            & \textbf{26.89 $\pm$ 8.94} 	         \\
	\bottomrule 
	\end{tabular}
	\end{center}
    \footnotesize{\textsuperscript{\emph{a}} ACTs computed for experiments with the random selection (random sel.) strategy are excluded from the analysis because different approaches may introduce systematic biases, making the comparison unreliable.}
\end{table*}

Generating training data for bulk material systems with large unit cells and multiple stable states poses a significant challenge in developing MLIPs. Therefore, we assess the performance of the bias-stress-driven AL applied to MIL-53(Al), a flexible MOF that undergoes reversible, large-amplitude volume changes under external stimuli, such as temperature and pressure (see \figref{fig:mil53_performance}). MIL-53(Al) features two stable phases: the closed-pore state with a unit cell volume of $V \sim 830$~\AA$^3$ and the large-pore state with $V \sim 1419$~\AA$^3$. For reference energy, force, and stress calculations, we use the CP2K simulation package (version 2023.1)\cite{Kuehne2020} and DFT at the PBE-D3(BJ) level.\cite{Perdew1996a, Grimme2010} Our baseline for generating candidate pools for AL involves unbiased MD and training data selected based on their uncertainty (and diversity) or at random. We also employ metadynamics,\cite{Vandenhaute2023} which uses an adaptive biasing strategy for cell parameters of MIL-53(Al), as a baseline. We assess the performance of MLIPs for MIL-53(Al) using the test data set presented by Ref.~\citenum{Vandenhaute2023}.

\Figsref{fig:mil53_performance} (a)--(c) demonstrate the performance of MLIPs developed for MIL-53(Al) depending on the number of acquired configurations. \Tabref{tab:mil53-results} presents error metrics evaluated for MLIPs at the end of each experiment. Here, we present results for the posterior-based uncertainty. The Supplementary Information presents equivalent results for other uncertainty methods and pressures. We observe that MLIPs trained with configurations generated using metadynamics outperform the others for data set sizes below $\sim$200 samples. This difference in performance can be attributed to how perturbed configurations are generated and the differing experimental settings between incremental learning and AL applied here. Bias-stress-driven AL outperforms metadynamics-based experiments after acquiring $\sim$200 configurations regarding force and stress RMSEs. 

Metadynamics-based experiments achieve performance on par with unbiased AL experiments conducted at 0~MPa after they reach a data set size of $\sim$200 configurations. For uncertainty-biased MD, the force RMSE improves by a factor of 1.14, and the stress RMSE improves by a factor of two compared to zero-pressure unbiased MD. Furthermore, AL experiments with biased MD simulations outperform unbiased MD simulations at 250~MPa regarding stress RMSE. Thus, bias-stress-driven MD generates a data set that better represents the relevant configurational space of flexible MOFs compared to MLIPs trained with conventional MD and metadynamics simulations. This improvement is achieved without significantly increasing the computational cost of data generation; see the Supplementary Information. Lastly, similar to the results obtained for alanine dipeptide, AL with a more advanced selection strategy outperforms experiments where training data is picked at random; see \tabref{tab:mil53-results}.

\Figsref{fig:mil53_performance} (d) and (e) show the main advantage of biased MD simulations over unbiased and metadynamics-based approaches. While exploring the large-pore state less frequently than metadynamics-based counterparts, bias-stress-driven MD spans a broader range of volumes and uniformly reduces energy, force, and stress RMSEs across the entire volume space. Compared to zero-pressure unbiased MD simulations, it promotes the exploration of the large-pore state. However, this state can be modeled using atomic environments from the closed-pore one. Thus bias stress does not excessively favor exploration of the former. Instead, it drives the dynamics more toward smaller volumes, for which all other approaches tend to predict energy, force, and stress values with larger errors. Note that, in \figref{fig:mil53_performance} (e), we reduce the temperature to 300~K and initiate AL experiments with 256 configurations, each having a unit cell volume below 1200~\AA$^3$ (drawn from the training data in Ref.~\citenum{Vandenhaute2023}). Using a lower temperature and learning the configurational space around the closed-pore state is required to decrease the probability of MD simulations exploring the large-pore stable state of MIL-53(Al). In contrast, we found that using randomly perturbed atomic configurations can lead to underestimated energy barriers by MLIPs, thus facilitating the transition between both stable phases in initial AL iterations. 

These results show that uncertainty-biased MD simulations aim to uniformly reduce errors across the relevant configurational space and promote the simultaneous exploration of extrapolative regions and transitions between stable states. Also, under selected physical conditions ($T=600$~K and $p=0$~MPa), the performance of our uncertainty-biased MD exhibits low sensitivity to stress biasing strength values for $\tau \geq 0.5$; see the Supplementary Information. Metadynamics, in contrast, may require longer simulation times to generate equivalent candidate pools as it focuses on generating configurations uniformly distributed in the CV space, which is unnecessary for developing MLIPs.

\subsection*{Exploration rates for cell parameters of MIL-53(Al)}

\begin{figure*}[t!]
	\includegraphics[width=\textwidth]{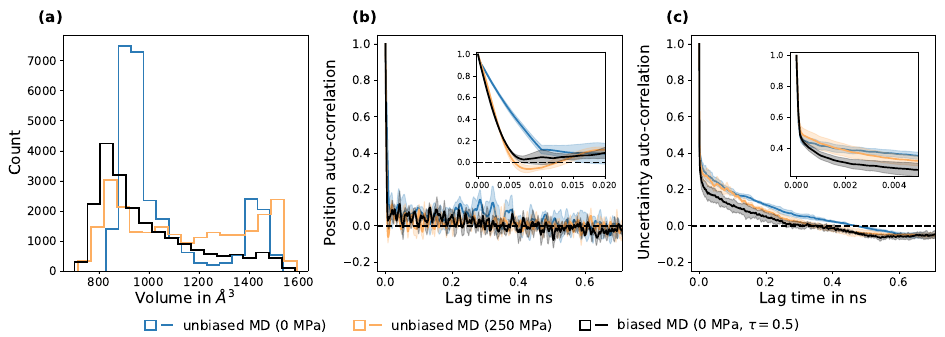}\hfill
	\includegraphics[width=\textwidth]{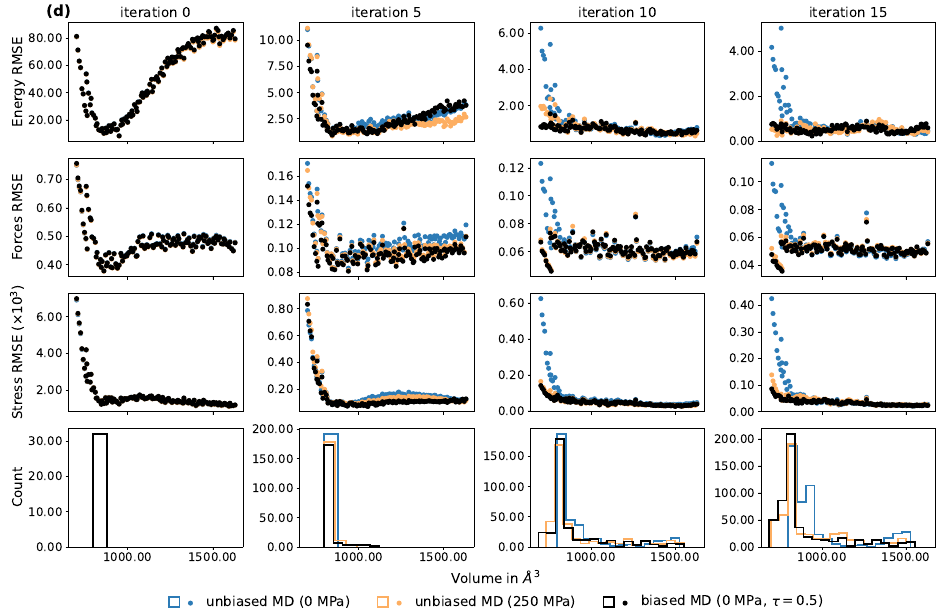}
	\caption{Evaluation of configurational space exploration rates for biased and unbiased MD simulations of MIL-53(Al). Here, MD simulations generate candidate pools of atomic configurations for AL algorithms. Results are provided for the posterior-based uncertainty quantification derived from sketched gradient features. Unlike unbiased MD simulations, which rely on atom-based uncertainties to terminate MD simulations, biased MD simulations use total and atom-based uncertainties to bias MD simulations and prompt their termination, respectively. We use three metrics to asses the exploration rates: \textbf{(a)} Volume distribution of configurations sampled throughout the experiment; \textbf{(b)} Auto-correlation functions for positions; and \textbf{(c)} Auto-correlation functions for atom-based uncertainties. Shaded areas denote the standard deviation across three independent runs. \textbf{(d)} Time evolution of the volume distribution of configurations acquired during training and of energy, force, and stress RMSEs evaluated on the test data set\cite{Vandenhaute2023} depending on the unit cell volume. Bias-stress-driven MD simulations achieve exploration rates comparable to those of high-pressure unbiased MD simulations. They aim to reduce RMSEs uniformly across the entire volume space, even in the early stages of AL, surpassing the performance of unbiased simulations.}
	\label{fig:mil53_sampling}
\end{figure*}

\Figref{fig:mil53_sampling} assesses the extent to which uncertainty-biased (bias stress) MD simulations enhance the exploration of the extensive volume space of MIL-53(Al). In \figref{fig:mil53_sampling} (a), we observe a higher frequency of transitions between stable phases for biased MD simulations than for zero-pressure counterparts. Additionally, uncertainty-biased simulations favor the exploration of smaller MIL-53(Al) volumes, in line with the results shown in \figref{fig:mil53_performance}. \Figsref{fig:mil53_sampling} (b) and (c) present ACFs for position and uncertainty spaces, with estimated ACTs provided in \tabref{tab:mil53-results}. Here, a faster decay of ACFs corresponds to shorter ACTs and indicates a faster exploration of the respective space. These results indicate that bias-stress-driven MD is at least as efficient as high-pressure MD simulations in exploring both spaces. \Figref{fig:mil53_sampling} (d) demonstrates the time evolution of energy, force, and stress RMSEs. It reveals that local atomic environments in the large-pore state are well represented by those in the closed-pore state, explaining the stronger preference for smaller volumes by biased MD; see \figref{fig:mil53_sampling} (a) and \figsref{fig:mil53_performance} (d) and (e). This effect is evident from the low force and stress RMSEs in the early AL iterations for the large-pore state, even though this state has not been explored yet. Furthermore, uncertainty-biased MD simulations surpass the performance of their counterparts already in the early stages by aiming to reduce errors across the test volume space uniformly.

\begin{figure}[t!]
	\includegraphics[width=\linewidth]{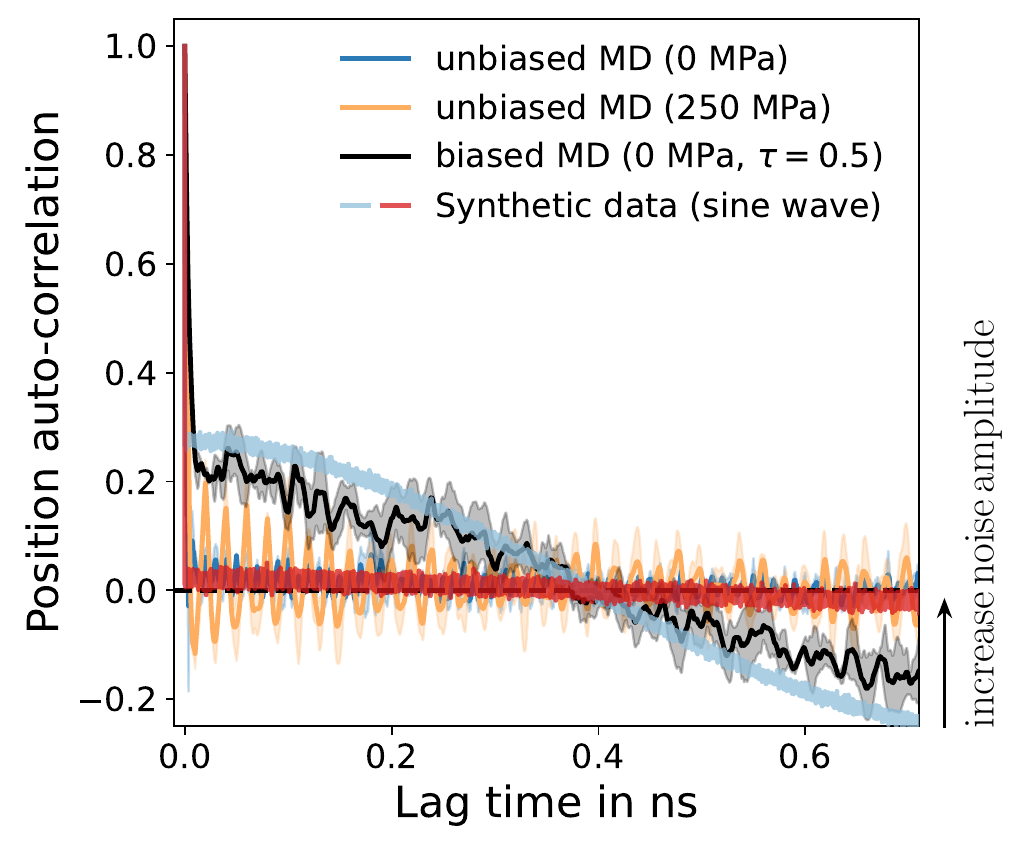}
	\caption{Position ACF obtained by running biased and unbiased MD simulations at 300~K for MIL-53(Al). Shaded areas denote the standard deviation across three independent runs. We employ a temperature of 300~K to reduce the probability of exploring the large-pore state of MIL-53(Al). The ACF exhibits strongly correlated motions attributed to volume fluctuations induced by the bias stress. These fluctuations can be modeled by a sine wave with a period twice the length of the simulation. The red line denotes a sine wave with a larger noise amplitude than the one denoted by the blue line.}
	\label{fig:mil53_positions_acfs_300K}
\end{figure}

From these results and the findings in \figref{fig:mil53_performance} (d), we conclude that bias-stress-driven MD significantly enhances the exploration of the relevant configurational space, including rare events (i.e., transitions between stable phases). However, in \tabref{tab:mil53-results}, we obtained longer ACTs for biased MD at 300~K compared to its unbiased counterparts, which contradicts our previous arguments. When examining the ACF shown in \figref{fig:mil53_positions_acfs_300K}, it becomes evident that a stronger correlation in the position space results from the volume fluctuations induced in MIL-53(Al) by the bias stress. These fluctuations can be represented by a sine wave with additive random noise and a period twice the simulation's length; see Methods. This observation implies that bias stress induces correlated motions in the MIL-53(Al) system, causing it to expand and contract alternately for half of the simulation time. This phenomenon results in periodic exploration of small and large volumes within the configurational space.

In contrast to the conventional approaches, including the bias-forces-driven MD simulations, which aim for uncorrelated random-walk-like behavior of predetermined CVs to capture configurational changes, our method introduces correlated motion that explores the entire configurational space. Increasing the amplitude of random noise in the sine wave reduces the amplitude of these fluctuations in the ACF, similar to raising the temperature in an atomic system. This decrease in the amplitude explains why this effect is not observed in \figref{fig:mil53_sampling} (b).

\section*{Discussion}

This work investigates a new paradigm for data set generation, facilitating the development of high-quality MLIPs for chemically complex atomic systems. We employ uncertainty-biased MD simulations to generate candidate pools for AL algorithms. Our results show that applying uncertainty bias facilitates simultaneous exploration of extrapolative regions and rare events. Efficient exploration of both is crucial in constructing comprehensive training data sets, enabling the development of uniformly accurate MLIPs. In contrast, classical enhanced sampling techniques (e.g., metadynamics) or unbiased MD simulations at elevated temperatures and pressures often cannot simultaneously explore extrapolative regions and rare events. Enhanced sampling techniques were designed to ensure the reconstruction of the underlying Boltzmann distribution. However, this property is unnecessary for data set generation and may limit their effectiveness in this context.

The performance of enhanced sampling techniques depends on the manual definition of hyper-parameters, e.g., CVs for metadynamics. Setting them requires expert knowledge because the wrong choice can limit the range of explored configurations. Uncertainty-biased MD only needs to define an uncertainty threshold and biasing strength. Both parameters influence the exploration rate of configurational space without constraining the space that can be explored. Under milder conditions, uncertainty-biased MD simulations outperform their unbiased counterparts for a broad range of biasing strength values, making the latter's choice more accessible. Yet, the dependence of the performance on the biasing strength value becomes more noticeable under extreme conditions, sometimes with no improvement by adding uncertainty bias to MD. A similar behavior can also be expected for metadynamics simulations.\cite{Nagyfalusi2017} Additionally, employing species-dependent biasing strength can restrict biasing in sensitive configurational regions, e.g., biasing hydrogen atoms.

Identifying extreme conditions like high temperatures and pressures can also accelerate phase space exploration in unbiased MD. However, a wrong choice of temperature and pressure may result in unphysical force predictions and degradation of the atomic system. In contrast, uncertainty-biased MD, conducted under milder conditions, explores relevant phase space at rates comparable to those obtained under extreme conditions and reduces the risk of degrading the atomic system. As mentioned, uncertainty-biased MD simulations outperform their unbiased counterparts for a broad range of biasing strength values in our setting. Furthermore, while evaluating uncertainty gradients increases the inference times by a factor of 1.4 to 1.7 compared to unbiased MD, applying uncertainty bias leads to, on average, shorter MD simulations. Thus, the difference in the computational cost between biased and unbiased MD is typically insignificant.

We compare uncertainty quantification methods, including the variance of an ensemble of MLIPs, and ensemble-free methods derived from sketched gradient features, focusing on configurational space exploration rates and generating uniformly accurate potentials; see the Supplementary Information. Overall, gradient-based approaches yield MLIPs with similar performance to those created using ensemble-based uncertainty while significantly reducing the computational cost of uncertainty quantification. For MIL-53(Al), we find that ensemble-based uncertainties overestimate the force error more strongly than gradient-based approaches, resulting in earlier termination of MD simulations and potentially worse configurational space exploration. For alanine dipeptide, using an ensemble of MLIPs improves their robustness during MD simulations, facilitating CV space exploration. Therefore, improving the robustness of a single MLIP during an MD simulation is a promising research direction,\cite{Ibayashi2023} combined with the proposed ensemble-free techniques.

While this study thoroughly investigates AL with uncertainty-biased MD for generating candidate pools, further research is still necessary. For example, one should analyze how well uncertainty-biased MD explores a configurational space with multiple stable states and how it identifies the respective slow modes using solely uncertainty bias. Also, assessing the uniform accuracy of resulting MLIPs and the enhanced exploration in higher-dimensional CV spaces remains challenging. Furthermore, the applicability of the proposed data generation approach to more complex molecular and material systems, such as biological polymers\cite{Zhao2022} and multicomponent alloys,\cite{Gubaev2023} is yet to be explored. Unlike MD, Monte Carlo simulations generally allow significant configurational changes, eliminating the need to explore intermediate transition paths. Combined with uncertainty bias, they might avoid exploring intermediate, low-uncertainty transition regions, improving the efficiency of uncertainty-driven data generation. Lastly, the extent to which MLIPs based on graph NNs can enhance the efficiency of the proposed data generation approach remains to be seen.

\section*{Methods} \label{sec:methods}

\subsection*{Machine-learned interatomic potentials}

We define an atomic configuration, $S = \{\mathbf{r}_i, Z_i\}_{i=1}^{N_\mathrm{at}}$, where $\mathbf{r}_i \in \mathbb{R}^3$ are Cartesian coordinates and $Z_i \in \mathbb{N}$ is the atomic number of atom $i$, with a total of $N_\mathrm{at}$ atoms. Our focus lies on interatomic NN potentials, which map an atomic configuration to a scalar energy $E$. The mapping is denoted as $f_{\boldsymbol{\theta}}: S \mapsto E \in \mathbb{R}$, where $\boldsymbol{\theta}$ denotes the trainable parameters. By assuming the locality of interatomic interactions, we decompose the total energy of the system into individual atomic contributions\cite{Behler2007}
\begin{equation}
 \label{eq:energy-model}
E\left(S, \boldsymbol{\theta}\right) = \sum_{i=1}^{N_\mathrm{at}} E_i\left(S_i, \boldsymbol{\theta}\right),
\end{equation}
where $S_i$ is the local environment of atom $i$, defined by the cutoff radius $r_\mathrm{c}$. The trainable parameters $\boldsymbol{\theta}$ are learned from atomic data sets containing atomic configurations and their energies, atomic forces, and stress tensors.

\subsection*{Gradient-based uncertainties}

We quantify the uncertainty of a trained MLIP by expanding its energy per atom $E_\mathrm{at} = E / N_\mathrm{at}$ around the locally optimal parameters $\bm{\theta}^\ast$\cite{Zaverkin2021a, Zaverkin2022d, Holzmueller2022}
\begin{equation}
 \label{eq:taylor-expansion}
 E_\mathrm{at}\left(S, \boldsymbol{\theta}\right) \approx E_\mathrm{at}\left(S, \boldsymbol{\theta}^\ast\right) + \left(\boldsymbol{\theta} - \boldsymbol{\theta}^\ast\right)^\top\underbrace{\nabla_{\boldsymbol{\theta}} E_\mathrm{at} \left(S, \boldsymbol{\theta}\right)\Big|_{\boldsymbol{\theta} = \boldsymbol{\theta}^\ast}}_{=\phi\left(S\right)},
\end{equation}
where $S$ denotes an atomic configuration as defined in the previous section. Gradient features $\phi\left(S\right) \in \mathbb{R}^{N_\mathrm{feat}}$ can be interpreted as the sensitivity of the energy to small parameter perturbations. Here, $N_\mathrm{feat}$ is the number of trainable parameters of the MLIP. We employ the energy per atom $E_\mathrm{at}$ in \eqref{eq:taylor-expansion}, as it accounts for the extensive nature of the energy, whose value depends on the system size. This choice ensures that uncertainties defined using gradient features do not favor the selection of larger structures. Gradient features can also be expressed as the mean of their atomic contributions: $\phi = \sum_{i=1}^{N_\mathrm{at}} \phi_i / N_\mathrm{at}$. For atomic gradient features $\phi_i$, using the energy per atom in \eqref{eq:taylor-expansion} is unnecessary. Here, we use $\phi=\phi\left(S\right)$ and $\phi_i=\phi_i\left(S_i\right)$, with $S_i$ denoting the local environment of an atom $i$, to simplify the notation. Thus, gradient features can be used to quantify uncertainties in total and atom-based properties of an atomic system, such as energy and atomic forces, respectively.

Particularly, we define the atom-based model's uncertainty (atomic forces) by employing squared distances between atomic gradient features
\begin{equation}
 \label{eq:distance-based-uncertainty}
 u_i^2 = \min_{\phi_j \in \Phi_\mathrm{train}} \lVert \phi_i - \phi_j \rVert_2^2.
\end{equation}
Alternatively, we consider Bayesian linear regression in \eqref{eq:taylor-expansion} and compute the posterior uncertainty as
\begin{equation}
 \label{eq:posterior-based-uncertainty}
 u_i^2 = \lambda^2 \phi_i^\top \left(\Phi_\mathrm{train}^\top\Phi_\mathrm{train} + \lambda^2\mathbf{I}\right)^{-1} \phi_i,
\end{equation}
where $\lambda$ is the regularization strength. Here, we define $\Phi_\mathrm{train} = \phi_j\left(\mathcal{X}_\mathrm{train}\right) \in \mathbb{R}^{\left(N_\mathrm{at}\cdot N_\mathrm{train}\right) \times N_\mathrm{feat}}$ with $\mathcal{X}_\mathrm{train}$ denoting the local atomic environments of configurations in the training set of size $N_\mathrm{train}$. In this work, we refer to our uncertainties as distance- and posterior-based uncertainties. Equivalent results can be obtained for total uncertainties (energy), employing gradient features $\phi = \sum_{i=1}^{N_\mathrm{at}} \phi_i / N_\mathrm{at}$ with $\Phi_\mathrm{train} = \phi\left(\mathcal{X}_\mathrm{train}\right) \in \mathbb{R}^{N_\mathrm{train} \times N_\mathrm{feat}}$.

Calculating uncertainties using gradient features is computationally challenging, especially for the posterior-based approach, for which a single uncertainty evaluation scales as $\mathcal{O}\left(N_\mathrm{feat}^2\right)$. Therefore, we employ the sketching technique\cite{Woodruff2014} to reduce the dimensionality of gradient features, i.e., $\phi_i^\mathrm{rp} = \mathbf{U}\phi_i \in \mathbb{R}^{N_\mathrm{rp}}$ with $N_\mathrm{rp}$ and $\mathbf{U} \in \mathbb{R}^{N_\mathrm{rp}\times N_\mathrm{feat}}$ denoting the number of random projections and a random matrix, respectively.\cite{Zaverkin2022d, Holzmueller2022} In previous work,\cite{Zaverkin2022d} we have observed that uncertainties derived from sketched gradient features demonstrate a better correlation with RMSEs of related properties than those based on last-layer features.\cite{Janet2019, Zaverkin2021a, Zhu2023} More details on sketched gradient features can be found in the following sections. Atom-based uncertainties, defined by the distances between gradient features, scale linearly with both the system size and the number of training structures, i.e., as $\mathcal{O}\left(N_\mathrm{at}N_\mathrm{train}\right)$. Consequently, they require an additional approximation to ensure computational efficiency. To address this, we employed the batch selection algorithm that maximizes distances within the training set, allowing us to identify the most representative subset of atomic gradient features; see the following sections.

\subsection*{Uncertainty-biased molecular dynamics}

Following previous work,\cite{Kulichenko2023, vanerOord2022} we define the biased energy as
\begin{equation}
E^\mathrm{biased}\left(S, \boldsymbol{\theta}\right) = E\left(S, \boldsymbol{\theta}\right) - \tau u\left(S, \boldsymbol{\theta}\right),
\end{equation}
where $\tau$ denotes the biasing strength. The negative sign ensures that negative uncertainty gradients with respect to atomic positions (bias forces) drive the system toward high uncertainty regions; see \figref{fig:scheme} (c). In this work, we use AD to compute bias forces acting on atom $i$, denoted as $-\nabla_{\mathbf{r}_i} u\left(S, \boldsymbol{\theta}\right)$ with atomic positions $\mathbf{r}_i$. The total biased force on atom $i$ reads
\begin{equation}
\mathbf{F}_i^\mathrm{biased}\left(S, \boldsymbol{\theta}\right) = \mathbf{F}_i\left(S, \boldsymbol{\theta}\right) + \tau \nabla_{\mathbf{r}_i} u\left(S, \boldsymbol{\theta}\right).
\end{equation}
These biased forces can be used for MD simulations in, e.g., canonical ($NVT$) statistical ensemble to bias the exploration of the configurational space.

In the case of bulk atomic systems, the configurational space often includes variations in cell parameters, which define the shape and size of the unit cell, necessitating enhanced exploration of them. For this purpose, we propose the concept of bias stress, defined by 
\begin{equation*}
 \frac{1}{V}\left.\nabla_{\boldsymbol{\epsilon}} u\left(S, \boldsymbol{\theta}\right)\right|_{\boldsymbol{\epsilon} = \mathbf{0}},
\end{equation*}
with $V$ denoting the volume of the periodic cell. This expression is motivated by the definition of the stress tensor.\cite{Knuth2015} Here, $u\left(S, \boldsymbol{\theta}\right)$ denotes the uncertainty after a strain deformation of the bulk atomic system with the symmetric tensor $\boldsymbol{\epsilon} \in \mathbb{R}^{3 \times 3}$, i.e., $\tilde{\mathbf{r}} = \left(\mathbf{1} + \boldsymbol{\epsilon}\right) \cdot \mathbf{r}$. The calculation of the bias stress is straightforward with AD. The total biased stress reads
\begin{equation}
 \label{eq:bias-stress}
\boldsymbol{\sigma}^\mathrm{biased}\left(S, \boldsymbol{\theta}\right) = \boldsymbol{\sigma}\left(S, \boldsymbol{\theta}\right) - \tau \frac{1}{V}\left.\nabla_{\boldsymbol{\epsilon}} u\left(S, \boldsymbol{\theta}\right)\right|_{\boldsymbol{\epsilon} = \mathbf{0}}.
\end{equation}
The bias stress tensor in \eqref{eq:bias-stress} effectively reduces the internal pressure in the bulk atomic system. We propose combining the bias stress tensor with MD simulations conducted in isothermal–isobaric ($NpT$) statistical ensemble to enhance the data-driven exploration of cell parameters and pressure-induced transitions in bulk materials.

Uncertainty gradients exhibit different magnitudes compared to energy gradients. Thus, re-scaling uncertainty gradients is necessary to ensure consistent driving toward uncertain regions. Building upon the approach introduced in Ref.~\citenum{vanerOord2022}, we implement a re-scaling technique that monitors the magnitudes of both actual and bias forces (alternatively, actual and bias stresses) over $N$ steps and then computes the ratio between them. To re-scale bias forces, we use the following expression
\begin{equation}
\tau_t = \tau_0 \times \frac{\sqrt{\sum_{n=0}^{N-1} \lVert \mathbf{F}_{t - n \Delta t}\rVert_2^2} }{ \sqrt{\sum_{n=0}^{N-1} \lVert \nabla_{\mathbf{r}_i} u_{t - n \Delta t}\rVert_2^2}}.
\end{equation}
An equivalent expression is applied for bias stresses.

The re-scaling of uncertainty gradients is reminiscent of the AdaGrad algorithm,\cite{Duchi2011} which dynamically adjusts the learning rate (analogous to the biasing strength) based on historical gradients from previous iterations. While incorporating momentum through exponential moving averages can improve the AdaGrad approach, treating all past gradients with equal weight is essential within the context of this study. Our attempts to damp learning along directions with high curvature (high-frequency oscillations), similar to the Adam optimizer,\cite{Adam2015} did not yield improved performance. We further find that employing species-dependent biasing strengths for bias forces, $\tau \rightarrow \tau_{Z_i}$, with a particular emphasis on damping biasing of hydrogen atoms, improves the efficiency of biased MD simulations.

We employ biased MD simulation to generate a candidate pool for AL, as depicted in \figref{fig:scheme} (a). We employ multiple parallel MD simulations to enhance the exploration of the configurational space further and improve the computational efficiency of AL. We expect biased MD simulations to have relatively short auto-correlation times (ACTs) obtained from position and uncertainty auto-correlation functions (ACFs). Short ACTs imply that the generated candidates will be less correlated than those generated with unbiased MD simulations. However, we cannot guarantee the generation of uncorrelated samples with biased MD simulations throughout AL, particularly in later AL iterations when the uncertainty level is reduced. Therefore, we propose to use batch selection algorithms (see later sections) that select $N_\mathrm{batch} > 1$ samples at once. These algorithms enforce the informativeness and diversity of the selected atomic configurations and the resulting training data set.

\subsection*{Gaussian moment neural network}

This work uses the Gaussian moment neural network (GM-NN) approach for modeling interatomic interactions.\cite{Zaverkin2020, Zaverkin2021b} GM-NN employs an artificial NN to map a local atomic environment $S_i$ to the atomic energy $E_i\left(S_i, \boldsymbol\theta\right)$; see \eqref{eq:energy-model}. It uses a fully-connected feed-forward NN with two hidden layers\cite{Zaverkin2020, Zaverkin2021b}
\begin{equation}
 \label{eq:neural-network}
\begin{split}
 y_i &= 0.1 \cdot \mathbf{b}^{(3)} + \frac{1}{\sqrt{d_2}} \mathbf{W}^{(3)} \phi\left(0.1 \cdot \mathbf{b}^{(2)} + \right. \\ 
 & \quad \left. \frac{1}{\sqrt{d_1}} \mathbf{W}^{(2)} \phi \left(0.1 \cdot \mathbf{b}^{(1)} + \frac{1}{\sqrt{d_0}} 
 \mathbf{W}^{(1)} \mathbf{G}_i\right)\right),
\end{split}
\end{equation}
with $\mathbf{W}^{(l+1)} \in \mathbb{R}^{d_{l+1} \times d_l}$ and $\mathbf{b}^{(l+1)} \in \mathbb{R}^{d_{l+1}}$ representing the weights and biases of layer $l+1$. In this work, we employ a NN with $d_0 = 910$ input neurons (corresponding to the dimension of the input feature vector $\mathbf{G}_i = \mathbf{G}_i\left(S_i\right)$), $d_1 = d_2 = 512$ hidden neurons, and a single output neuron, $d_3 = 1$. The network's weights $\mathbf{W}^{(l+1)}$ are initialized by selecting entries from a normal distribution with zero mean and unit variance. The trainable bias vectors $\mathbf{b}^{(l+1)}$ are initialized to zero. To improve the accuracy and convergence of the GM-NN model, we implement a neural tangent parameterization (factors of $0.1$ and $1/\sqrt{d_l}$).\cite{Jacot2018} For the activation function $\phi$, we use the Swish/SiLU function.\cite{Elfwing2018, Ramachandran2018}

To aid the training process, we scale and shift the output of the NN
\begin{equation}
E_{i}\left(S_i, \boldsymbol{\theta}\right) = c \cdot (\rho_{Z_i} y_i + \mu_{Z_i}),
\end{equation}
where the trainable shift parameters $\mu_{Z_i}$ are initialized by solving a linear regression problem, and the trainable scale parameters $\rho_{Z_i}$ are initialized to one. The per-atom RMSE of the regression solution determines the constant $c$.\cite{Zaverkin2021b}

GM-NN models employ the Gaussian moment (GM) representation to encode the invariance of total energy with respect to translations, rotations, and permutations of the same species.\cite{Zaverkin2020} By computing pairwise distance vectors $\mathbf{r}_{ij} = \mathbf{r}_i - \mathbf{r}_j$ and then splitting them into radial and angular components, denoted as $r_{ij} = \lVert \mathbf{r}_{ij} \rVert_2$ and $\hat{\mathbf{r}}_{ij} = \mathbf{r}_{ij}/r_{ij}$, respectively, we obtain GMs as follows
\begin{equation}
 \label{eq:local-tensor}
 \boldsymbol{\Psi}_{i, L, s} = \sum_{j \neq i} R_{Z_i, Z_j, s}\left(r_{ij}, \boldsymbol{\beta}\right) \hat{\mathbf{r}}_{ij}^{\otimes L},
\end{equation}
where $\hat{\mathbf{r}}_{ij}^{\otimes L} = \hat{\mathbf{r}}_{ij} \otimes \cdots \otimes \hat{\mathbf{r}}_{ij}$ is the $L$-fold outer product. The nonlinear radial functions $R_{Z_i, Z_j, s}\left(r_{ij}, \boldsymbol{\beta}\right)$ are defined as a sum of Gaussian functions $\Phi_{s^\prime}\left(r_{ij}\right)$ ($N_\mathrm{Gauss} = 9$ for this work)\cite{Zaverkin2021b}
\begin{equation}
 \label{eq:radial_function}
 R_{Z_i, Z_j, s}\left(r_{ij}, \boldsymbol{\beta}\right) = \frac{1}{\sqrt{N_\mathrm{Gauss}}}\sum_{s^\prime=1}^{N_\mathrm{Gauss}}\beta_{Z_i, Z_j, s, s^\prime} \Phi_{s^\prime}\left(r_{ij}\right).
\end{equation}
The factor $1/\sqrt{N_\mathrm{Gauss}}$ impacts the effective learning rate inspired by neural tangent parameterization.\cite{Jacot2018} The radial functions are centered at equidistantly spaced grid points ranging from $r_\mathrm{min}=0.5$~\AA{} to $r_\mathrm{c}$, set to 5.0~\AA{} and 6.0~\AA{} for alanine dipeptide and MIL-53(Al), respectively. The radial functions are re-scaled by a cosine cutoff function,\cite{Behler2007} to ensure a smooth dependence on the number of atoms within the cutoff sphere. Chemical information is embedded in the GM representation through trainable parameters $\beta_{Z_i, Z_j, s, s^\prime}$, with the index $s$ iterating over the number of independent radial basis functions ($N_\mathrm{basis} = 7$ for this work). 

Features invariant to rotations, $\mathbf{G}_i$, are obtained by computing full tensor contractions of tensors defined in \eqref{eq:local-tensor}, e.g.,\cite{Zaverkin2020, Zaverkin2021b}
\begin{equation}
 \label{eq:example_contraction}
 G_{i, s_1, s_2, s_3} = \left(\boldsymbol{\Psi}_{i,1,s_1}\right)_{a} \left(\boldsymbol{\Psi}_{i,1,s_2}\right)_{b}\left(\boldsymbol{\Psi}_{i,2,s_3}\right)_{a,b},
\end{equation}
where we use Einstein notation, i.e., the right-hand side is summed over $a, b \in \{1, 2, 3\}$. Specific full tensor contractions are defined by using generating graphs.\cite{Suk2011} In a practical implementation, we compute all GMs at once and reduce the number of invariant features based on the permutational symmetries of the respective graphs.

All parameters $\boldsymbol{\theta}=\{ \mathbf{W}, \textbf{b}, \boldsymbol{\beta}, \boldsymbol{\rho}, \boldsymbol{\mu} \}$ of the NN are optimized by minimizing the combined squared loss on training data $\mathcal{D}_\mathrm{train} = \left(\mathcal{X}_\mathrm{train}, \mathcal{Y}_\mathrm{train}\right)$, with $\mathcal{X}_\mathrm{train} = \{S^{(k)}\}_{k=1}^{N_\mathrm{train}}$ and $\mathcal{Y}_\mathrm{train} = \{E_k^\mathrm{ref}, \{\mathbf{F}_{i,k}^\mathrm{ref}\}_{i=1}^{N_\mathrm{at}}, \boldsymbol{\sigma}_k^\mathrm{ref}\}_{k=1}^{N_\mathrm{train}}$,
\begin{equation}
 \label{eq:loss}
 \begin{split}
 \mathcal{L}\left( \boldsymbol{\theta}, \mathcal{D}_\mathrm{train}\right) = \sum_{k=1}^{N_\mathrm{train}} \Bigg[ C_\mathrm{e} & \Big\lVert E_k^\mathrm{ref} - E(S^{(k)}, \boldsymbol{\theta})\Big\rVert_2^2 + \\ C_\mathrm{f} \sum_{i=1}^{N_\mathrm{at}^{(k)}} & \Big\lVert \mathbf{F}_{i,k}^\mathrm{ref} - \mathbf{F}_i\left(S^{(k)}, \boldsymbol{\theta}\right)\Big\rVert_2^2 + \\ C_\mathrm{s} &\Big\lVert V_k \boldsymbol{\sigma}_{k}^\mathrm{ref} - V_k \boldsymbol{\sigma}\left(S^{(k)}, \boldsymbol{\theta}\right)\Big\rVert_2^2\Bigg].
 \end{split}
\end{equation}
We have chosen $C_\mathrm{e}=1.0$, $C_\mathrm{f} = 4.0~\AA^2$, and $C_\mathrm{s} = 0.01$ to balance the relative contributions of energies, forces, and stresses, respectively. 

Using AD, we compute atomic forces as negative gradients of total energy with respect to atomic coordinates
\begin{equation}
 \mathbf{F}_i\left(S^{(k)}, \boldsymbol{\theta}\right) = -\nabla_{\mathbf{r}_i} E\left(S^{(k)}, \boldsymbol{\theta}\right).
\end{equation}
Furthermore, we use AD to compute stress tensor, defined by\cite{Knuth2015}
\begin{equation}
 \boldsymbol{\sigma}\left(S^{(k)}, \boldsymbol{\theta}\right) = \frac{1}{V_k}\left.\nabla_{\boldsymbol{\epsilon}} E\left(S^{(k)}, \boldsymbol{\theta}\right)\right|_{\boldsymbol{\epsilon} = \mathbf{0}},
\end{equation}
where $E\left(S^{(k)}, \boldsymbol{\theta}\right)$ is total energy after a strain deformation with symmetric tensor $\boldsymbol{\epsilon} \in \mathbb{R}^{3 \times 3}$, i.e., $\tilde{\mathbf{r}} = \left(\mathbf{1} + \boldsymbol{\epsilon}\right) \cdot \mathbf{r}$. As the stress tensor is symmetric, we use only its upper triangular part in the loss function. Here, $V_k$ is the volume of the periodic cell.

We employ the Adam optimizer\cite{Adam2015} to minimize the loss function. The respective parameters of the optimizer are $\beta_1=0.9$, $\beta_2=0.999$, and $\epsilon=10^{-7}$. Usually, we work with a mini-batch of 32 molecules. However, smaller mini-batches were used in the initial AL iterations because the training data sizes were less than 32. The layer-wise learning rates are decayed linearly. The initial values are set to 0.03 for the parameters of the fully connected layers, 0.02 for the trainable representation, as well as 0.05 and 0.001 for the shift and scale parameters of atomic energies, respectively. The training is performed for 1000 training epochs. To prevent overfitting during training, we employ the early stopping technique.\cite{Prechelt2012} All models are trained using PyTorch.\cite{Paszke2019}

\subsection*{Sketched gradient features}

We obtain atomic gradient features by computing gradients of \eqref{eq:energy-model} with respect to the parameters of the fully connected layers in \eqref{eq:neural-network}. Particularly, we make use of the product structure of atomic gradient features. To obtain the latter, we re-write the network in \eqref{eq:neural-network} as follows
\begin{equation}
 \begin{split}
 & \mathbf{z}_i^{(l+1)} = \tilde{\mathbf{W}}^{(l+1)} \tilde{\mathbf{x}}_i^{(l)} \in \mathbb{R}^{d_{l+1}}, \\
 & \tilde{\mathbf{W}}^{(l+1)} = \left(\mathbf{W}^{(l+1)}, \mathbf{b}^{(l+1)}\right) \in \mathbb{R}^{d_{l+1} \times \left(d_{l}+1\right)}, \\
 & \tilde{\mathbf{x}}_i^{(l)} = \left(\frac{1}{\sqrt{d_l}}\mathbf{x}_i^{(l)}, 0.1\right)^\top \in \mathbb{R}^{d_{l}+1},
 \end{split}
\end{equation}
where $\mathbf{z}^{(l)}$ and $\mathbf{x}^{(l)}$ denote the pre- and post-activation vectors of layer $l$. Thus, atomic gradient features read
\begin{equation}
 \label{eq:full-gradient-features}
 \begin{split}
 \phi_i(S_i) & = \left(\frac{\partial \mathbf{z}_i^{(L)}}{\partial \tilde{\mathbf{W}}^{(1)}}, \cdots, \frac{\partial \mathbf{z}_i^{(L)}}{\partial \tilde{\mathbf{W}}^{(L)}}\right) \\
 & = \left(\frac{\partial \mathbf{z}_i^{(L)}}{\partial \mathbf{z}_i^{(1)}} \otimes \tilde{\mathbf{x}}_i^{(0)}, \cdots, \frac{\partial \mathbf{z}_i^{(L)}}{\partial \mathbf{z}_i^{(L)}} \otimes \tilde{\mathbf{x}}_i^{(L-1)} \right).
 \end{split} 
\end{equation}

To make the calculation of gradient features computationally tractable, we employ the random projections (sketching) technique,\cite{Woodruff2014} as proposed in Refs.~\citenum{Holzmueller2022, Zaverkin2022d}. For atomic gradient features $\phi_i\left(S_i\right) \in \mathbf{R}^{N_\mathrm{feat}}$ and a random matrix $\mathbf{U} \in \mathbb{R}^{N_\mathrm{rp} \times N_\mathrm{feat}}$---with $N_\mathrm{feat}$ and $N_\mathrm{rp}$ denoting the number of atomic features and random projections, respectively---we can define randomly projected atomic gradient features as
\begin{equation}
 \phi_i^\mathrm{rp}\left(S_i\right) = \mathbf{U} \phi_i\left(S_i\right) \in \mathbb{R}^{N_\mathrm{rp}}.
\end{equation}
While a Gaussian sketch could be employed, where the elements of $\mathbf{U}$ are drawn from standard normal distributions, we use a tensor sketching approach that is more runtime and memory efficient.\cite{Holzmueller2022} Specifically, denoting the element-wise or Hadamard product as $\odot$, we compute
\begin{equation}
 \label{eq:sketched-atomic-features}
 \phi_i^\mathrm{rp}(S_i) = \sum_{l=1}^L \left(\mathbf{U}_\mathrm{out}^{(l)} \phi_{i, \mathrm{out}}^{(l)}(S_i)\right) \odot \left(\mathbf{U}_\mathrm{in}^{(l-1)} \phi_{i, \mathrm{in}}^{(l-1)}(S_i)\right),
\end{equation}
with $\phi_{i, \mathrm{out}}^{(l)}(S_i) = \partial \mathbf{z}_i^{(L)}/\partial \mathbf{z}_i^{(l)}$ and $\phi_{i, \mathrm{in}}^{(l)}(S_i) = \tilde{\mathbf{x}}_i^{(l)}$. All entries of $\mathbf{U}_\mathrm{in}^{(l)}$ and $\mathbf{U}_\mathrm{out}^{(l)}$ are sampled independently from a standard normal distribution.

For atom-based uncertainties, we can directly use the sketched atomic gradient features. For (total) uncertainties per atom, we need to work with a mean $\phi(S) = \sum_{i=1}^{N_{\mathrm{at}}} \phi_i(S_i) / N_\mathrm{at}$. Thus, we use that the individual projections (rows of \eqref{eq:sketched-atomic-features}) are linear in the features and obtain for the (total) gradient features\cite{Zaverkin2022d}
\begin{equation}
 \phi^\mathrm{rp}(S) = \frac{1}{N_\mathrm{at}}\sum_{i=1}^{N_\mathrm{at}} \sum_{l=1}^L \left(\mathbf{U}_\mathrm{out}^{(l)} \phi_{i, \mathrm{out}}^{(l)}(S_i)\right) \odot \left(\mathbf{U}_\mathrm{in}^{(l-1)} \phi_{i, \mathrm{in}}^{(l-1)}(S_i)\right),
\end{equation}
given that all of the individual random projections use the same random matrices.

\subsection*{Ensemble-based uncertainty quantification}

The variance of the predictions of individual models in an ensemble of MLIPs can be used to quantify their uncertainty. Thus, we define the variance of predicted energy as
\begin{equation}
 u^2 = \frac{1}{M}\sum_{j=1}^M\lVert E_{j} - \bar{E}\rVert_2^2,
\end{equation}
where $M$ is the number of models in the ensemble. The variance of atomic forces reads
\begin{equation}
 u_i^2 = \frac{1}{3M}\sum_{j=1}^M\lVert\mathbf{F}_{i,j} - \bar{\mathbf{F}}_{i}\rVert_2^2,
\end{equation}
Here, $\bar{E}$ and $\bar{\mathbf{F}}_{i}$ denote the arithmetic mean of the predictions from individual models. Our experiments demonstrated that $M=3$ is sufficient to obtain good performance. Using larger ensembles would make the ensemble-based uncertainty quantification even more computationally inefficient than gradient-based alternatives.

\subsection*{Batch selection methods}

The simplest batch selection method is based on querying points only by their uncertainty values. Specifically, given the already selected structures $\mathcal{X}_\mathrm{batch}$ from an unlabeled pool $\mathcal{X}_\mathrm{pool}$ we select the next point by
\begin{equation}
 \label{eq:max_diag}
 S = \underset{S \in \mathcal{X}_\mathrm{pool}\backslash\mathcal{X}_\mathrm{batch}}{\argmax}\,u\left(S\right),
\end{equation}
until $N_\mathrm{batch} > 1$ structures are selected. In this work, we use this selection method combined with ensemble-based uncertainties.

For the posterior-based uncertainty, we can constrain the diversity of the selected batch by using the posterior covariance between structures
\begin{equation}
 \mathrm{Cov}\left(S, S^\prime\right) = \lambda^2 \phi\left(S\right)^\top \left(\Phi_\mathrm{train}^\top\Phi_\mathrm{train} + \lambda^2\mathbf{I}\right)^{-1} \phi\left(S^\prime\right),
\end{equation}
with $\Phi_\mathrm{train} = \phi\left(\mathcal{X}_\mathrm{train}\right)$. The corresponding method greedily selects structures, i.e., one structure per iteration, such that the determinant of the covariance matrix is maximized\cite{Kirsch2019, Zaverkin2022d, Holzmueller2022}
\begin{equation}
 \begin{split}
 & S = \\ & \underset{S \in \mathcal{X}_\mathrm{pool}\backslash\mathcal{X}_\mathrm{batch}}{\argmax}\,\det \left[ \mathrm{Cov}\left(\mathcal{X}_\mathrm{batch} \cup \{S\}, \mathcal{X}_\mathrm{batch} \cup \{S\}\right)\right].
 \end{split}
\end{equation}

For the distance-based uncertainty, we ensure the diversity of the acquired batch by greedily selecting structures with a maximum distance to all previously selected and training data points. The respective selection method reads\cite{Sener2017, Zaverkin2022d, Holzmueller2022}
\begin{equation}
 \begin{split}
 & S = \\ & \underset{S \in \mathcal{X}_\mathrm{pool}\backslash\mathcal{X}_\mathrm{batch}}{\argmax}\,\underset{S^\prime \in \mathcal{X}_\mathrm{train} \cup \mathcal{X}_\mathrm{batch}}{\min}\, \lVert \phi\left(S\right) - \phi\left(S^\prime\right) \rVert_2^2.
 \end{split}
\end{equation}
We also applied this batch selection method to define the most representative subset of atomic gradient features when calculating atom-based uncertainty using feature space distances.

Lastly, to compare the performance of uncertainty-based data generation approaches with conventional random sampling from an ab initio MD, we employ a random selection strategy combined with posterior-based uncertainty to terminate MD simulations. We define random selection as
\begin{equation}
    S \sim \mathcal{U}\left(\mathcal{X}_\mathrm{pool}\right),
\end{equation}
where $\mathcal{U}$ is the uniform distribution over $\mathcal{X}_\mathrm{pool}$.

\subsection*{Conformal prediction}

Conformal prediction methods offer distribution-free uncertainty quantification with guaranteed finite sample coverage,\cite{Vovk2005, Lei2017, Romano2019, Angelopoulos2022, Hu2022} thus ensuring calibration. Finite sample coverage can be defined as
\begin{equation}
 \mathbb{P}\{y_\mathrm{test} \in C\left(x_\mathrm{test}\right)\} \geq 1-\alpha.
\end{equation}
Here, $\left(x_\mathrm{test}, y_\mathrm{test}\right)$ are the newly observed data, while $C$ defines the prediction set based on previous observations $\{\left(x_k, y_k\right)\}_{k=1}^{N_\mathrm{calibr}}$. The user determines the hyper-parameter $\alpha$ and defines the desired confidence level. CP methods guarantee that the prediction set contains the true label with a probability of almost $1-\alpha$.

We employ inductive CP, which comprises the following steps:\cite{Vovk2005, Hu2022} (1) A subset of calibration data, sized $N_\mathrm{calibr}$, is selected, and the corresponding errors are computed on this subset. For atomic forces, we employ RMSEs $\Delta \mathbf{F}_i^2 = \lVert\mathbf{F}_{i} - \mathbf{F}_{i}^\mathrm{ref}\rVert_2^2 / 3$, while for total energies the respective energy absolute errors per atom, $\Delta e = \lvert E - E^\mathrm{ref}\rvert / N_\mathrm{at}$, are used. (2) The uncertainty $u\left(S\right)$ is calculated for this subset of data. (3) The ratio $\Delta e / u\left(S\right)$ or $\Delta \mathbf{F}_i / u\left(S_i\right)$ is computed. (4) Utilizing quantile regression, the $\left(1-\alpha\right)\left(N_\mathrm{calibr}+1\right)/N_\mathrm{calibr}$-th quantile, denoted as $s$, is determined. (5) This $s$ value is applied to new observations, resulting in the re-scaled and calibrated uncertainty, $\tilde{u} = s \cdot u$.

\subsection*{Coverage of collective variable space}

To measure how well different methods explore the (bounded) space of interest, we implement a tree-based weighted recursive partitioning of a $d$-dimensional Euclidean space, which is reminiscent of quadtrees \cite{finkel1974quad} and matrix-based octrees \cite{meagher1980octree} but allows to choose how many times $n$ to split each dimension. Thus, the variety of the tree is $k = n^d$. Each node of this complete k-ary tree encodes a generalized hypercube of $d$ dimensions, where each side length depends on the boundaries of the original space. The root node represents the full bounded space. A tree of height $L$ has total number of partitions equal to $(k^{L+1} - 1)/(k-1)$, and each level $\ell$ has $k^\ell$ nodes. The hyper-parameters we choose in this paper are $n=2$, $d=2$ (for the CVs $\phi$ and $\psi$ of alanine dipeptide), and $L=5$, for a total of $1365$ partitions of the space of interest.

Our proposed surface coverage metric uses this data structure as a proxy to capture how many space partitions a method can explore in the least amount of iterations. At the same time, we need to penalize methods that get stuck in a region of the space, exploring partitions of smaller volumes, that is, those represented by nodes at deeper levels in the tree. For this reason, each node at level $\ell$ is associated with a reward (or weight) of $1/k^{\ell}$, so each level of the tree has a cumulative reward of $1$. The optimal strategy would be to perform a breadth-first search of the nodes of this tree, which translates into observing the largest partitions of unobserved space first. In addition, partitions that are revisited by the methods give no additional reward, so there is no gain in getting stuck in a certain partition. We visually represent the idea of the algorithm in the Supplementary Information for the simple case of $d=2$.

\subsection*{Auto-correlation analysis}

We evaluate the performance of uncertainty-biased MD simulations by investigating the auto-correlation between subsequent time frames of the MD trajectory. The auto-correlation function (ACF) is defined as\cite{Janke2013}
\begin{equation}
 A_\mathcal{O}\left(k\right) = \frac{\langle\mathcal{O}_i\mathcal{O}_{i+k}\rangle - \langle\mathcal{O}_i\rangle^2}{\langle\mathcal{O}_i^2\rangle - \langle\mathcal{O}_i\rangle^2},
\end{equation}
where $\langle\cdots\rangle$ denotes the thermodynamic expectation value, $k$ is the lag time, and $\mathcal{O}$ is an observable, e.g., atomic positions or atom-based uncertainties. From ACF, we can calculate the auto-correlation time (ACT) for an MD trajectory of length $N$
\begin{equation}
    \mathrm{ACT}_\mathcal{O} = \frac{1}{2} + \sum_{k=1}^N A_\mathcal{O}\left(k\right)\left(1 - \frac{k}{N}\right).
\end{equation}
ACT is related to effective sample size (ESS) by
\begin{equation}
 \mathrm{ESS}_\mathcal{O} = \frac{N}{2 \cdot \mathrm{ACT}_\mathcal{O}}.
\end{equation}
In this work, we calculate ESS as implemented in TensorFlow\cite{Dillon2017} and use it to estimate the ACT.

\subsection*{Test data set for alanine dipeptide}

The test data set for alanine dipeptide comprises 2000 configurations randomly selected from an MD trajectory at 1200~K. This trajectory was generated within the ASE simulation package\cite{Hjorth2017} by running an MD simulation in the canonical ($NVT$) statistical ensemble using the Langevin thermostat. We have used a time step of 0.5~fs and a total simulation time of 1~ns. The AMBER ff19SB force field has provided forces,\cite{Tian2020} as implemented in the TorchMD package using PyTorch.\cite{Doerr2021, Paszke2019} The data set effectively covers the relevant configurational space of alanine dipeptide, representing an upper boundary in exploring its collective variables (CVs).

\subsection*{MLIP learning details for alanine dipeptide}

Each AL experiment starts with training an MLIP with eight alanine dipeptide configurations randomly perturbed from its initial configuration in the $\mathrm{C}_\mathrm{7eq}$ state. Trained MLIPs are then used to run eight parallel MD simulations, initialized from the initial configuration or configurations selected in later iterations. Each MD simulation runs until reaching an empirically defined uncertainty threshold of 1.5~eV/{\AA}. A lower threshold value may result in slower CV space exploration, while a larger one would lead to the exploration of unphysical configurations. The maximum data set size, comprising training and validation data, is limited to 512 configurations. The Supplementary Information presents the scaling of the presented AL experiments to larger data set sizes, acquiring data sets of 1024 samples. Biased (bias-forces-driven) and unbiased MD simulations are performed using the canonical ($NVT$) statistical ensemble within the ASE simulation package.\cite{Hjorth2017} Unbiased MD simulations are run with the Langevin thermostat at temperatures of 300~K, 600~K, and 1200~K, whereas biased simulations are performed at a constant temperature of 300~K. We have chosen an integration time step of 0.5~fs and set a maximum of 20,000 steps for an MD simulation. A biasing strength of $\tau=0.25$ was also chosen for biased AL experiments. In reference calculations, we employ a force threshold of 20~eV/\AA{} to exclude unphysical structures, potentially expected at high biasing strengths (equivalently, a smaller integration time step could be used). All AL experiments have been repeated five times.

\subsection*{Reference DFT calculations for MIL-53(Al)}

DFT calculations for MIL-53(Al) were performed using the CP2K simulation package (version 2023.1).\cite{Kuehne2020} To ensure consistency with incremental learning experiments,\cite{Vandenhaute2023} we employed the PBE functional\cite{Perdew1996a} with Grimme D3 dispersion correction.\cite{Grimme2010} A hybrid basis set, combining TZVP Gaussian basis functions and plane waves, was employed.\cite{Lippert1997} GTH pseudopotentials were used to smoothen the electron density near the nuclei.\cite{Goedecker1996} To ensure the convergence of force and stress calculations, a plane wave cutoff energy of 1000~Ry was selected.

\subsection*{MLIP learning details for MIL-53(Al)}

In each AL experiment, we start with 32 MIL-53(Al) configurations randomly perturbed around its closed-pore state, with 90~\% reserved for training. Trained MLIPs are then used to perform 32 parallel MD simulations, each running until it reaches an uncertainty threshold of 1.0~eV/\AA. The maximum data set size is limited to 512 configurations, comprising training and validation data. The Supplementary Information presents the scaling of the presented AL experiments to larger data set sizes, acquiring data sets of 1024 samples. Both biased (bias-stress-driven) and unbiased MD simulations use the isothermal–isobaric form of the Nos{\'e}--Hoover dynamics.\cite{Melchionna1993, Melchionna2000} Unbiased MD simulations are carried out at 600~K and 0~MPa, as well as $\pm$ 250~MPa (half of the simulations each), while biased simulations are performed at 600~K and 0~MPa. The characteristic time scales of the thermostat and barostat are set to 0.1~ps and 1~ps, respectively. We have chosen an integration time step of 0.5~fs and set a maximum of 20,000 MD steps for an MD simulation. A stress-biasing strength of $\tau=0.5$ is used in biased AL experiments. In reference calculations, we employ a force threshold of 20~eV/\AA{} to exclude strongly distorted structures. We use the data set from Ref.~\citenum{Vandenhaute2023} as a metadynamics-generated baseline and select the first 500 sequentially generated configurations. All AL experiments are repeated three times, except for metadynamics, which was run once.\cite{Vandenhaute2023} For metadynamics, we train three MLIPs initialized using different random seeds.

\subsection*{Random perturbation of atomic configurations}

We obtain randomly perturbed atomic configurations by adding atomic shifts, denoted as $\boldsymbol{\delta}_i$, to the original atomic positions $\mathbf{r}_i$
\begin{equation}
    \tilde{\mathbf{r}}_i = \mathbf{r}_i + \boldsymbol{\delta}_i.
\end{equation}
The components of $\boldsymbol{\delta}_i$ are sampled independently from a uniform distribution: for alanine dipeptide, the range is between $-0.02$~\AA{} and $0.02$~\AA{}, and for MIL-53(Al), it is between $-0.08$~\AA{} and $0.08$~\AA{}. Additionally, for MIL-53(Al), we introduce random perturbations to its periodic cell $\mathbf{B}$ using a strain deformation $\boldsymbol{\epsilon}=\left(\mathbf{A} + \mathbf{A}^\top\right)/2$, where the components of $\mathbf{A}$ are sampled independently from a uniform distribution between $-0.02$ and $0.02$. This transformation can be expressed as
\begin{equation}
    \tilde{\mathbf{B}} = \mathbf{B} \left(\mathbf{I} + 2\boldsymbol{\epsilon}\right)^{1/2}.
\end{equation}
The shifted atomic positions are re-scaled according to
\begin{equation}
    \tilde{\tilde{\mathbf{r}}}_i = \left(\mathbf{I} + 2\boldsymbol{\epsilon}\right)^{1/2} \tilde{\mathbf{r}}_i.
\end{equation}

\subsection*{Sine wave with additive random noise}

We model large-amplitude volume fluctuations in MIL-53(Al) induced by the bias stress using a sine wave with period $T_0$ and additive random noise $N\left(t\right)$
\begin{equation*}
    A \sin\left(\frac{2 \pi t}{T_0}\right) + B N\left(t\right),
\end{equation*}
where $A$ and $B$ denote the sine wave's amplitude and random noise, respectively. In this work, $N\left(t\right) \sim \mathcal{N}\left(0, 1\right)$ represents random noise following a normal distribution with zero mean and unit variance. We chose $A = 1.0$ and $B=0.5$ for the blue line in \figref{fig:mil53_positions_acfs_300K}. For the red line, we increase the noise amplitude to $B=2.0$. To represent the volume fluctuations induced in MIL-53(Al) (see \figref{fig:mil53_positions_acfs_300K}), a sine wave with the period twice the length of the MD simulation, i.e., $T_0 = 3.2$~ns is required.

\section*{Data Availability}

The data sets generated during this study are available in the Zenodo repository: \url{https://doi.org/10.5281/zenodo.10776838}. The MIL-53(Al) test data set is available at \url{https://doi.org/10.5281/zenodo.6359970} (Ref.~\citenum{Vandenhaute2023}).

\section*{Code Availability}

The source code for this study is available on GitHub and can be accessed via this link: \url{https://github.com/nec-research/alebrew}.

\section*{Acknowledgements}

Funded by Deutsche Forschungsgemeinschaft (DFG, German Research Foundation) under Germany's Excellence Strategy - EXC 2075 -- 390740016. We acknowledge the support by the Stuttgart Center for Simulation Science (SimTech). The authors thank the International Max Planck Research School for Intelligent Systems (IMPRS-IS) for supporting David Holzm{\"u}ller.

\section*{Author contributions}

All authors designed the project, discussed the results, and wrote the manuscript. V.Z. performed the calculations.

\section*{Competing interests}

The authors declare no competing interests.

\section*{Additional information}

\textbf{Supplementary Information} accompanies.

\bibliography{references}


\clearpage

\begin{appendices}

\setcounter{equation}{0}
\setcounter{figure}{0}
\setcounter{table}{0}

\renewcommand{\theequation}{\arabic{equation}}

\renewcommand{\figurename}{Supplementary Figure}
\renewcommand{\tablename}{Supplementary Table}

\section*{Overview}

In this Supplementary Information, we complement the results presented in the main text. We follow the structure of the latter to align our arguments and observations.

\section*{Supplementary Results}

\subsection*{Calibrating uncertainties with conformal prediction}

\begin{figure*}[htbp]
\centering
\includegraphics[width=\textwidth]{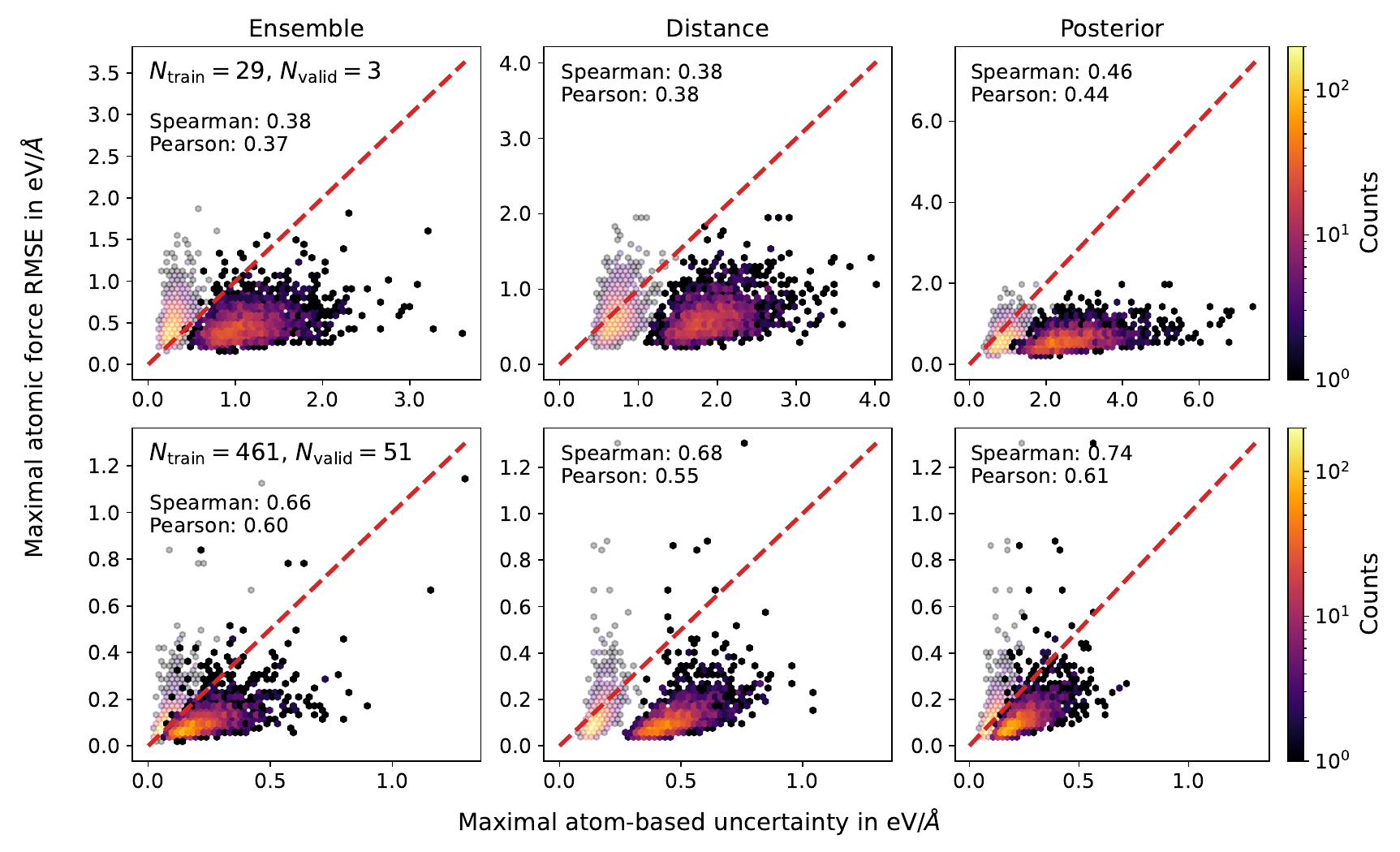}
\caption{Correlation of \textit{maximal atom-based uncertainties} with \textit{maximal atomic force RMSEs} for \textit{alanine dipeptide}. The results are presented for the alanine dipeptide test data set; see Methods. All uncertainty quantification methods are calibrated using CP and atomic force RMSEs. The top row shows the results of MLIPs trained using 29 atomic configurations, while three are additionally used for early stopping and uncertainty calibration. The bottom row shows the results obtained with $461$ and $51$ atomic configurations, respectively. The training and validation data are drawn from the same MD trajectory as the test data; see Methods. Transparent hexbin points represent uncertainties calibrated with $\alpha=0.5$ (low confidence; see Methods), while opaque ones denote uncertainties calibrated with $\alpha=0.05$ (high confidence).}
\label{fig:diala_uncertainty}
\end{figure*}

\Figref{fig:diala_uncertainty} presents the correlation between maximal atom-based uncertainties and maximal atomic force RMSEs for alanine dipeptide. Atom-based uncertainties are calibrated using CP with atomic force RMSEs evaluated on the calibration data. For the top row of \figref{fig:diala_uncertainty}, we draw 29 training and three validation (and calibration) configurations from the MD trajectory used for generating the test data; see Methods. We observe that posterior-based uncertainties outperform distance- and ensemble-based ones regarding Pearson and Spearman correlation coefficients. The same trend is observed for the bottom row of \figref{fig:diala_uncertainty}, where 461 configurations have been used for training and 51 for validation (and calibration).

\Figref{fig:diala_uncertainty} demonstrates that using CP with a higher confidence level prevents MLIPs from underestimating force errors, which differs from low confidence. For MLIPs trained with 29 structures, calibrated gradient-based uncertainties overestimate the actual force error more strongly than ensemble-based ones. This observation may explain a better performance of MD simulations with ensemble-based uncertainties regarding CV space exploration, as the overestimation of actual force errors by ensemble-free uncertainties may lead to premature termination of corresponding MD simulations. However, we find that ensemble-based AL experiments feature, on average, shorter MD trajectories than their gradient-based counterparts. Thus, we relate the improvement in the CV space coverage and the respective exploration rates to the enhanced robustness of an MLIP ensemble during MD simulations. The robustness of MLIP ensembles is associated with averaging out prediction errors of individual models, thus facilitating the exploration of extrapolative but physically meaningful regions.

\begin{figure*}[htbp]
\centering
\includegraphics[width=\textwidth]{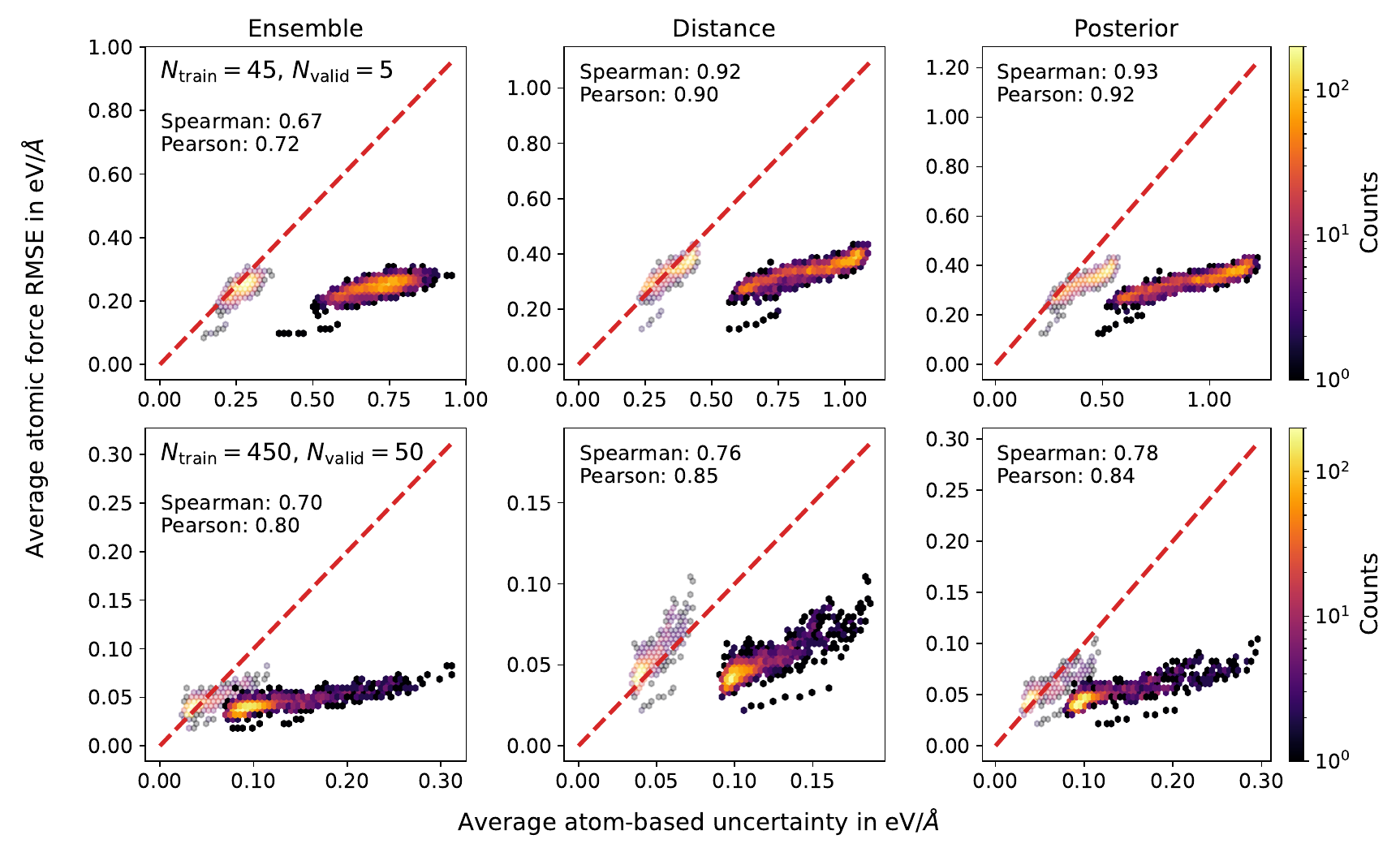}
\caption{Correlation of \textit{average atom-based uncertainties} with \textit{average atomic force RMSEs} for \textit{MIL-53(Al)}. The results are presented for the test data set from Supplementary Reference~\citenum{Vandenhaute2023}. All uncertainty quantification methods are calibrated using CP and atomic force RMSEs. The top row shows the results of MLIPs trained using 45 atomic configurations, while five are additionally used for early stopping and uncertainty calibration. The bottom row shows the results obtained with 450 and 50 MIL-53(Al) configurations, respectively. The training and validation data are taken from Supplementary Reference~\citenum{Vandenhaute2023}. Transparent hexbin points represent uncertainties calibrated with $\alpha=0.5$ (low confidence; see Methods), while opaque ones denote uncertainties calibrated with $\alpha=0.05$ (high confidence). The offset in hexbin points observed with $\alpha=0.05$ arises from calibrating atom-based uncertainties with atomic force RMSEs. However, this offset does not impact correlation coefficients.}
\label{fig:mil53_avg_uncertainty}
\end{figure*}

\begin{figure*}[htbp]
\centering
\includegraphics[width=\textwidth]{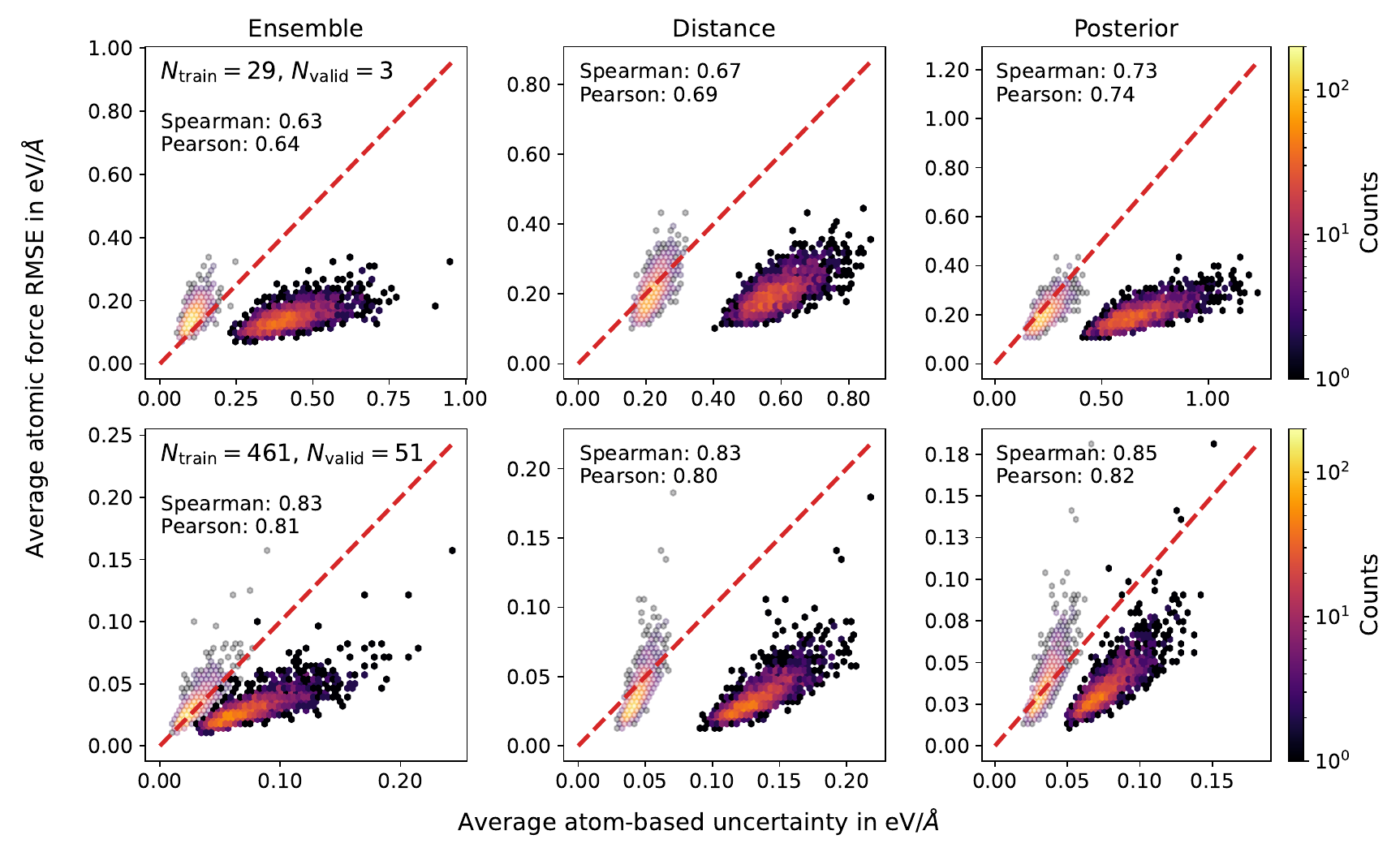}
\caption{Correlation of \textit{average atom-based uncertainties} with \textit{average atomic force RMSEs} for \textit{alanine dipeptide}. The results are presented for the alanine dipeptide test data set; see Methods. All uncertainty quantification methods are calibrated using CP and atomic force RMSEs. The top row shows the results of MLIPs trained using 29 atomic configurations, while three are additionally used for early stopping and uncertainty calibration. The bottom row shows the results obtained with $461$ and $51$ atomic configurations, respectively. The training and validation data are drawn from the same MD trajectory as the test data; see Methods. Transparent hexbin points represent uncertainties calibrated with $\alpha=0.5$ (low confidence; see Methods), while opaque ones denote uncertainties calibrated with $\alpha=0.05$ (high confidence). The offset in hexbin points observed with $\alpha=0.05$ arises from calibrating atom-based uncertainties with atomic force RMSEs. However, this offset does not impact correlation coefficients.}
\label{fig:diala_avg_uncertainty}
\end{figure*}

Supplementary Figures~\ref{fig:mil53_avg_uncertainty} and \ref{fig:diala_avg_uncertainty} examine the correlation between average atom-based uncertainties and average atomic force RMSEs for MIL-53(Al) and alanine dipeptide, respectively. These figures highlight that average atom-based uncertainties correlate more strongly with average atomic force RMSEs than maximal atom-based uncertainties with maximal atomic force RMSEs. Thus, assessing the predictive power of atom-based uncertainties using maximal atomic force RMSEs is crucial, as they underestimate the latter stronger than average force RMSEs. Furthermore, average atomic force RMSEs and average atom-based uncertainties can be less sensitive to large errors as these can be averaged out and go unnoticed during MD simulation. Consequently, an MD simulation may explore unphysical regions before the respective uncertainties can terminate the simulation.

\subsection*{Performance of bias-forces-driven active learning}

\begin{figure*}[htbp]
	\centering
	\includegraphics[width=\textwidth]{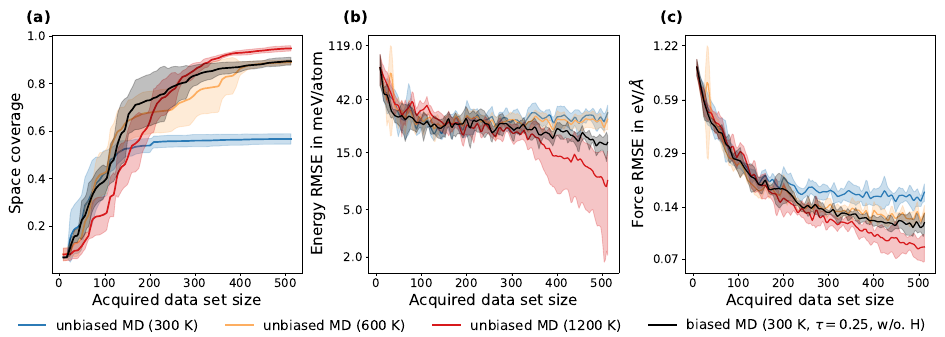}
	\caption{Comparison of AL approaches employing biased and unbiased MD simulations to generate the candidate pool of atomic configurations for alanine dipeptide. Results are provided for the \textit{distance-based uncertainty} quantification derived from sketched gradient features. Unlike unbiased MD simulations, which rely on atom-based uncertainties to terminate MD simulations, biased MD simulations use total and atom-based uncertainties to bias MD simulations and prompt their termination, respectively. We use three metrics to assess the performance of our AL approaches: \textbf{(a)} Coverage of the CV space; \textbf{(b)} RMSEs in predicted energies; and \textbf{(c)} RMSEs in atomic forces. All RMSEs are evaluated on the alanine dipeptide test data set; see Methods. Shaded areas denote the standard deviation across five independent runs.}
	\label{fig:diala_distance_performance}
\end{figure*}

\begin{figure*}[htbp]
	\centering
	\includegraphics[width=\textwidth]{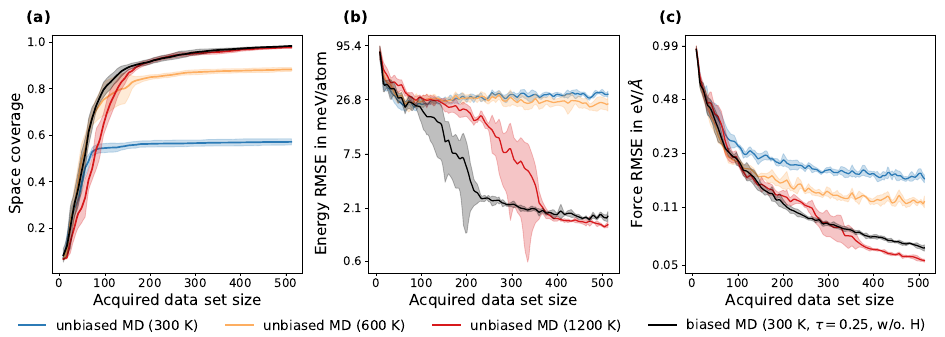}
	\caption{Comparison of AL approaches employing biased and unbiased MD simulations to generate the candidate pool of atomic configurations for alanine dipeptide. Results are provided for the \textit{ensemble-based uncertainty} quantification. Unlike unbiased MD simulations, which rely on atom-based uncertainties to terminate MD simulations, biased MD simulations use total and atom-based uncertainties to bias MD simulations and prompt their termination, respectively. We use three metrics to assess the performance of our AL approaches: \textbf{(a)} Coverage of the CV space; \textbf{(b)} RMSEs in predicted energies; and \textbf{(c)} RMSEs in atomic forces. All RMSEs are evaluated on the alanine dipeptide test data set; see Methods. Shaded areas denote the standard deviation across five independent runs.}
	\label{fig:diala_ensemble_performance}
\end{figure*}

Supplementary Figures~\ref{fig:diala_distance_performance} and \ref{fig:diala_ensemble_performance} demonstrate the performance of MLIPs trained with AL using distance- and ensemble-based uncertainties. Supplementary Tables~\ref{tab:ala2-results-distance} and \ref{tab:ala2-results-ensemble} present error metrics evaluated for MLIPs at the end of each experiment. We observe a better performance of MLIPs trained with AL that uses ensemble-based uncertainties, already during initial AL iterations indicated by a greater slope of the curve representing the CV space coverage. This improvement is also observed for unbiased MD simulations at milder conditions. Thus, it cannot be attributed to a better performance of the uncertainty bias compared to gradient-based methods but to the enhanced robustness of an MLIP ensemble. We find that uncertainty-biased MD simulations, which use MLIP ensembles, feature, on average, shorter MD trajectories than single MLIP models.

\subsection*{Exploration rates for collective variables of alanine dipeptide}

\begin{figure*}[htbp]
	\centering
	\includegraphics[width=\textwidth]{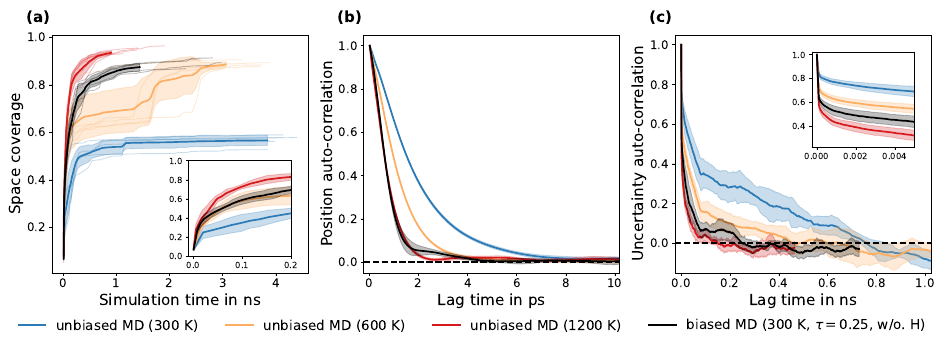}
	\caption{Evaluation of configurational space exploration rates for biased and unbiased MD simulations of alanine dipeptide. Here, MD simulations generate candidate pools of atomic configurations for AL algorithms. Results are provided for the alanine dipeptide molecule and use \textit{distance-based uncertainties} derived from sketched gradient features. Unlike unbiased MD simulations, which rely on atom-based uncertainties to terminate MD simulations, biased MD simulations use total and atom-based uncertainties to bias MD simulations and prompt their termination, respectively. We use three metrics to asses the exploration rates: \textbf{(a)} Coverage of the CV space; \textbf{(b)} Auto-correlation functions of atomic positions; and \textbf{(c)} Auto-correlation functions of atom-based uncertainties. Shaded areas denote the standard deviation across five independent runs.}
	\label{fig:diala_exploration_distance}
\end{figure*}

\begin{figure*}[htbp]
	\centering
	\includegraphics[width=\textwidth]{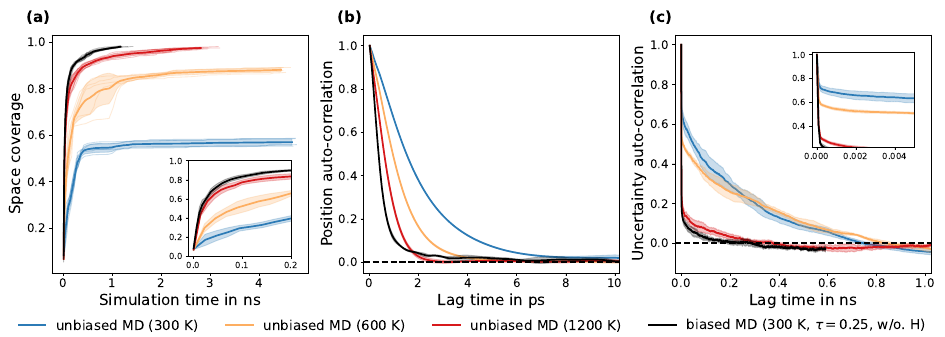}
	\caption{Evaluation of configurational space exploration rates for biased and unbiased MD simulations of alanine dipeptide. Here, MD simulations generate candidate pools of atomic configurations for AL algorithms. Results are provided for the \textit{ensemble-based uncertainty} quantification. Unlike unbiased MD simulations, which rely on atom-based uncertainties to terminate MD simulations, biased MD simulations use total and atom-based uncertainties to bias MD simulations and prompt their termination, respectively. We use three metrics to asses the exploration rates: \textbf{(a)} Coverage of the CV space; \textbf{(b)} Auto-correlation functions of atomic positions; and \textbf{(c)} Auto-correlation functions of atom-based uncertainties. Shaded areas denote the standard deviation across five independent runs.}
	\label{fig:diala_exploration_ensemble}
\end{figure*}

Supplementary Figures~\ref{fig:diala_exploration_distance} and \ref{fig:diala_exploration_ensemble} demonstrate the exploration rate analysis for uncertainty-biased and unbiased MD simulations that utilize distance- and ensemble-based uncertainties. Supplementary Tables~\ref{tab:ala2-results-distance} and \ref{tab:ala2-results-ensemble} present ACTs for the corresponding AL experiments. We find that ensemble-based MD simulations feature, on average, shorter MD trajectories and ACTs than their counterparts. Thus, we suggest that MLIP ensembles drive MD simulations more robustly toward unexplored regions for alanine dipeptide, leading to an even faster termination of MD simulations.

\subsection*{Performance of bias-stress-driven active learning}

\begin{figure*}[htbp]
	\centering
	\includegraphics[width=\textwidth]{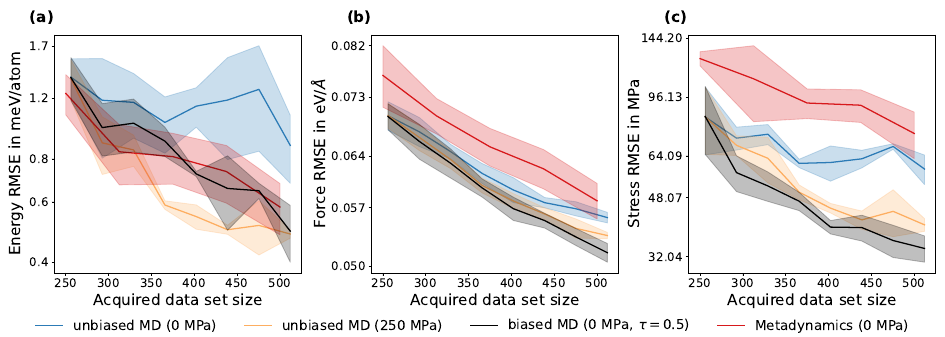}
	\caption{Comparison of AL approaches employing biased and unbiased \textit{MD simulations at 300~K} to generate the candidate pool of atomic configurations for MIL-53(Al). Results are provided for the \textit{posterior-based uncertainty} quantification derived from sketched gradient features. Unlike unbiased MD simulations, which rely on atom-based uncertainties to terminate MD simulations, biased MD simulations use total and atom-based uncertainties to bias MD simulations and prompt their termination, respectively. We use three metrics to assess the performance of our AL approaches: \textbf{(a)} Energy RMSE; \textbf{(b)} Force RMSE; and \textbf{(c)} Stress RMSE. All RMSEs are evaluated on the MIL-53(Al) test data set.\cite{Vandenhaute2023} Shaded areas denote the standard deviation across three independent runs, except for metadynamics. For it, shaded areas denote standard deviation across three randomly initialized MLIPs. All results are obtained for MD simulations run at 300~K, and AL experiments initialized using MLIPs trained with 256 closed-pore configurations drawn from the training data provided by Supplementary Reference~\citenum{Vandenhaute2023}.}
	\label{fig:mil53_performance_posterior_300K}
\end{figure*}

\Figref{fig:mil53_performance_posterior_300K} presents the performance of MLIPs trained with AL and MD simulations at 300~K and 0~MPa (or $\pm$250~MPa), a setting with suppressed large-pore phase exploration. The figure demonstrates that even unbiased MD simulations outperform metadynamics-based experiments in atomic force and stress RMSEs. We find that MLIPs can model the large-pore state, not explored during unbiased MD simulations, using the local environments from the closed-pore state. We enhance the MLIP performance by employing bias stress and observe that uncertainty-biased MD simulations outperform their high-pressure counterparts, exploring a larger portion of the configurational space; see Fig.~5 (e).

\begin{figure*}[htbp]
	\centering
	\includegraphics[width=\textwidth]{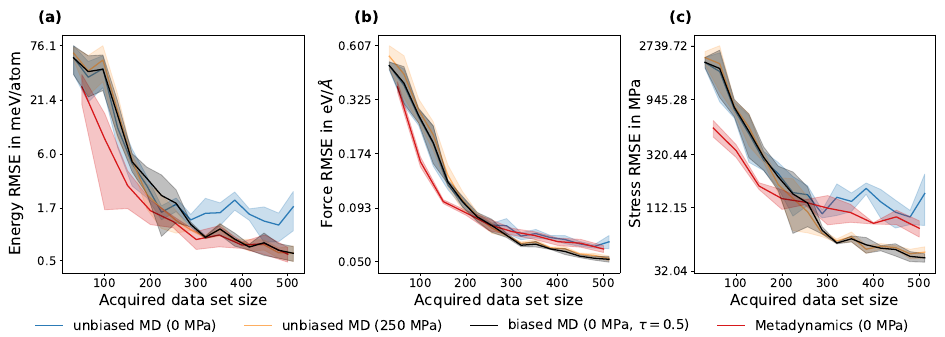}
	\caption{Comparison of AL approaches employing biased and unbiased \textit{MD simulations at 600~K} to generate the candidate pool of atomic configurations for MIL-53(Al). Results are provided for the \textit{distance-based uncertainty} quantification derived from sketched gradient features. Unlike unbiased MD simulations, which rely on atom-based uncertainties to terminate MD simulations, biased MD simulations use total and atom-based uncertainties to bias MD simulations and prompt their termination, respectively. We use three metrics to assess the performance of our AL approaches: \textbf{(a)} Energy RMSE; \textbf{(b)} Force RMSE; and \textbf{(c)} Stress RMSE. All RMSEs are evaluated on the MIL-53(Al) test data set.\cite{Vandenhaute2023} Shaded areas denote the standard deviation across three independent runs, except for metadynamics. For it, shaded areas denote standard deviation across three randomly initialized MLIPs. All results are obtained for MD simulations run at 600~K, and AL experiments initialized using MLIPs trained with 32 closed-pore configurations obtained by randomly distorting the initial MIL-53(Al) configuration.}
	\label{fig:mil53_performance_distance_600K}
\end{figure*}

\begin{figure*}[htbp]
	\centering
	\includegraphics[width=\textwidth]{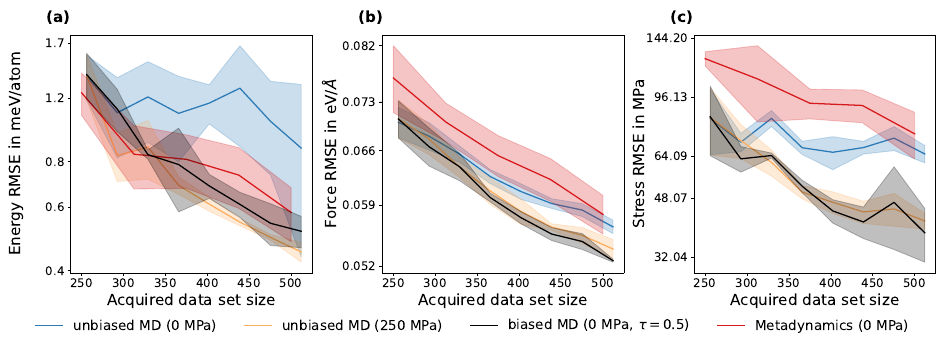}
	\caption{Comparison of AL approaches employing biased and unbiased \textit{MD simulations at 300~K} to generate the candidate pool of atomic configurations for MIL-53(Al). Results are provided for the \textit{distance-based uncertainty} quantification derived from sketched gradient features. Unlike unbiased MD simulations, which rely on atom-based uncertainties to terminate MD simulations, biased MD simulations use total and atom-based uncertainties to bias MD simulations and prompt their termination, respectively. We use three metrics to assess the performance of our AL approaches: \textbf{(a)} Energy RMSE; \textbf{(b)} Force RMSE; and \textbf{(c)} Stress RMSE. All RMSEs are evaluated on the MIL-53(Al) test data set.\cite{Vandenhaute2023} Shaded areas denote the standard deviation across three independent runs, except for metadynamics. For it, shaded areas denote standard deviation across three randomly initialized MLIPs. All results are obtained for MD simulations run at 300~K, and AL experiments initialized using MLIPs trained with 256 closed-pore configurations drawn from the training data provided by Supplementary Reference~\citenum{Vandenhaute2023}.}
	\label{fig:mil53_performance_distance_300K}
\end{figure*}

\begin{figure*}[htbp]
	\centering
	\includegraphics[width=\textwidth]{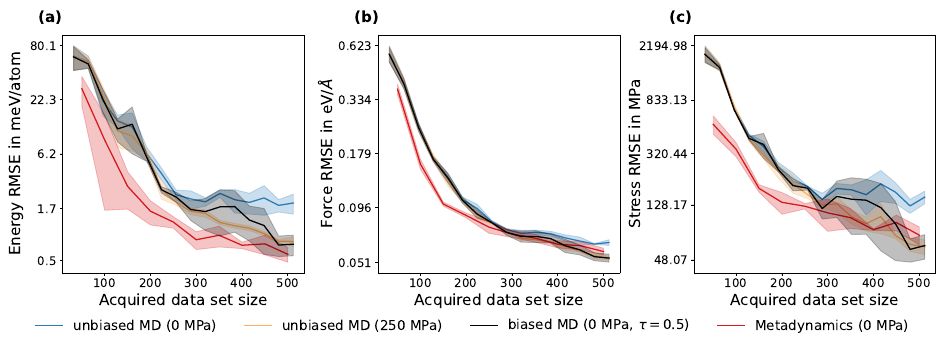}
	\caption{Comparison of AL approaches employing biased and unbiased \textit{MD simulations at 600~K} to generate the candidate pool of atomic configurations for MIL-53(Al). Results are provided for the \textit{ensemble-based uncertainty} quantification. Unlike unbiased MD simulations, which rely on atom-based uncertainties to terminate MD simulations, biased MD simulations use total and atom-based uncertainties to bias MD simulations and prompt their termination, respectively. We use three metrics to assess the performance of our AL approaches: \textbf{(a)} Energy RMSE; \textbf{(b)} Force RMSE; and \textbf{(c)} Stress RMSE. All RMSEs are evaluated on the MIL-53(Al) test data set.\cite{Vandenhaute2023} Shaded areas denote the standard deviation across three independent runs, except for metadynamics. For it, shaded areas denote standard deviation across three randomly initialized MLIPs. All results are obtained for MD simulations run at 600~K, and AL experiments initialized using MLIPs trained with 32 closed-pore configurations obtained by randomly distorting the initial MIL-53(Al) configuration.}
	\label{fig:mil53_performance_ensemble_600K}
\end{figure*}

\begin{figure*}[htbp]
	\centering
	\includegraphics[width=\textwidth]{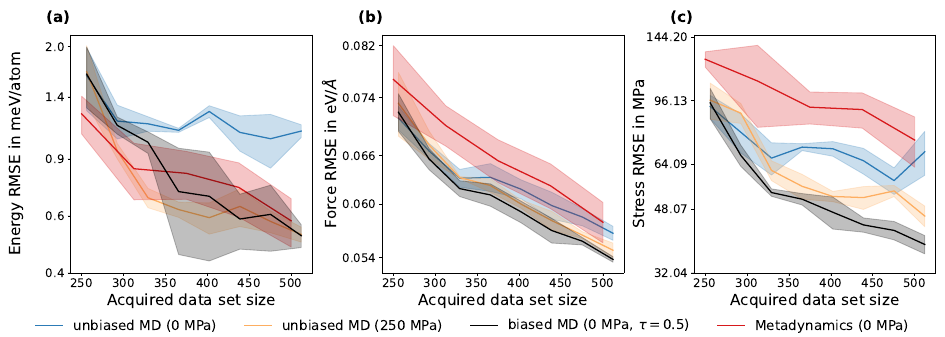}
	\caption{Comparison of AL approaches employing biased and unbiased \textit{MD simulations at 300~K} to generate the candidate pool of atomic configurations for MIL-53(Al). Results are provided for the \textit{ensemble-based uncertainty} quantification. Unlike unbiased MD simulations, which rely on atom-based uncertainties to terminate MD simulations, biased MD simulations use total and atom-based uncertainties to bias MD simulations and prompt their termination, respectively. We use three metrics to assess the performance of our AL approaches: \textbf{(a)} Energy RMSE; \textbf{(b)} Force RMSE; and \textbf{(c)} Stress RMSE. All RMSEs are evaluated on the MIL-53(Al) test data set.\cite{Vandenhaute2023} Shaded areas denote the standard deviation across three independent runs, except for metadynamics. For it, shaded areas denote standard deviation across three randomly initialized MLIPs. All results are obtained for MD simulations run at 300~K, and AL experiments initialized using MLIPs trained with 256 closed-pore configurations drawn from the training data provided by Supplementary Reference~\citenum{Vandenhaute2023}.}
	\label{fig:mil53_performance_ensemble_300K}
\end{figure*}

Supplementary Figures~\ref{fig:mil53_performance_distance_600K} and \ref{fig:mil53_performance_distance_300K} demonstrate the results for biased and unbiased MD simulations at 600~K and 300~K with distance-based uncertainties. Supplementary Figures~\ref{fig:mil53_performance_ensemble_600K} and \ref{fig:mil53_performance_ensemble_300K} show the corresponding results for the ensemble-based uncertainty quantification. All uncertainty-biased MD simulations outperform metadynamics regarding the atomic force and stress RMSEs. Ensemble-based uncertainty quantification leads to the worst performance across the chosen uncertainty methods, explained by the premature termination of MD simulations, in line with the results in Fig.~2.

\subsection*{Exploration rates for cell parameters of MIL-53(Al)}

\begin{figure*}[htbp]
	\centering
	\includegraphics[width=\textwidth]{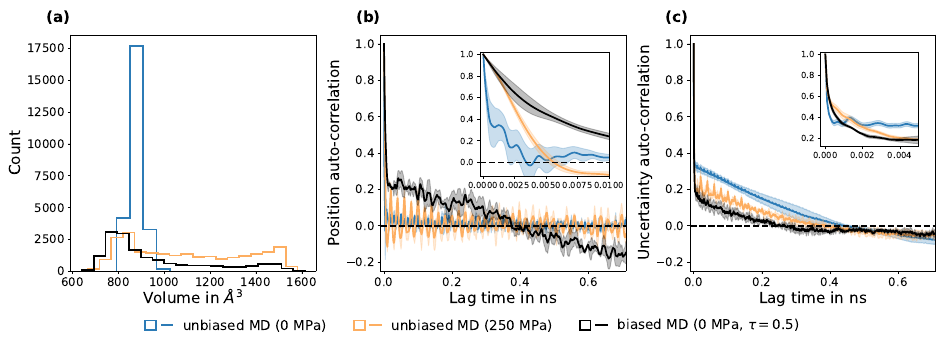}\vfill
	\includegraphics[width=\textwidth]{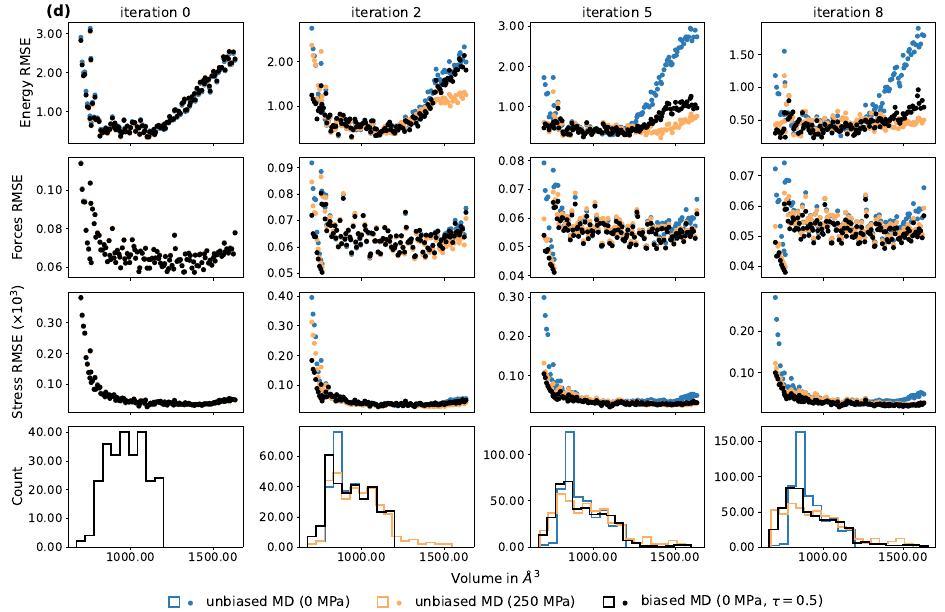}
	\caption{Evaluation of configurational space exploration rates for biased and unbiased \textit{MD simulations at 300~K} of MIL-53(Al). Here, MD simulations generate candidate pools of atomic configurations for AL algorithms. Results are provided for the \textit{posterior-based uncertainty} quantification derived from sketched gradient features. Unlike unbiased MD simulations, which rely on atom-based uncertainties to terminate MD simulations, biased MD simulations use total and atom-based uncertainties to bias MD simulations and prompt their termination, respectively. We use three metrics to asses the exploration rates: \textbf{(a)} Volume distribution of configurations sampled throughout the experiment; \textbf{(b)} Auto-correlation functions for positions; and \textbf{(c)} Auto-correlation functions for atom-based uncertainties. Shaded areas denote the standard deviation across three independent runs. \textbf{(d)} Time evolution of the volume distribution of configurations acquired during training and of energy, forces, and stress RMSEs evaluated on the test data set depending on the unit cell volume.}
	\label{fig:mil53_exploration_posterior_300K}
\end{figure*}

\Figref{fig:mil53_exploration_posterior_300K} (a)--(c) complements the investigation of the exploration rates in the main text with MD simulations at 300~K. We identify strongly correlated moves in position ACFs as large amplitude volume fluctuations. However, no correlated moves are observed for atom-based uncertainty ACF because biased MD simulations explore high uncertainty regions in each AL iteration. \Figref{fig:mil53_exploration_posterior_300K} (d) demonstrates that uncertainty bias leads to improved RMSE values already at early AL iteration. However, biased MD simulations are stronger driven toward smaller volumes already in the early iterations, different from high-pressure unbiased MD simulations and the results obtained at 600~K.

\begin{figure*}[htbp]
	\centering
	\includegraphics[width=\textwidth]{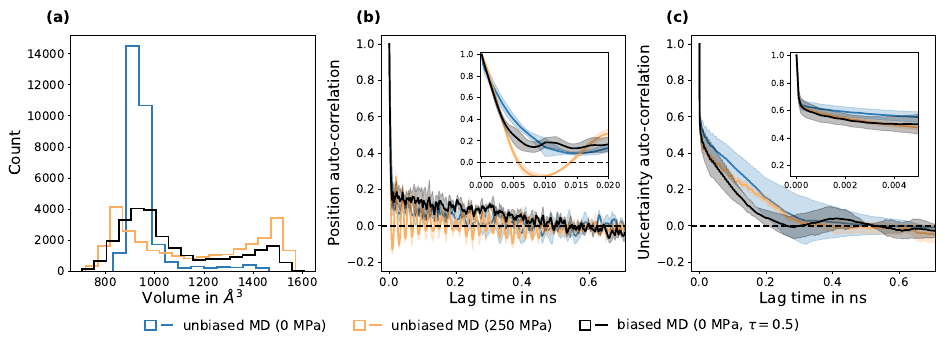}
	\caption{Evaluation of configurational space exploration rates for biased and unbiased \textit{MD simulations at 600~K} of MIL-53(Al). Here, MD simulations generate candidate pools of atomic configurations for AL algorithms. Results are provided for the \textit{distance-based uncertainty} quantification derived from sketched gradient features. Unlike unbiased MD simulations, which rely on atom-based uncertainties to terminate MD simulations, biased MD simulations use total and atom-based uncertainties to bias MD simulations and prompt their termination, respectively. We use three metrics to asses the exploration rates: \textbf{(a)} Volume distribution of configurations sampled throughout the experiment; \textbf{(b)} Auto-correlation functions for positions; and \textbf{(c)} Auto-correlation functions for atom-based uncertainties. Shaded areas denote the standard deviation across three independent runs.}
	\label{fig:mil53_exploration_distance_600K}
\end{figure*}

\begin{figure*}[htbp]
	\centering
	\includegraphics[width=\textwidth]{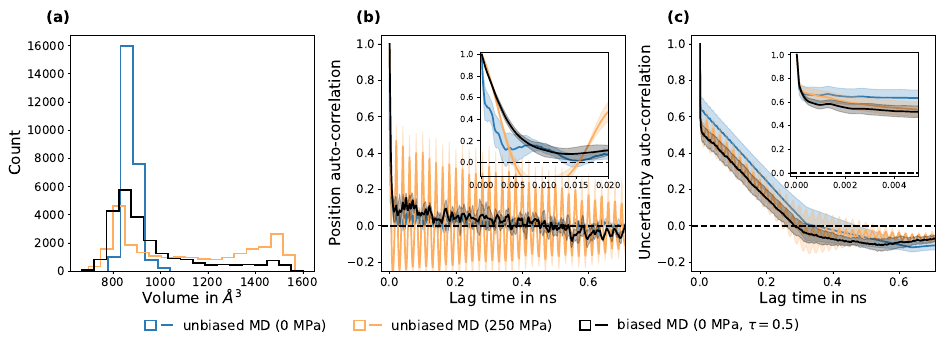}
	\caption{Evaluation of configurational space exploration rates for biased and unbiased \textit{MD simulations at 300~K} of MIL-53(Al). Here, MD simulations generate candidate pools of atomic configurations for AL algorithms. Results are provided for the \textit{distance-based uncertainty} quantification derived from sketched gradient features. Unlike unbiased MD simulations, which rely on atom-based uncertainties to terminate MD simulations, biased MD simulations use total and atom-based uncertainties to bias MD simulations and prompt their termination, respectively. We use three metrics to asses the exploration rates: \textbf{(a)} Volume distribution of configurations sampled throughout the experiment; \textbf{(b)} Auto-correlation functions for positions; and \textbf{(c)} Auto-correlation functions for atom-based uncertainties. Shaded areas denote the standard deviation across three independent runs.}
	\label{fig:mil53_exploration_distance_300K}
\end{figure*}

\begin{figure*}[htbp]
	\centering
	\includegraphics[width=\textwidth]{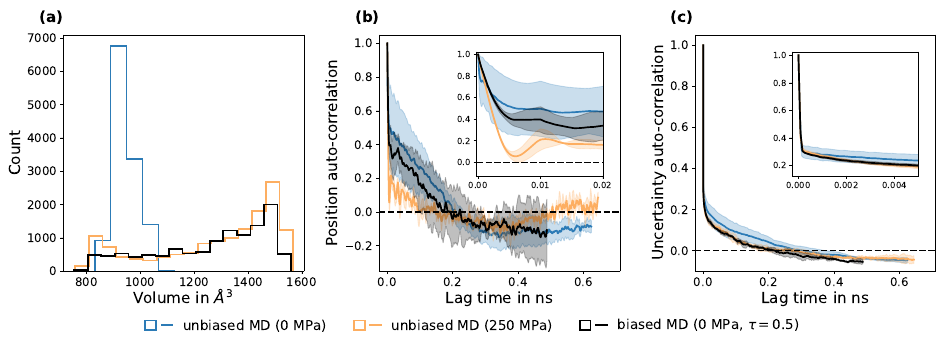}
	\caption{Evaluation of configurational space exploration rates for biased and unbiased \textit{MD simulations at 600~K} of MIL-53(Al). Here, MD simulations generate candidate pools of atomic configurations for AL algorithms. Results are provided for the \textit{ensemble-based uncertainty} quantification. Unlike unbiased MD simulations, which rely on atom-based uncertainties to terminate MD simulations, biased MD simulations use total and atom-based uncertainties to bias MD simulations and prompt their termination, respectively. We use three metrics to asses the exploration rates: \textbf{(a)} Volume distribution of configurations sampled throughout the experiment; \textbf{(b)} Auto-correlation functions for positions; and \textbf{(c)} Auto-correlation functions for atom-based uncertainties. Shaded areas denote the standard deviation across three independent runs.}
	\label{fig:mil53_exploration_ensemble_600K}
\end{figure*}

\begin{figure*}[htbp]
	\centering
	\includegraphics[width=\textwidth]{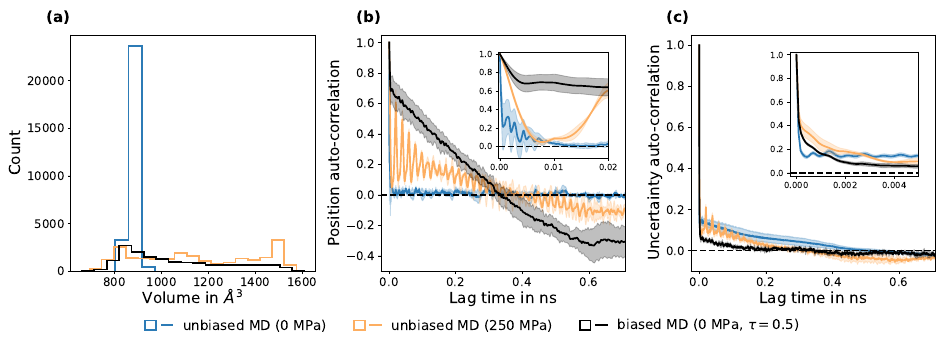}
	\caption{Evaluation of configurational space exploration rates for biased and unbiased \textit{MD simulations at 300~K} of MIL-53(Al). Here, MD simulations generate candidate pools of atomic configurations for AL algorithms. Results are provided for the \textit{ensemble-based uncertainty} quantification. Unlike unbiased MD simulations, which rely on atom-based uncertainties to terminate MD simulations, biased MD simulations use total and atom-based uncertainties to bias MD simulations and prompt their termination, respectively. We use three metrics to asses the exploration rates: \textbf{(a)} Volume distribution of configurations sampled throughout the experiment; \textbf{(b)} Auto-correlation functions for positions; and \textbf{(c)} Auto-correlation functions for atom-based uncertainties. Shaded areas denote the standard deviation across three independent runs.}
	\label{fig:mil53_exploration_ensemble_300K}
\end{figure*}

Supplementary Figures~\ref{fig:mil53_exploration_distance_600K} and \ref{fig:mil53_exploration_distance_300K} present the distribution of volumes as well as position and uncertainty ACFs for MD simulations at 300~K and 600~K, which use distance-based uncertainty. We observe that uncertainty bias facilitates the exploration of the large-pore state of MIL-53(Al) at 300~K. At 600~K, we obtain position ACFs for uncertainty-biased MD that decay slower than for unbiased MD simulations, explained similarly to Fig.~7. Supplementary Figures~\ref{fig:mil53_exploration_ensemble_600K} and \ref{fig:mil53_exploration_ensemble_300K} demonstrate that MD simulations biased with ensemble-based uncertainties explore the closed-pore state with a lower frequency {than with other uncertainty methods}. Thus, configurations with smaller unit cell volumes are predicted with larger energy, force, and stress RMSEs. Finally, ensemble-based uncertainty reaches high-uncertainty regions faster than its counterparts.

\subsection*{Scaling to larger data set sizes}

\begin{figure*}[htbp]
	\centering
	\includegraphics[width=\textwidth]{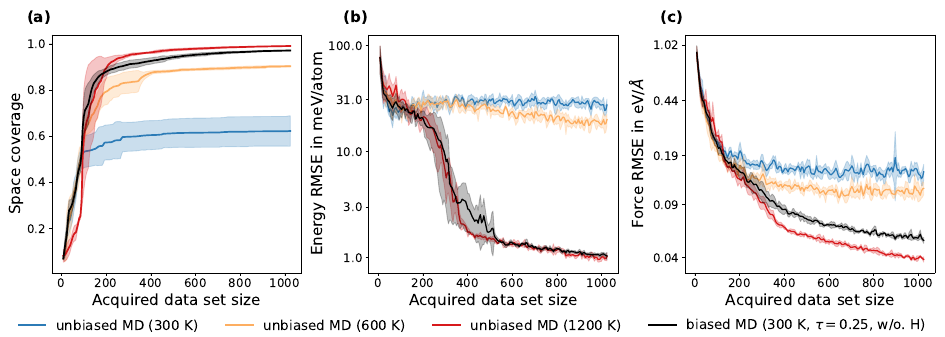}
	\caption{Comparison of AL approaches employing biased and unbiased MD simulations to generate the candidate pool of atomic configurations for alanine dipeptide. Results are provided for the \textit{posterior-based uncertainty} quantification derived from sketched gradient features and a \textit{maximal number of acquired data points of 1024}. Unlike unbiased MD simulations, which rely on atom-based uncertainties to terminate MD simulations, biased MD simulations use total and atom-based uncertainties to bias MD simulations and prompt their termination, respectively. We use three metrics to assess the performance of our AL approaches: \textbf{(a)} Coverage of the CV space; \textbf{(b)} RMSEs in predicted energies; and \textbf{(c)} RMSEs in atomic forces. All RMSEs are evaluated on the alanine dipeptide test data set; see Methods. Shaded areas denote the standard deviation across five independent runs.}
	\label{fig:diala_1024_performance}
\end{figure*}

\begin{figure*}[t!]
	\includegraphics[width=0.99\textwidth]{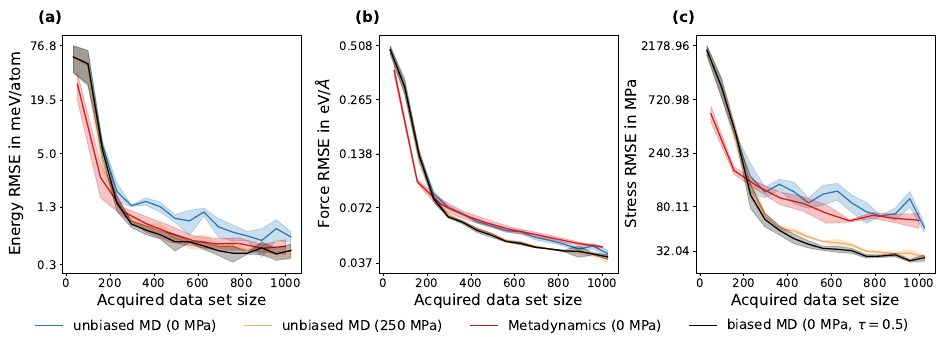}
	\caption{Comparison of AL approaches employing biased and unbiased MD simulations to generate the candidate pool of atomic configurations for MIL-53(Al). Results are provided for the \textit{posterior-based uncertainty} quantification derived from sketched gradient features and a \textit{maximal number of acquired data points of 1024}. Unlike unbiased MD simulations, which rely on atom-based uncertainties to terminate MD simulations, biased MD simulations use total and atom-based uncertainties to bias MD simulations and prompt their termination, respectively. We use three metrics to assess the performance of our AL approaches: \textbf{(a)} Energy RMSE; \textbf{(b)} Force RMSE; and \textbf{(c)} Stress RMSE. All RMSEs are evaluated on the MIL-53(Al) test data set.\cite{Vandenhaute2023} Shaded areas denote the standard deviation across three independent runs, except for metadynamics. For it, shaded areas denote standard deviation across three randomly initialized MLIPs.}
	\label{fig:mil53_1024_performance}
\end{figure*}

The main text presents the exceptional performance of our AL approaches when acquiring a data set of 512 samples for alanine dipeptide and MIL-53(Al), achieving high accuracy in energy, atomic force, and stress predictions (close to or better than, e.g., the desired 0.043~eV/\AA{} for atomic forces). Practical applications, however, may demand larger data sets. Supplementary Figures~\ref{fig:diala_1024_performance} and \ref{fig:mil53_1024_performance}, as well as Supplementary Tables~\ref{tab:ala2-results-1024} and \ref{tab:mil53-results-1024}, demonstrate scaling of our AL approach, using a posterior-based uncertainty method, to larger data set sizes, acquiring data sets of 1024 samples. The results show improved MLIP performance  for both systems as the data set size increases; compare Tables~1 and 2 with Supplementary Tables~\ref{tab:ala2-results-1024} and \ref{tab:mil53-results-1024}. An overall lower uncertainty level and a larger portion of explored phase space can explain longer position and uncertainty ACTs.

\subsection*{Comparison of uncertainty methods}

\Tabref{tab:method-comparison} compares numerical results obtained for different uncertainty quantification methods. For alanine dipeptide, MLIPs that use ensemble-based uncertainties outperform those relying on gradient-based uncertainties. The improved ensemble robustness during MD simulations explains the improved CV space coverage, coupled with improved energy and atomic force RMSEs. The robustness of MLIP ensembles facilitates the CV space exploration already in early AL iteration. For MIL-53(Al), gradient-based uncertainties outperform the ensemble-based counterpart, mainly because the latter overestimates the atomic force errors to a greater extent (see Fig.~2). The overestimation of force errors leads to premature termination of MD simulations. Using larger ensembles could improve their performance while significantly increasing the computational cost. Overall, gradient-based uncertainty methods yield MLIPs with similar or, sometimes, even better performance than those created using ensemble-based approaches while significantly reducing the computational cost of uncertainty quantification.

\subsection*{Runtime analysis}

\begin{figure*}[htbp]
	\centering
	\includegraphics[width=\textwidth]{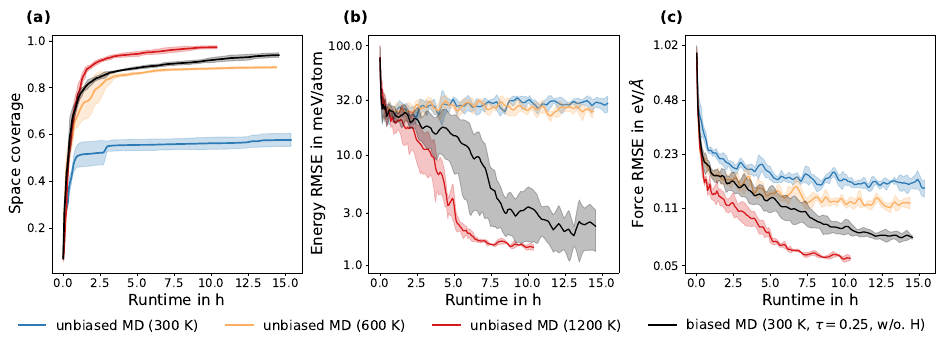}
	\caption{Runtime comparison for AL approaches employing biased and unbiased MD simulations to generate the candidate pool of atomic configurations for alanine dipeptide. Results are provided for the \textit{posterior-based uncertainty} quantification. Unlike unbiased MD simulations, which rely on atom-based uncertainties to terminate MD simulations, biased MD simulations use total and atom-based uncertainties to bias MD simulations and prompt their termination, respectively. We use three metrics to assess the performance of our AL approaches: \textbf{(a)} Coverage of the CV space; \textbf{(b)} RMSEs in predicted energies; and \textbf{(c)} RMSEs in atomic forces. All RMSEs are evaluated on the alanine dipeptide test data set; see Methods. Shaded areas denote the standard deviation across five independent runs. Runtime comprises the time required for reference AMBER calculations, MLIP training, batch selection from MD trajectories, and running the respective MD simulations.}
	\label{fig:diala_runtime_performance}
\end{figure*}

\begin{figure*}[htbp]
	\centering
	\includegraphics[width=\textwidth]{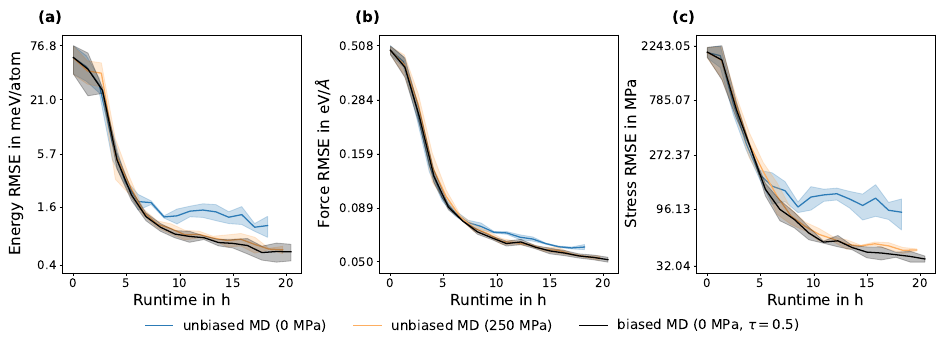}
	\caption{Runtime comparison for AL approaches employing biased and unbiased \textit{MD simulations at 600~K} to generate the candidate pool of atomic configurations for MIL-53(Al). Results are provided for the \textit{posterior-based uncertainty} quantification. Unlike unbiased MD simulations, which rely on atom-based uncertainties to terminate MD simulations, biased MD simulations use total and atom-based uncertainties to bias MD simulations and prompt their termination, respectively. We use three metrics to assess the performance of our AL approaches: \textbf{(a)} Energy RMSE; \textbf{(b)} Force RMSE; and \textbf{(c)} Stress RMSE. All RMSEs are evaluated on the MIL-53(Al) test data set.\cite{Vandenhaute2023} Shaded areas denote the standard deviation across three independent runs. All results are obtained for MD simulations run at 600~K, and AL experiments initialized using MLIPs trained with 32 closed-pore configurations obtained by randomly distorting the initial MIL-53(Al) configuration. Runtime comprises the time required for reference DFT calculations, MLIP training, batch selection from MD trajectories, and running the respective MD simulations.}
	\label{fig:mil53_runtime_performance}
\end{figure*}

When comparing inference times between unbiased and uncertainty-biased MD simulations for MIL-53(Al), we observed that computing the uncertainty gradient results in values larger by 1.4 for distance-based, 1.5 for ensemble-based, and 1.7 for posterior-based uncertainty quantification. However, applying uncertainty bias leads to, on average, shorter MD simulations until a high-uncertainty configuration is explored. Therefore, considering the overall runtime for acquiring a certain amount of training data by each method provides a more appropriate metric for assessing computational efficiency. Supplementary Figures~\ref{fig:diala_runtime_performance} and \ref{fig:mil53_runtime_performance} illustrate CV space coverage, as well as energy, atomic force, and stress RMSEs as a function of the runtime, comprising the time required for reference AMBER (alanine dipeptide) or DFT (MIL-53(Al)) calculations, MLIP training, batch selection from MD trajectories, and running the respective MD simulations. The maximal acquired training data size is set to 512 samples. \Tabref{tab:results-runtime} presents the numerical values for overall runtime obtained by conducting unbiased and uncertainty-biased AL experiments.

For alanine dipeptide, uncertainty-biased MD simulations at 300~K show no computational overhead compared to unbiased counterparts at 300~K and 600~K. On average, unbiased MD simulations at 1200~K perform the same number of steps as our biased MD at 300~K, making them 1.4 times more computationally efficient than uncertainty-biased experiments at 300~K. We found similar runtime values for high-pressure unbiased and zero-pressure uncertainty-biased AL experiments with MIL-53(Al). However, zero-pressure unbiased MD simulations required less time by a factor of 1.09–1.13 to generate 512 samples. This difference is attributed to more self-consistent field (SCF) iterations performed for structures obtained during high-pressure unbiased and zero-pressure uncertainty-biased MD simulations than those obtained during zero-pressure unbiased MD.

\subsection*{Biasing strength ablation studies}

\begin{figure*}[htbp]
	\centering
	\includegraphics[width=0.75\textwidth]{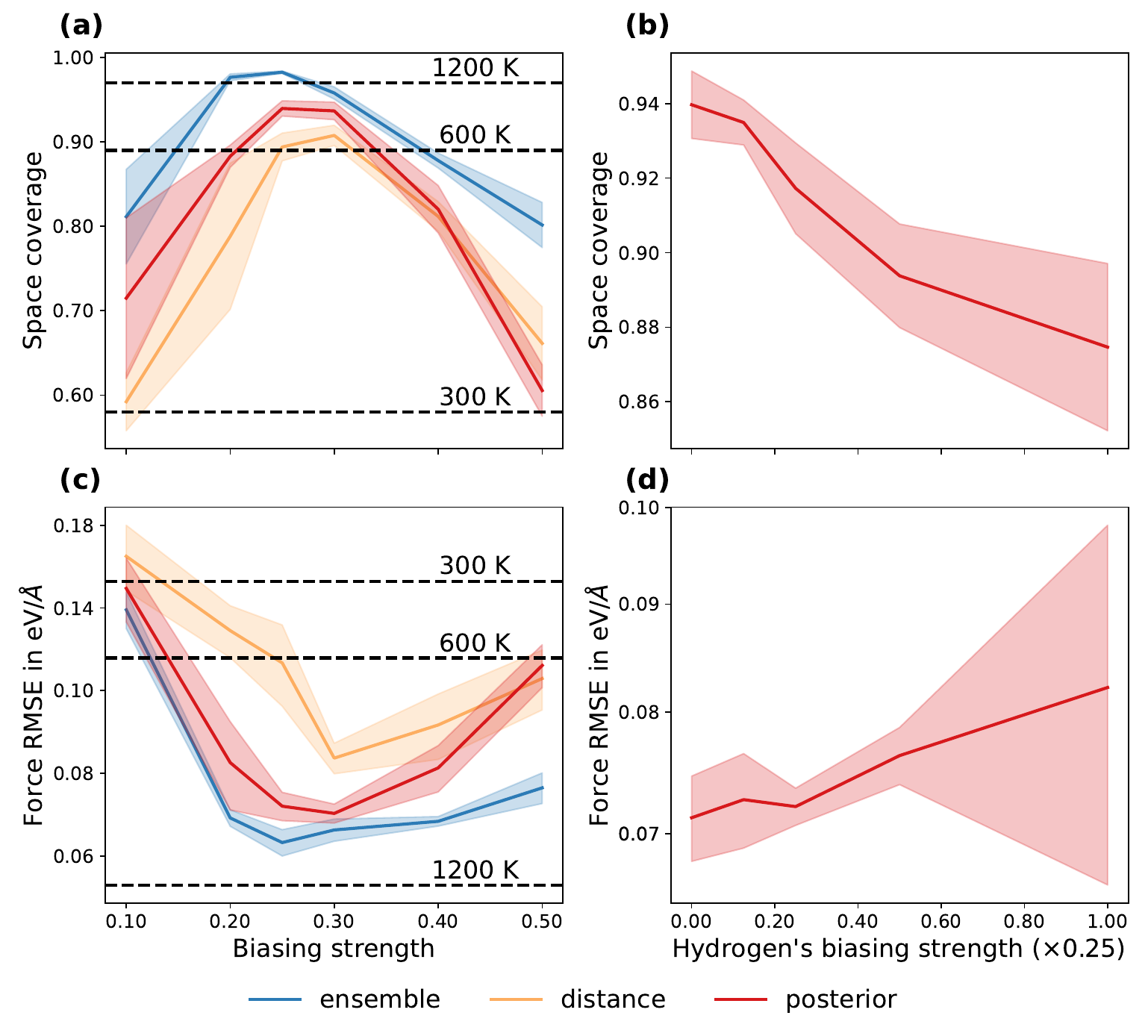}
	\caption{Dependence of the MLIP performance on the force biasing strength used in \textit{uncertainty-biased MD simulations of alanine dipeptide at 300~K}. \textbf{(a)} Dependence of CV space coverage on the force biasing strength. \textbf{(b)} Dependence of CV space coverage on the hydrogen's force biasing strength. \textbf{(c)} Dependence of force RMSE in eV/\AA{} on the force biasing strength. \textbf{(d)} Dependence of force RMSE in eV/\AA{} on the hydrogen's force biasing strength. For the experiments with hydrogen's force biasing strength, \textit{posterior-based uncertainty} quantification derived from sketched gradient features has been used. Shaded areas denote the standard deviation across five independent runs. Black dashed lines represent the results obtained for unbiased MD with \textit{posterior-based uncertainty} quantification.}
	\label{fig:diala_hyperparameters}
\end{figure*}

\Figsref{fig:diala_hyperparameters} (a) and (c) demonstrate that uncertainty-biased AL experiments at 300~K, using posterior-based uncertainty quantification, outperform their unbiased counterparts at 300~K for a broad range of biasing strength values, $\tau \leq 0.5$. They also outperform the experiments at 600~K and approach the performance of those at 1200~K for biasing strength values of $0.2 \lesssim \tau \lesssim 0.4$. \Figsref{fig:diala_hyperparameters} (b) and (d) show the results depending on the hydrogen's biasing strength. Changing the hydrogen's biasing strength from 0.25 to 0.0 improves the performance of the posterior-based AL experiments at 300~K by a factor of 1.08 and 1.15 for CV space coverage and atomic force RMSE, respectively.

\begin{figure*}[htbp]
	\centering
	\includegraphics[width=0.75\textwidth]{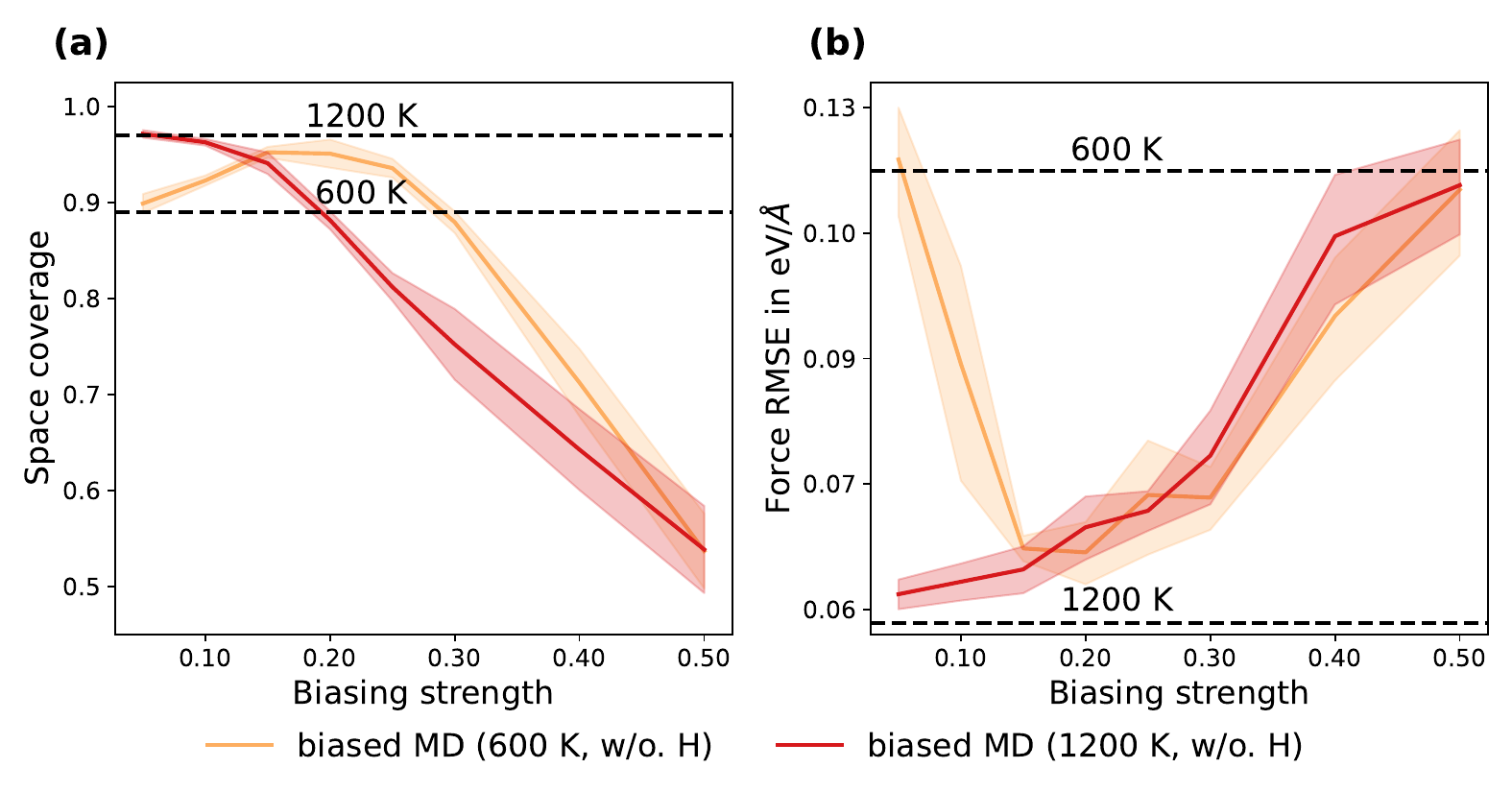}
	\caption{Dependence of the MLIP performance on the force biasing strength used in \textit{uncertainty-biased MD simulations of alanine dipeptide at 600~K and 1200~K}. All results are provided for \textit{posterior-based uncertainty} quantification derived from sketched gradient features. \textbf{(a)} Dependence of CV space coverage on the force biasing strength. \textbf{(b)} Dependence of force RMSE in eV/\AA{} on the force biasing strength. Shaded areas denote the standard deviation across five independent runs. Black dashed lines represent the results obtained for unbiased MD with \textit{posterior-based uncertainty} quantification.}
	\label{fig:diala_hyperparameters_600K_1200K}
\end{figure*}

\begin{figure*}[htbp]
	\centering
	\includegraphics[width=\textwidth]{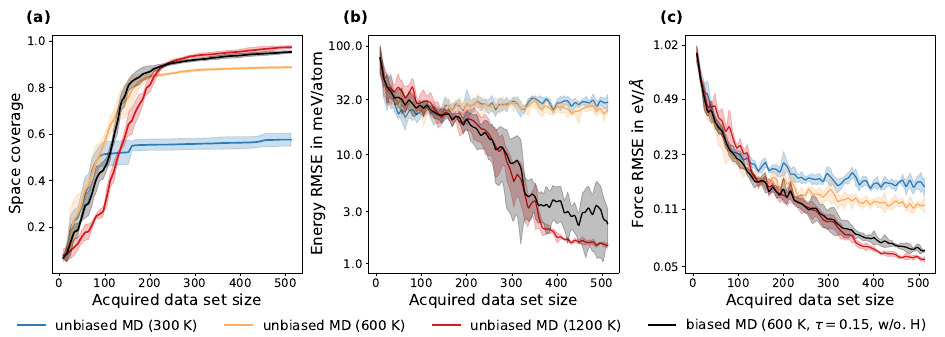}
	\caption{Comparison of AL approaches employing biased and unbiased MD simulations to generate the candidate pool of atomic configurations for alanine dipeptide. Results are provided for the \textit{posterior-based uncertainty} quantification derived from sketched gradient features and a \textit{temperature of 600~K}. Unlike unbiased MD simulations, which rely on atom-based uncertainties to terminate MD simulations, biased MD simulations use total and atom-based uncertainties to bias MD simulations and prompt their termination, respectively. We use three metrics to assess the performance of our AL approaches: \textbf{(a)} Coverage of the CV space; \textbf{(b)} RMSEs in predicted energies; and \textbf{(c)} RMSEs in atomic forces. All RMSEs are evaluated on the alanine dipeptide test data set; see Methods. Shaded areas denote the standard deviation across five independent runs.}
	\label{fig:diala_600K_performance}
\end{figure*}

\begin{figure*}[htbp]
	\centering
	\includegraphics[width=\textwidth]{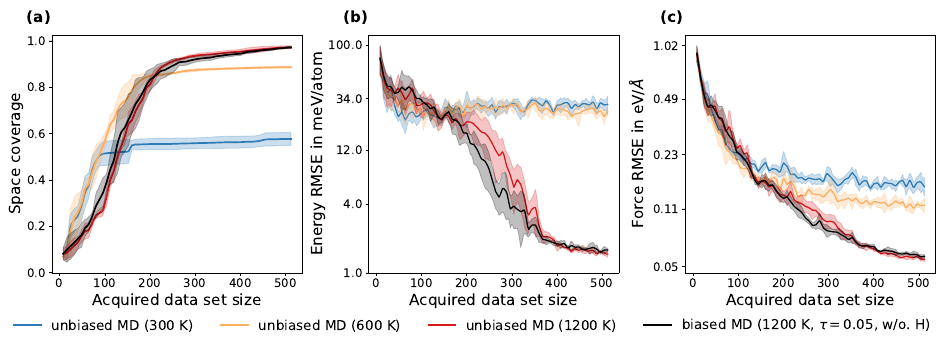}
	\caption{Comparison of AL approaches employing biased and unbiased MD simulations to generate the candidate pool of atomic configurations for alanine dipeptide. Results are provided for the \textit{posterior-based uncertainty} quantification derived from sketched gradient features and a \textit{temperature of 1200~K}. Unlike unbiased MD simulations, which rely on atom-based uncertainties to terminate MD simulations, biased MD simulations use total and atom-based uncertainties to bias MD simulations and prompt their termination, respectively. We use three metrics to assess the performance of our AL approaches: \textbf{(a)} Coverage of the CV space; \textbf{(b)} RMSEs in predicted energies; and \textbf{(c)} RMSEs in atomic forces. All RMSEs are evaluated on the alanine dipeptide test data set; see Methods. Shaded areas denote the standard deviation across five independent runs.}
	\label{fig:diala_1200K_performance}
\end{figure*}

The range of meaningful biasing strength values depends on the temperature at which MD simulations are conducted. \Figref{fig:diala_hyperparameters_600K_1200K} provides a similar analysis to those in \figsref{fig:diala_hyperparameters} (a) and (c) but for MD at 600~K and 1200~K. Moreover, Supplementary Figures~\ref{fig:diala_600K_performance} and \ref{fig:diala_1200K_performance} demonstrate results depending on the number of acquired configurations obtained for uncertainty-biased AL experiments at 600~K and 1200~K with biasing strength values of 0.15 and 0.05, respectively. Applying uncertainty bias to MD simulations at 600~K enhances their performance for a broad range of biasing strength values. We find that uncertainty bias improves the performance of unbiased AL experiments at 600~K in CV space coverage and atomic force RMSE for $\tau \lesssim 0.3$ and $\tau \leq 0.5$, respectively. Furthermore, uncertainty-biased MD simulations at 600~K and with $\tau=0.15$ outperform their counterparts at 300~K and $\tau=0.25$.

Employing uncertainty bias at extreme temperatures, e.g., 1200~K for alanine dipeptide, does not improve the performance of the respective AL experiments. Moreover, it may worsen the performance of MD simulations compared to unbiased counterparts when slightly increasing biasing strength values, e.g., already for $\tau=0.05$ employed in \figref{fig:diala_1200K_performance}. We attribute this observation to the fact that uncertainty bias applied to MD simulations at extreme temperatures causes even stronger distortion of the atomic system than during unbiased simulations. Thus, in this case, uncertainty bias limits the exploration of phase space by reaching high-uncertainty regions before large changes in positions occur. It is also related to the fact that bias forces should converge to smaller values with increasing temperatures, similar to what is observed for metadynamics simulations.\cite{Nagyfalusi2017}

\begin{figure*}[htbp]
	\centering
	\includegraphics[width=0.75\textwidth]{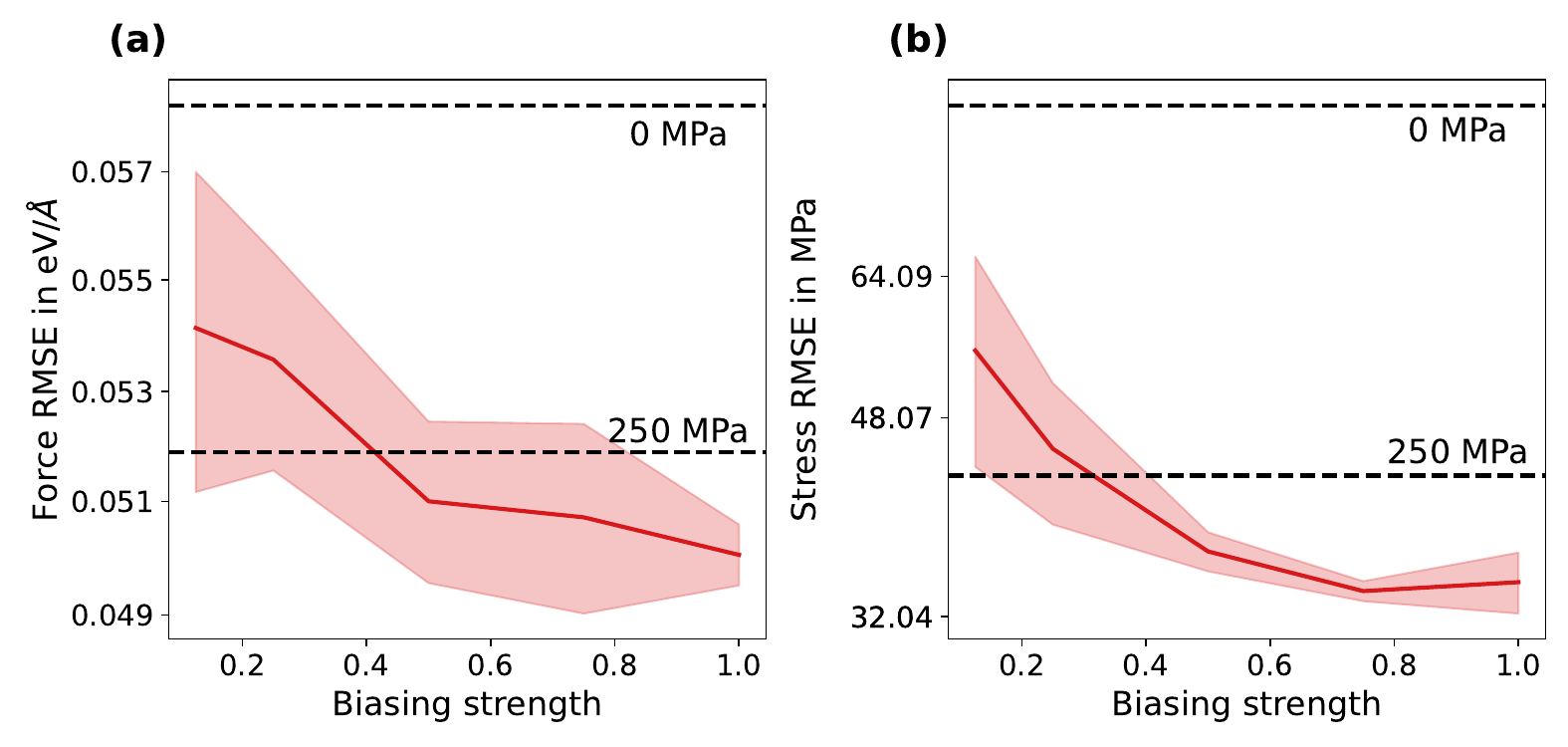}
	\caption{Dependence of the MLIP performance on the stress biasing strength used in \textit{uncertainty-biased MD simulations of MIL-53(Al) at 600~K}. All results are provided for \textit{posterior-based uncertainty} quantification derived from sketched gradient features. \textbf{(a)} Dependence of force RMSE in eV/\AA{} on the stress biasing strength. \textbf{(b)} Dependence of stress RMSE in MPa on the stress biasing strength. Shaded areas denote the standard deviation across three independent runs. Black dashed lines represent the results obtained for unbiased MD with \textit{posterior-based uncertainty} quantification.}
	\label{fig:mil53_hyperparameters}
\end{figure*}

\Figref{fig:mil53_hyperparameters} compares the performance of MLIPs trained with uncertainty-biased AL experiments, which use posterior-based uncertainty quantification, depending on the stress basing strength. Here, uncertainty-biased MD simulations employ bias stress to drive MD simulations toward unexplored regions. By using bias stress in MD simulations, we effectively reduce the internal pressure in the system. This bias is less extreme in perturbing local atomic environments than bias forces, allowing atoms to adjust to the new cell without exploring high-uncertainty regions. Thus, we can use larger biasing strength values, as shown in \figref{fig:mil53_hyperparameters}. Moreover, we observe a continuous improvement of atomic force and stress RMSEs with increasing biasing strength. Applying bias stress to high-pressure AL experiments at 600~K, we have observed only a negligible improvement compared to experiments conducted at zero pressure. We obtained 0.56 $\pm$ 0.03, 0.051 $\pm$ 0.001, and 36.38 $\pm$ 2.10 for energy, atomic force, and stress RMESs, respectively.

\subsection*{Comparison with random selection}

\begin{figure*}[htbp]
	\centering
	\includegraphics[width=\textwidth]{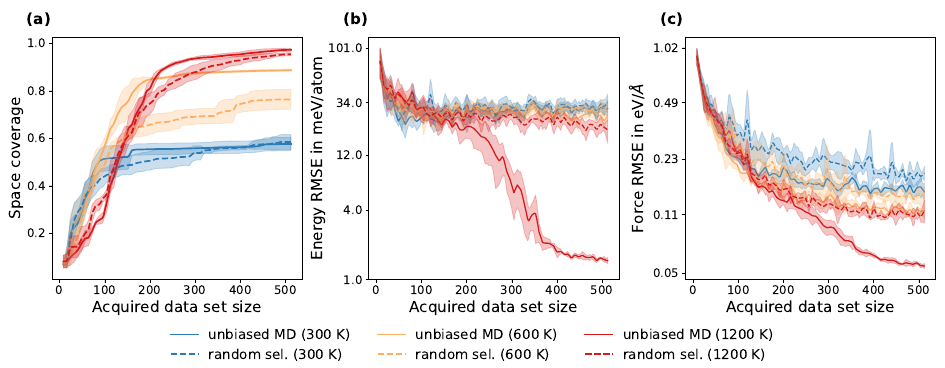}
	\caption{Comparison of batch selection strategies applied to candidate pools generated by running unbiased MD simulations for alanine dipeptide. Results are provided for the \textit{posterior-based uncertainty} quantification derived from sketched gradient features. Unlike unbiased MD simulations, which select training data based on their uncertainty and diversity, experiments that randomly choose training data rely on atom-based uncertainties only to terminate MD simulations. We use three metrics to assess the performance of our approaches: \textbf{(a)} Coverage of the CV space; \textbf{(b)} RMSEs in predicted energies; and \textbf{(c)} RMSEs in atomic forces. All RMSEs are evaluated on the alanine dipeptide test data set; see Methods. Shaded areas denote the standard deviation across five independent runs.}
	\label{fig:diala_random_performance}
\end{figure*}

\begin{figure*}[htbp]
	\centering
	\includegraphics[width=\textwidth]{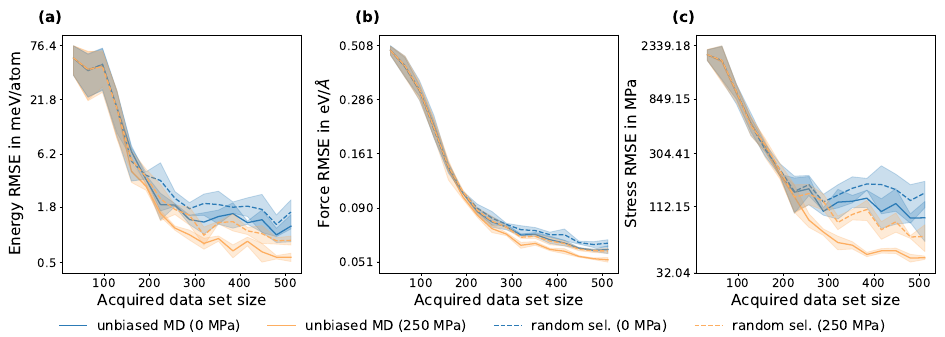}
	\caption{Comparison of batch selection strategies applied to candidate pools generated by running unbiased \textit{MD simulations at 600~K} for MIL-53(Al). Results are provided for the \textit{posterior-based uncertainty} quantification. Unlike unbiased MD simulations, which select training data based on their uncertainty and diversity, experiments that choose training data at random rely on atom-based uncertainties only to terminate MD simulations. We use three metrics to assess the performance of our AL approaches: \textbf{(a)} Energy RMSE; \textbf{(b)} Force RMSE; and \textbf{(c)} Stress RMSE. All RMSEs are evaluated on the MIL-53(Al) test data set.\cite{Vandenhaute2023} Shaded areas denote the standard deviation across three independent runs. All results are obtained for MD simulations run at 600~K, and AL experiments initialized using MLIPs trained with 32 closed-pore configurations obtained by randomly distorting the initial MIL-53(Al) configuration.}
	\label{fig:mil53_random_performance}
\end{figure*}

Supplementary Figures~\ref{fig:diala_random_performance} and \ref{fig:mil53_random_performance} complement the results in the main text (see Tables~1 and 2), comparing learning curves of approaches that employ uncertainty-based selection (particularly, greedy determinant maximization; see Methods) with those that use random selection. The figures illustrate that employing uncertainty-based selection strategies enhances the performance of MLIPs compared to random selection. For example, despite covering the same CV space at 1200~K for alanine dipeptide, experiments based on advanced selection strategies outperform their counterparts by a factor of 13.5 and 2.1 in energy and atomic force RMSEs. See our previous work for a more detailed comparison of batch selection methods.\cite{Zaverkin2022d}

\subsection*{Adversarial attacks for alanine dipeptide}

This section compares our uncertainty-biased MD simulations with adversarial attacks introduced in Supplementary Reference~\citenum{Schwalbe-Koda2021}. \Figref{fig:diala_adversarial_005_performance} and \tabref{tab:ala2-results-adversarial-posterior} demonstrate that adversarial attacks conducted at 300~K and 1200~K, with learning rates set to 0.005 and 0.01, are outperformed by our uncertainty-biased MD simulations at 300~K by factors of 2.3, 12.1, and 3.7 in CV space coverage, energy RMSE, and atomic force RMSE, respectively. All results are obtained for posterior-based uncertainty quantification. \Tabref{tab:adversarial-method-comparison} compares different uncertainty quantification methods.

Adversarial attacks often perform worse than unbiased MD simulations at 300~K or experiments conducted with the random selection strategy; see \tabref{tab:ala2-results-adversarial-posterior}. Note that the approach in Supplementary Reference~\citenum{Schwalbe-Koda2021} was developed and optimized for a setting where new configurations are obtained by performing uncertainty-driven adversarial attacks on data sets containing a few thousand atomic configurations. In Supplementary Reference~\citenum{Schwalbe-Koda2021}, 10,000 structures drawn from MD at 1200~K were used as an initial training data set for alanine dipeptide. In contrast, we use only eight randomly perturbed configurations at the beginning of AL for alanine dipeptide. Thus, adversarial attacks have been designed to augment training data sets rather than generate them from scratch---the setting of the present work. Further modifications to the approach proposed in Supplementary Reference~\citenum{Schwalbe-Koda2021} may be necessary to enhance its performance in the investigated setting, which is beyond the scope of this work.

Finally, the method proposed in Supplementary Reference~\citenum{Schwalbe-Koda2021} can be considered a local optimization algorithm in the uncertainty domain. Thus, it can easily converge to local uncertainty maxima. These maxima often feature low uncertainty values and atomic structures, which are very similar. In Supplementary Reference~\citenum{Schwalbe-Koda2021}, this issue is mitigated by constraining the diversity of selected configurations according to the root mean square deviation (RMSD) between them. However, their approach necessitates using more initial configurations for an AL iteration, e.g., 700 samples (see Supplementary Note 2 of the original work\cite{Schwalbe-Koda2021}), than eight samples employed in this work. In contrast, uncertainty-biased MD simulations employ thermostats instead of a simple constraint for the energy in \eqref{eq:adversarial-probability}, which proves beneficial for avoiding local uncertainty maxima and facilitating the exploration of the uncertainty landscape.

\section*{Supplementary Methods}

\subsection*{Coverage of collective variable space}

\Figref{fig:tree} demonstrates the idea of the tree-based weighted recursive partitioning algorithm for the simple case of $d=2$.

\subsection*{Uncertainty-driven adversarial attacks}

Following the original work,\cite{Schwalbe-Koda2021} we define the adversarial objective as
\begin{equation}
     \label{eq:adversarial-objective}
     \underset{\boldsymbol{\delta}}{\mathrm{max}}\,p\left(S_{\boldsymbol{\delta}}, \boldsymbol{\theta}\right)u\left(S_{\boldsymbol{\delta}}, \boldsymbol{\theta}\right),
\end{equation}
where $S_{\boldsymbol{\delta}}$ denotes an atomic structure which positions have been displaced by $\boldsymbol{\delta}$, i.e., $S_{\boldsymbol{\delta}} = \{\mathbf{r}_i + \boldsymbol{\delta}_i, Z_i\}_{i=1}^{N_\mathrm{at}}$. While in Supplementary Reference~\citenum{Schwalbe-Koda2021} the average variance in predicted forces has been used, we chose $u\left(S_{\boldsymbol{\delta}}, \boldsymbol{\theta}\right)$ to be the uncertainty of the energy predicted by an MLIP (ensemble-free or ensemble-based), similar to our AL experiments that use uncertainty-biased MD simulations. The probability $p$ of the structure $S_{\boldsymbol{\delta}}$ to be explored at temperature $T$ can be approximated by\cite{Schwalbe-Koda2021}
\begin{equation}
    \label{eq:adversarial-probability}
    p\left(S_{\boldsymbol{\delta}}, \boldsymbol{\theta}\right) = \frac{1}{Q\left(\mathcal{D}_\mathrm{train} \cup \mathcal{D}_\mathrm{valid}\right)} \exp\left(-\frac{E\left(S_{\boldsymbol{\delta}}, \boldsymbol{\theta}\right)}{k_\mathrm{B} T}\right),
\end{equation}
with the partition function $Q$ approximated by
\begin{equation}
    Q\left(\mathcal{D}_\mathrm{train} \cup \mathcal{D}_\mathrm{valid}\right) = \sum_{E \in \mathcal{D}_\mathrm{train} \cup \mathcal{D}_\mathrm{valid}} \exp\left(-\frac{E}{k_\mathrm{B} T}\right),
\end{equation}
where $\mathcal{D}_\mathrm{train}$ and $\mathcal{D}_\mathrm{valid}$ denote the training and validation data sets, respectively.

Different from the original work,\cite{Schwalbe-Koda2021} we aid the optimization process by defining
\begin{equation}
    \boldsymbol{\delta} = \hat{\boldsymbol{\delta}} + \boldsymbol{\delta}_\mathrm{rnd},
\end{equation}
where $\hat{\boldsymbol{\delta}}$ is the trainable displacement parameter initialized to zeros and $\boldsymbol{\delta}_\mathrm{rnd}$ is the not trainable random displacement initialized by selecting entries from a normal distribution $\mathcal{N}\left(0, \sigma_{\boldsymbol{\delta}}\mathbf{I}\right)$ with $\sigma_{\boldsymbol{\delta}} = 0.01$~\AA.

\subsection*{Details of adversarial attacks for alanine dipeptide}

Each AL experiment utilizes candidate pools generated through adversarial attacks, initiated with the training of an MLIP using eight alanine dipeptide configurations randomly perturbed from its initial configuration in the $\mathrm{C}_\mathrm{7eq}$ state. The MLIP's energy uncertainties are then used to conduct eight parallel adversarial attacks, initialized from the initial configuration or configurations selected in later iterations. Adversarial attacks continue until an empirically defined uncertainty threshold of 1.5~eV/\AA{} is reached. The maximum data set size, comprising training and validation data, is limited to 512 configurations. We employ the Adam optimizer\cite{Adam2015} to obtain the optimal parameter $\hat{\boldsymbol{\delta}}$, maximizing the adversarial objective in \eqref{eq:adversarial-objective}. We chose two learning rates $\alpha$ for Adam, 0.005 and 0.01, and set a maximum of 1000 steps. We employ two temperatures, 300~K and 1200~K, to limit the energy of explored configurations. All AL experiments have been repeated five times.

\clearpage

\clearpage

\begin{table*}[htbp]
	\caption{CV space coverage, atomic energy (E-) and atomic force (F-) RMSEs, as well as position (Pos.) and uncertainty (Unc.) auto-correlation times (ACTs) for alanine dipeptide experiments conducted with \textit{distance-based uncertainties}. E- and F-RMSEs are reported for MLIPs obtained at the end of each experiment, while CV space coverage and ACTs are computed using the entire trajectory obtained throughout the experiment. E-RMSE is given in meV/atom, while F-RMSE is in eV/\AA. All E-RMSE and F-RMSE values are computed for the test data set obtained from a long MD trajectory at 1200~K; see Methods. ACTs are provided in ps. For biased MD, we demonstrate results obtained without (w/o.) biasing hydrogen atoms. The best performance is highlighted in bold, and the second-best performance is underlined.
	\label{tab:ala2-results-distance}
	}
	\begin{center}
	\begin{tabular}{lccccc}
	\toprule 
	Experiment		    						& CV space cov.						& E-RMSE 						      & F-RMSE 						            & Pos. ACT 	& Unc. ACT                                                          \\
	\midrule 
	unbiased MD (300~K)							& 0.57 $\pm$ 0.02					& 32.46 $\pm$ 5.27				       & 0.172 $\pm$ 0.020				        & 2.06 $\pm$ 0.11				        & 247.94 $\pm$ 39.43				    \\
	unbiased MD (600~K)    						& \emph{0.89 $\pm$ 0.01}			& 28.71 $\pm$ 5.34				       & 0.130 $\pm$ 0.014				        & 1.25 $\pm$ 0.06				        & 122.44 $\pm$ 23.02			        \\
	unbiased MD (1200~K)   						& \textbf{0.95 $\pm$ 0.01}			& \textbf{8.76 $\pm$ 6.70} 		       & \textbf{0.083 $\pm$ 0.015}	            & \emph{0.79 $\pm$ 0.05}		        & \textbf{19.01	$\pm$ 6.17}		        \\
	biased MD (300~K, $\tau=0.25$, w/o. H)      & \emph{0.89 $\pm$ 0.02}			& \emph{18.30 $\pm$ 2.62}  	           & \emph{0.114 $\pm$ 0.015}	            & \textbf{0.78 $\pm$ 0.04}		        & \emph{57.54	$\pm$ 25.03}			\\
	\bottomrule 
	\end{tabular}
	\end{center}
\end{table*}

\clearpage

\begin{table*}[htbp]
	\caption{CV space coverage, atomic energy (E-) and atomic force (F-) RMSEs, as well as position (Pos.) and uncertainty (Unc.) auto-correlation times (ACTs) for alanine dipeptide experiments conducted with \textit{ensemble-based uncertainties}. E- and F-RMSEs are reported for MLIPs obtained at the end of each experiment, while CV space coverage and ACTs are computed using the entire trajectory obtained throughout the experiment. E-RMSE is given in meV/atom, while F-RMSE is in eV/\AA. All E-RMSE and F-RMSE values are computed for the test data set obtained from a long MD trajectory at 1200~K; see Methods. ACTs are provided in ps. For biased MD, we demonstrate results obtained without (w/o.) biasing hydrogen atoms. The best performance is highlighted in bold, and the second-best performance is underlined.
	\label{tab:ala2-results-ensemble}
	}
	\begin{center}
	\begin{tabular}{lccccc}
	\toprule 
	Experiment		    						& CV space cov.						& E-RMSE 						      & F-RMSE 						            & Pos. ACT 	& Unc. ACT                                                          \\
	\midrule 
	unbiased MD (300~K)							& 0.57 $\pm$ 0.01					& 30.53 $\pm$ 2.19				       & 0.162 $\pm$ 0.009				        & 2.29 $\pm$ 0.12				        & 252.74 $\pm$ 18.32				    \\
	unbiased MD (600~K)    						& \emph{0.88 $\pm$ 0.01}			& 24.28 $\pm$ 3.76				       & 0.119 $\pm$ 0.010				        & 1.19 $\pm$ 0.03				        & 253.46 $\pm$ 9.31			            \\
	unbiased MD (1200~K)   						& \textbf{0.98 $\pm$ 0.00}			& \textbf{1.42 $\pm$ 0.09} 		       & \textbf{0.053 $\pm$ 0.001}	            & \emph{0.72 $\pm$ 0.03}		        & \emph{31.59	$\pm$ 7.10}		        \\
	biased MD (300~K, $\tau=0.25$, w/o. H)      & \textbf{0.98 $\pm$ 0.00}			& \emph{1.73 $\pm$ 0.20}  	           & \emph{0.063 $\pm$ 0.003}	            & \textbf{0.61 $\pm$ 0.09}		        & \textbf{16.39	$\pm$ 4.62}				\\
	\bottomrule 
	\end{tabular}
	\end{center}
\end{table*}

\clearpage

\clearpage

\clearpage

\clearpage

\begin{table*}[t!]
	\caption{Atomic energy (E-), atomic force (F-), and stress (S-) RMSEs, as well as position (Pos.) and uncertainty (Unc.) auto-correlation times (ACTs) for MIL-53(Al) experiments conducted with \textit{distance-based uncertainties}. E-, F-, and S-RMSEs are reported for MLIPs obtained at the end of each experiment, while ACTs are computed using the entire trajectory sampled throughout the experiment. E-RMSE is given in meV/atom, F-RMSE in eV/\AA, and S-RMSE in MPa. All E-RMSE, F-RMSE, and S-RMSE values are computed for the test data set obtained based on first principles MD trajectories at 600~K; see Supplementary Reference~\citenum{Vandenhaute2023}. ACTs are provided in ps. The best performance is highlighted in bold, and the second-best performance is underlined.
	\label{tab:mil53-results-distance}
	}
	\begin{center}
	\begin{tabular}{lccccc}
	\toprule 
	Experiment		    						& E-RMSE						& F-RMSE 						& S-RMSE 						& Pos. ACT						& Unc. ACT						\\
	\midrule 
	\multicolumn{6}{c}{$T=600$~K} \\
	\midrule
	unbiased MD (0~MPa)							& 1.76 $\pm$ 0.76				& 0.063 $\pm$ 0.005				& 148.73 $\pm$ 69.62			& \emph{5.14 $\pm$ 1.46}		& 129.04 $\pm$ 43.56			\\
	unbiased MD (250~MPa)    					& 0.60 $\pm$ 0.11		        & \emph{0.052 $\pm$ 0.001}	    & \emph{46.72 $\pm$ 5.19}		& \textbf{2.78 $\pm$ 0.97}		& \emph{99.00 $\pm$ 6.42}		\\
	Metadynamics (0~MPa)   						& \textbf{0.58 $\pm$ 0.10}		& 0.058 $\pm$ 0.002 			& 74.83 $\pm$ 11.89 			& --							& --							\\
	biased MD (0~MPa, $\tau=0.5$) 				& \emph{0.59 $\pm$ 0.10} 		& \textbf{0.051 $\pm$ 0.002} 	& \textbf{41.79 $\pm$ 3.43}		& 21.04 $\pm$ 11.58				& \textbf{78.47 $\pm$ 30.85} 	\\
	\midrule 
	\multicolumn{6}{c}{$T=300$~K} \\
	\midrule
	unbiased MD (0~MPa)							& 0.87 $\pm$ 0.44				& 0.056 $\pm$ 0.001				& 65.18 $\pm$ 3.86				&  \emph{3.01 $\pm$ 0.84}		& 127.89 $\pm$ 5.87				\\
	unbiased MD (250~MPa)    					& \textbf{0.45 $\pm$ 0.03}		& \emph{0.054 $\pm$ 0.001}	    & \emph{41.15 $\pm$ 2.24}		& \textbf{1.85 $\pm$ 0.06}		&  \emph{95.72 $\pm$ 14.12}		\\
	biased MD (0~MPa, $\tau=0.5$) 				& \emph{0.51 $\pm$ 0.05} 		& \textbf{0.053 $\pm$ 0.000} 	& \textbf{37.94 $\pm$ 6.91}		& 16.22 $\pm$ 12.70				& \textbf{84.12 $\pm$ 13.61} 	\\
	\bottomrule 
	\end{tabular}
	\end{center}
\end{table*}

\begin{table*}[t!]
	\caption{Atomic energy (E-), atomic force (F-), and stress (S-) RMSEs, as well as position (Pos.) and uncertainty (Unc.) auto-correlation times (ACTs) for MIL-53(Al) experiments conducted with \textit{ensemble-based uncertainties}. E-, F-, and S-RMSEs are reported for MLIPs obtained at the end of each experiment, while ACTs are computed using the entire trajectory sampled throughout the experiment. E-RMSE is given in meV/atom, F-RMSE in eV/\AA, and S-RMSE in MPa. All E-RMSE, F-RMSE, and S-RMSE values are computed for the test data set obtained based on first principles MD trajectories at 600~K; see Supplementary Reference~\citenum{Vandenhaute2023}. ACTs are provided in ps. The best performance is highlighted in bold, and the second-best performance is underlined.
	\label{tab:mil53-results-distance}
	}
	\begin{center}
	\begin{tabular}{lccccc}
	\toprule 
	Experiment		    						& E-RMSE						& F-RMSE 						& S-RMSE 						& Pos. ACT						& Unc. ACT						\\
	\midrule 
	\multicolumn{6}{c}{$T=600$~K} \\
	\midrule
	unbiased MD (0~MPa)							& 1.95 $\pm$ 0.45				& 0.064 $\pm$ 0.002				& 147.47 $\pm$ 16.51			& 35.57 $\pm$ 21.45				& 41.70 $\pm$ 16.44				\\
	unbiased MD (250~MPa)    					& 0.79 $\pm$ 0.09		        & \textbf{0.054 $\pm$ 0.001}	& \textbf{60.82 $\pm$ 8.07}		& \textbf{6.69 $\pm$ 3.15}		& \emph{23.91 $\pm$ 4.04}		\\
	Metadynamics (0~MPa)   						& \textbf{0.58 $\pm$ 0.10}		&  \emph{0.058 $\pm$ 0.002} 	& 74.83 $\pm$ 11.89 			& --							& --							\\
	biased MD (0~MPa, $\tau=0.5$) 				&  \emph{0.74 $\pm$ 0.18} 		& \textbf{0.054 $\pm$ 0.002} 	&  \emph{62.08 $\pm$ 13.41}	    & \emph{20.38 $\pm$ 10.48}		& \textbf{21.64 $\pm$ 1.66} 	\\
	\midrule 
	\multicolumn{6}{c}{$T=300$~K} \\
	\midrule
	unbiased MD (0~MPa)							& 1.10 $\pm$ 0.05				& 0.057 $\pm$ 0.001				& 69.43 $\pm$ 9.68				& \emph{4.59 $\pm$ 3.91}		& 50.66 $\pm$ 11.47				\\
	unbiased MD (250~MPa)    					&  \emph{0.53 $\pm$ 0.03}		&  \emph{0.055 $\pm$ 0.001}	    & \emph{46.08	$\pm$ 3.03}		& \textbf{29.82 $\pm$ 6.55}		& \emph{18.45 $\pm$ 3.23}		\\
	biased MD (0~MPa, $\tau=0.5$) 				& \textbf{0.52 $\pm$ 0.04} 		& \textbf{0.054 $\pm$ 0.000} 	& \textbf{38.44 $\pm$ 2.24}		& 94.00 $\pm$ 19.62				& \textbf{6.67 $\pm$ 0.78} 	    \\
	\bottomrule 
	\end{tabular}
	\end{center}
\end{table*}

\clearpage

\begin{table*}[htbp]
	\caption{CV space coverage, atomic energy (E-) and atomic force (F-) RMSEs, as well as position (Pos.) and uncertainty (Unc.) auto-correlation times (ACTs) for alanine dipeptide experiments conducted with \textit{posterior-based uncertainties} and a \textit{maximal acquired data set size of 1024 samples}. E- and F-RMSEs are reported for MLIPs obtained at the end of each experiment, while CV space coverage and ACTs are computed using the entire trajectory obtained throughout the experiment. E-RMSE is given in meV/atom, while F-RMSE is in eV/\AA. All E-RMSE and F-RMSE values are computed for the test data set obtained from a long MD trajectory at 1200~K; see Methods. ACTs are provided in ps. For biased MD, we demonstrate results obtained without (w/o.) biasing hydrogen atoms. The best performance is highlighted in bold, and the second-best performance is underlined.
	\label{tab:ala2-results-1024}
	}
	\begin{center}
	\begin{tabular}{lccccc}
	\toprule 
	Experiment		    						& CV space cov.						& E-RMSE 						      & F-RMSE 						            & Pos. ACT 	                            & Unc. ACT                              \\
    \midrule 
	unbiased MD (300~K)							& 0.62 $\pm$ 0.07					& 27.58 $\pm$ 3.69				       & 0.148 $\pm$ 0.018				        & 2.08 $\pm$ 0.17				        & 353.12 $\pm$ 244.13				    \\
	unbiased MD (600~K)    						& 0.90 $\pm$ 0.00					& 20.07 $\pm$ 2.52				       & 0.115 $\pm$ 0.011				        & 1.20 $\pm$ 0.01				        & 423.23 $\pm$ 213.76			        \\
	unbiased MD (1200~K)   						& \textbf{0.99 $\pm$ 0.00}			& \textbf{1.02 $\pm$ 0.09} 		       & \textbf{0.039 $\pm$ 0.002}	            & \textbf{0.71 $\pm$ 0.01}		        & \emph{167.60 $\pm$ 62.90}		        \\
	biased MD (300~K, $\tau=0.25$, w/o. H)      & \emph{0.97 $\pm$ 0.00}			& \emph{1.04 $\pm$ 0.08}  	           & \emph{0.052 $\pm$ 0.002}	            & \emph{0.75 $\pm$ 0.15}		        & \textbf{140.16 $\pm$ 70.85}		    \\
	\bottomrule 
	\end{tabular}
	\end{center}
\end{table*}

\begin{table*}[t!]
	\caption{Atomic energy (E-), atomic force (F-), and stress (S-) RMSEs, as well as position (Pos.) and uncertainty (Unc.) auto-correlation times (ACTs) for MIL-53(Al) experiments conducted with \textit{posterior-based uncertainties} and a \textit{maximal acquired data set size of 1024 samples}. E-, F-, and S-RMSEs are reported for MLIPs obtained at the end of each experiment, while ACTs are computed using the entire trajectory sampled throughout the experiment. E-RMSE is given in meV/atom, F-RMSE in eV/\AA, and S-RMSE in MPa. All E-RMSE, F-RMSE, and S-RMSE values are computed for the test data set obtained based on first principles MD trajectories at 600~K; see Supplementary Reference~\citenum{Vandenhaute2023}. ACTs are provided in ps. The best performance is highlighted in bold, and the second-best performance is underlined.
	\label{tab:mil53-results-1024}
	}
	\begin{center}
	\begin{tabular}{lccccc}
	\toprule 
	Experiment		    						& E-RMSE						& F-RMSE 						& S-RMSE 						& Pos. ACT 						& Unc. ACT						\\
	\midrule 
	\multicolumn{6}{c}{$T=600$~K}                                                                                                                                                                               \\
    \midrule
	unbiased MD (0~MPa)							& 0.61 $\pm$ 0.08				& 0.042 $\pm$ 0.002				& 52.01 $\pm$ 4.63				& 60.78 $\pm$ 28.67				& 308.72 $\pm$ 23.89			\\
	unbiased MD (250~MPa)    					& \textbf{0.43 $\pm$ 0.09}		& \textbf{0.039 $\pm$ 0.002}	& \emph{28.63 $\pm$ 0.33}		& \textbf{9.30 $\pm$ 8.17}		& \emph{222.89 $\pm$ 33.29}		\\
	Metadynamics (0~MPa)   						& \emph{0.48 $\pm$ 0.09}		& 0.045 $\pm$ 0.001 			& 60.49 $\pm$ 9.10 			    & --							& --							\\
	biased MD (0~MPa, $\tau=0.5$) 				& \textbf{0.43 $\pm$ 0.08} 		& \emph{0.040 $\pm$ 0.001} 	    & \textbf{27.98 $\pm$ 1.83}		& \emph{39.94 $\pm$ 24.17}		& \textbf{135.27 $\pm$ 19.65} 	\\
	\bottomrule 
	\end{tabular}
	\end{center}
\end{table*}

\clearpage

\begin{table*}[htbp]
	\caption{Performance comparison of various uncertainty quantification methods for alanine dipeptide and MIL-53(Al). We evaluate CV space coverage, atomic energy (E-), force (F-), and stress (S-) RMSEs, position (Pos.) and uncertainty (Unc.) auto-correlation times (ACTs), as well as training and inference times. E-, F-, and S-RMSEs are reported for MLIPs obtained at the end of each experiment, while CV space coverage and ACTs are computed using the entire trajectory obtained throughout the experiment. E-RMSE is given in meV/atom, F-RMSE in eV/\AA, S-RMSE in MPa. ACTs are provided in ps, while training and inference times are in h and ms/atom, respectively. The best performance is highlighted in bold.
	\label{tab:method-comparison}
	}
	\begin{center}
	\begin{tabular}{lccc}
	\toprule 
			    				& ensemble						& distance 						& posterior 					\\
	\midrule 
	\multicolumn{4}{c}{alanine dipeptide}                                                                                       \\
	\midrule
	CV space cov.				& \textbf{0.98 $\pm$ 0.00}		& 0.89 $\pm$ 0.02				& 0.94 $\pm$ 0.01		        \\
	E-RMSE    					& \textbf{1.73 $\pm$ 0.20}		& 18.30 $\pm$ 2.62				& 1.97 $\pm$ 0.88		        \\
	F-RMSE   					& \textbf{0.063 $\pm$ 0.003}	& 0.114 $\pm$ 0.015 			& 0.071 $\pm$ 0.003 		    \\
	Pos. ACT					& \textbf{0.61 $\pm$ 0.09}		& 0.78 $\pm$ 0.04				& 0.69 $\pm$ 0.04		        \\
	Unc. ACT					& \textbf{16.39 $\pm$ 4.62}		& 57.54 $\pm$ 25.03				& 52.79 $\pm$ 19.40		        \\
	training time 				& 7.07 $\pm$ 0.17 				& \textbf{3.38 $\pm$ 0.11} 		& 3.71 $\pm$ 0.16		        \\
	inference time 				& 0.77 $\pm$ 0.01				& \textbf{0.39 $\pm$ 0.02} 		& \textbf{0.39 $\pm$ 0.02}		\\
	\midrule 
	\multicolumn{4}{c}{MIL-53(Al)}                                                                                              \\
	\midrule
	E-RMSE    					& 0.74 $\pm$ 0.18				& 0.59 $\pm$ 0.10		        & \textbf{0.57 $\pm$ 0.08}		\\
	F-RMSE   					& 0.054 $\pm$ 0.002				& \textbf{0.051 $\pm$ 0.002} 	& \textbf{0.051 $\pm$ 0.001}	\\
	S-RMSE   					& 62.08 $\pm$ 13.41				& 41.79 $\pm$ 3.43 	            & \textbf{36.60 $\pm$ 1.46} 	\\
	Pos. ACT					& 20.38 $\pm$ 10.48		        & 21.04 $\pm$ 11.58				& \textbf{2.75 $\pm$ 0.46}		\\
	Unc. ACT					& \textbf{21.64 $\pm$ 1.66}		& 78.47 $\pm$ 30.85				& 44.86 $\pm$ 14.08             \\
	training time 				& 8.92 $\pm$ 0.08 				& \textbf{3.34 $\pm$ 0.04} 		& 3.43 $\pm$ 0.032				\\
	inference time 				& 0.038 $\pm$ 0.002 			& 0.020 $\pm$ 0.000 			& \textbf{0.017 $\pm$ 0.001}	\\
	\bottomrule 
	\end{tabular}
	\end{center}
\end{table*}

\clearpage

\begin{table*}[htbp]
	\caption{Runtime in h measured for alanine dipeptide and MIL-53(Al) experiments performed with \textit{posterior-based uncertainties}. The best performance is highlighted in bold.
	\label{tab:results-runtime}
	}
	\begin{center}
	\begin{tabular}{lc}
	\toprule 
	Experiment		    						& Runtime                      \\
	\midrule 
	\multicolumn{2}{c}{alanine dipeptide}                                      \\
	\midrule
	unbiased MD (300~K)							& 15.37 $\pm$ 0.19	           \\
	unbiased MD (600~K)    						& 14.39 $\pm$ 0.24	           \\
	unbiased MD (1200~K)   						& \textbf{10.36	$\pm$ 0.40}	   \\
	biased MD (300~K, $\tau=0.25$, w/o. H)      & 14.56	$\pm$ 0.88	           \\
    \midrule 
	\multicolumn{2}{c}{MIL-53(Al)}                                             \\
	\midrule
    unbiased MD (0~MPa)							& \textbf{18.59 $\pm$ 0.29}	   \\
	unbiased MD (250~MPa)    					& 20.19 $\pm$ 0.41	           \\
	biased MD (0~MPa, $\tau=0.5$)               & 20.99	$\pm$ 0.52	           \\
	\bottomrule 
	\end{tabular}
	\end{center}
\end{table*}

\clearpage

\clearpage

\begin{table*}[htbp]
	\caption{CV space coverage, atomic energy (E-) and atomic force (F-) RMSEs, as well as position (Pos.) and uncertainty (Unc.) auto-correlation times (ACTs) for alanine dipeptide experiments conducted with \textit{posterior-based uncertainties at 600~K and 1200~K}. E- and F-RMSEs are reported for MLIPs obtained at the end of each experiment, while CV space coverage and ACTs are computed using the entire trajectory obtained throughout the experiment. E-RMSE is given in meV/atom, while F-RMSE is in eV/\AA. All E-RMSE and F-RMSE values are computed for the test data set obtained from a long MD trajectory at 1200~K; see Methods. ACTs are provided in ps. We demonstrate results obtained without (w/o.) biasing hydrogen atoms. The best performance is highlighted in bold.
	\label{tab:ala2-results-600K-1200K}
	}
	\begin{center}
	\begin{tabular}{lccccc}
	\toprule 
	Experiment		    						& CV space cov.						& E-RMSE 						      & F-RMSE 						            & Pos. ACT 	& Unc. ACT                                                          \\
	\midrule
	biased MD (600~K, $\tau=0.15$, w/o. H)      & 0.95 $\pm$ 0.01			        & 2.33 $\pm$ 0.85  	                  & 0.062 $\pm$ 0.001	                    & 0.73 $\pm$ 0.04		        & 58.48	$\pm$ 15.72				                \\
    biased MD (1200~K, $\tau=0.05$, w/o. H)     & \textbf{0.97 $\pm$ 0.00}			& \textbf{1.59 $\pm$ 0.06}  	      & \textbf{0.058 $\pm$ 0.001}	            & \textbf{0.68 $\pm$ 0.03}		& \textbf{16.62	$\pm$ 3.84}				        \\
	\bottomrule 
	\end{tabular}
	\end{center}
\end{table*}

\clearpage

\clearpage

\begin{figure*}[htbp]
	\centering
	\includegraphics[width=\textwidth]{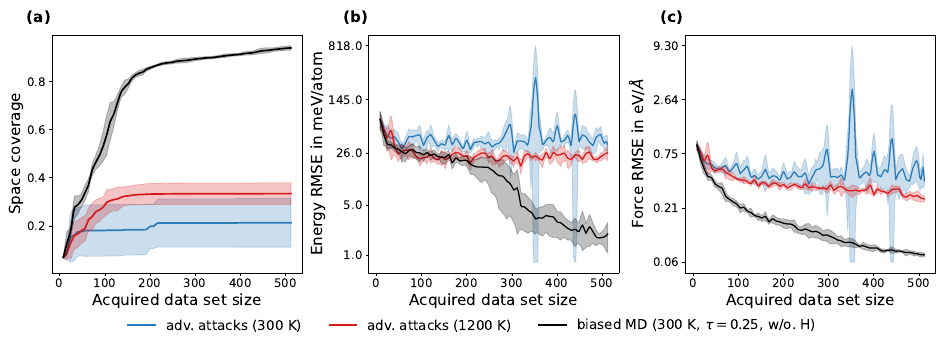}
	\caption{Comparison of AL approaches that use uncertainty-biased MD simulations and adversarial attacks (see Supplementary Methods) to generate the candidate pool of atomic configurations for alanine dipeptide. Results are provided for the \textit{posterior-based uncertainty} quantification derived from sketched gradient features. For adversarial attacks, we employ the Adam optimizer\cite{Adam2015} and the \textit{learning rate of 0.005}. Uncertainty-biased MD simulations and adversarial attacks use total and atom-based uncertainties to bias the respective atomistic simulations and prompt their termination, respectively. We use three metrics to assess the performance of our AL approaches: \textbf{(a)} Coverage of the CV space; \textbf{(b)} RMSEs in predicted energies; and \textbf{(c)} RMSEs in atomic forces. All RMSEs are evaluated on the alanine dipeptide test data set; see Methods. Shaded areas denote the standard deviation across five independent runs.}
	\label{fig:diala_adversarial_005_performance}
\end{figure*}

\begin{table*}[htbp]
	\caption{CV space coverage, atomic energy (E-) and atomic force (F-) RMSEs, as well as position (Pos.) and uncertainty (Unc.) auto-correlation times (ACTs) for alanine dipeptide experiments conducted with \textit{posterior-based uncertainties}. Results are provided for unbiased and biased MD simulations, as well as for \textit{adversarial attacks} (adv. attacks). E- and F-RMSEs are reported for MLIPs obtained at the end of each experiment, while CV space coverage and ACTs are computed using the entire trajectory obtained throughout the experiment. E-RMSE is given in meV/atom, while F-RMSE is in eV/\AA. All E-RMSE and F-RMSE values are computed for the test data set obtained from a long MD trajectory at 1200~K; see Methods. ACTs are provided in ps. For biased MD, we compare two cases: one with (w.) biasing hydrogen atoms and one without (w/o.). For adversarial sampling, we demonstrate results obtained with biasing hydrogen atoms. We also compare adversarial attacks to experiments that involve the random selection (random sel.) strategy for acquiring training data. The best performance is highlighted in bold, and the second-best performance is underlined.
	\label{tab:ala2-results-adversarial-posterior}
	}
	\begin{center}
	\begin{tabular}{lccccc}
	\toprule 
	Experiment		    					& CV space cov.				& E-RMSE 					& F-RMSE                        & ACT\textsuperscript{\emph{a}} 	    & Unc. ACT\textsuperscript{\emph{a}}          \\
	\midrule
    random sel. (300~K)					    & 0.58 $\pm$ 0.03			& 34.09 $\pm$ 6.29		    & 0.191 $\pm$ 0.019		        & --		                            & --		                                  \\
	unbiased MD (300~K)					    & 0.58 $\pm$ 0.03			& 30.29 $\pm$ 5.47			& 0.149 $\pm$ 0.019				& \emph{2.07 $\pm$ 0.11}			    & 327.11 $\pm$ 8.69				              \\
	adv. attacks (300~K, $\alpha=0.005$)    & 0.21 $\pm$ 0.10			& 34.40 $\pm$ 8.74  	    & 0.400 $\pm$ 0.032	            & --                                    & --			                              \\
    adv. attacks (1200~K, $\alpha=0.005$)   & 0.33 $\pm$ 0.05			& 26.61 $\pm$ 6.00  	    & 0.262 $\pm$ 0.018	            & --                                    & --			                              \\
    adv. attacks (300~K, $\alpha=0.01$)     & 0.22 $\pm$ 0.04			& 35.65 $\pm$ 3.54  	    & 0.348 $\pm$ 0.097	            & --                                    & --			                              \\
    adv. attacks (1200~K, $\alpha=0.01$)    & 0.41 $\pm$ 0.09			& 23.79 $\pm$ 6.05  	    & 0.279 $\pm$ 0.035	            & --                                    & --			                              \\
	biased MD (300~K, $\tau=0.25$, w. H) 	& \emph{0.87 $\pm$ 0.02} 	& \emph{5.09 $\pm$ 5.40} 	& \emph{0.082 $\pm$ 0.016}		& 2.08 $\pm$ 0.13			            & \textbf{19.38 $\pm$ 7.42}                   \\
	biased MD (300~K, $\tau=0.25$, w/o. H)  & \textbf{0.94 $\pm$ 0.01}	& \textbf{1.97 $\pm$ 0.88}  & \textbf{0.071 $\pm$ 0.003}    & \textbf{0.69 $\pm$ 0.04}              & \emph{52.79	$\pm$ 19.40}				  \\
    \bottomrule 
	\end{tabular}
	\end{center}
    \footnotesize{\textsuperscript{\emph{a}} ACTs computed for experiments with the random selection (random sel.) strategy are excluded from the analysis because different approaches may introduce systematic biases, making the comparison unreliable. We also exclude ACTs obtained for adversarial attacks (adv. attacks) as the corresponding lag time is unitless, different from MD simulations.}
\end{table*}

\begin{table*}[htbp]
	\caption{Comparison of the three uncertainty quantification methods used to perform \textit{adversarial attacks} for alanine dipeptide. We evaluate CV space coverage, atomic energy (E-), and force (F-) RMSEs. E- and F-RMSEs are reported for MLIPs obtained at the end of each experiment, while CV space coverage is computed using the entire trajectory. E-RMSE is given in meV/atom, while F-RMSE is in eV/\AA. The best performance is highlighted in bold.
	\label{tab:adversarial-method-comparison}
	}
	\begin{center}
	\begin{tabular}{lccc}
	\toprule 
			    				& ensemble				 & distance 					& posterior 					\\
    \midrule 
	\multicolumn{4}{c}{$\alpha=0.005$, $T=300$}                                                                         \\
	\midrule
	CV space cov.				& 0.14 $\pm$ 0.05	     & 0.11 $\pm$ 0.06				& \textbf{0.21 $\pm$ 0.10}		\\
	E-RMSE    					& 39.1 $\pm$ 21.26		 & \textbf{28.73 $\pm$ 3.79}	& 34.4 $\pm$ 8.74				\\
	F-RMSE   					& 0.623 $\pm$ 0.221	     & 0.466 $\pm$ 0.059 			& \textbf{0.400 $\pm$ 0.032} 	\\
    \midrule 
	\multicolumn{4}{c}{$\alpha=0.005$, $T=1200$}                                                                        \\
	\midrule
	CV space cov.				& 0.31 $\pm$ 0.10		& 0.22 $\pm$ 0.06				& \textbf{0.33 $\pm$ 0.05}		\\
	E-RMSE    					& 29.24 $\pm$ 5.74		& \textbf{24.66 $\pm$ 4.97}		& 26.61 $\pm$ 6.00				\\
	F-RMSE   					& 0.359 $\pm$ 0.051	    & 0.367 $\pm$ 0.042 			& \textbf{0.262 $\pm$ 0.018} 	\\
    \midrule 
	\multicolumn{4}{c}{$\alpha=0.01$, $T=300$\textsuperscript{\emph{a}}}                                                \\
	\midrule
	CV space cov.				& 0.16 $\pm$ 0.04		& 0.15 $\pm$ 0.04				& \textbf{0.22 $\pm$ 0.04}		\\
	E-RMSE    					& 51.31 $\pm$ 5.22		& 58.28 $\pm$ 17.14				& \textbf{35.65 $\pm$ 3.54}		\\
	F-RMSE   					& 0.454 $\pm$ 0.089	    & 0.577 $\pm$ 0.285 			& \textbf{0.348 $\pm$ 0.097} 	\\
    \midrule 
	\multicolumn{4}{c}{$\alpha=0.01$, $T=1200$}                                                                         \\
	\midrule
	CV space cov.				& 0.32 $\pm$ 0.03		& 0.22 $\pm$ 0.06				& \textbf{0.41 $\pm$ 0.09}		\\
	E-RMSE    					& 25.92 $\pm$ 1.11		& 27.29 $\pm$ 5.93				& \textbf{23.79 $\pm$ 6.05}		\\
	F-RMSE   					& 0.333 $\pm$ 0.023	    & 0.381 $\pm$ 0.048 			& \textbf{0.279 $\pm$ 0.035} 	\\
	\bottomrule 
	\end{tabular}
	\end{center}
    \footnotesize{\textsuperscript{\emph{a}} For ensemble-based uncertainty quantification, results are averaged over three independent runs; other methods use five runs. For two of five experiments with the ensemble-based method, the algorithm could not generate new configurations before reaching the maximal data set size.}
\end{table*}

\clearpage

\begin{figure*}[htbp]
    \centering
    \includegraphics[width=\textwidth]{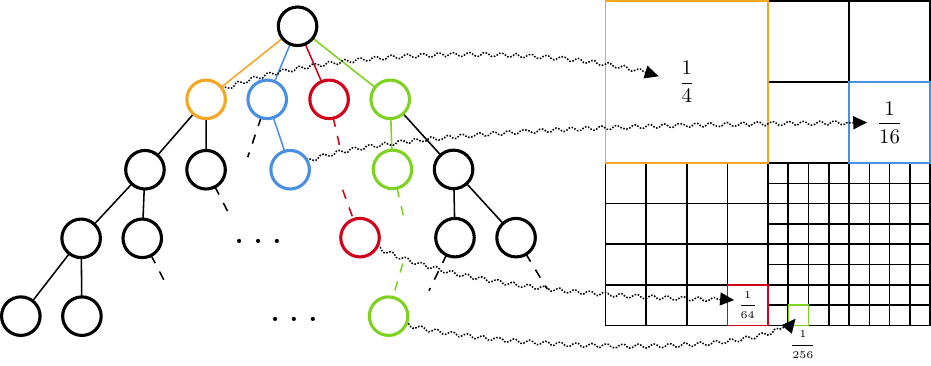}
	\caption{Example of a two-dimensional space partitioning with weights attached to each sub-partition.}
	\label{fig:tree}
\end{figure*}

\end{appendices}

\end{document}